\pdfoutput=1
\documentclass[fleqn,useAMS,usenatbib]{mnras}

\usepackage[T1]{fontenc}
\usepackage{ae,aecompl}

\usepackage{graphicx}
\usepackage{subfigure}

\usepackage{epstopdf}

\usepackage{amsmath}	
\usepackage{amssymb}	

\usepackage{txfonts}

\title[The First Spectra of Wolf Rayet Stars in M101]{The First
  Optical Spectra of Wolf Rayet Stars in M101 Revealed with Gemini/GMOS} 

\author[J.L.Pledger et al.] {J.L. Pledger $^{1}$\thanks{E-mail:
    jpledger@uclan.ac.uk}, M.M. Shara$^{2}$, M. Wilde$^{2}$,
  P.A. Crowther$^{3}$, K.S.Long$^{4}$, D. Zurek$^{2}$, \newauthor A.F.J. Moffat$^{5}$ \\
  $^{1}$Jeremiah Horrocks Institute for Mathematics, Physics \&
  Astronomy, University of Central
  Lancashire, Preston, PR1 2HE, UK. \\
  $^{2}$Department of Astrophysics, American Museum of Natural
  History,
  Central Park West and 79th Street, New York, NY 10024-5192 \\
  $^{3}$Department of Physics \& Astronomy, University of Sheffield,
  Hounsfield Road, Sheffield, S3 7RH, UK \\
$^{4}$ Space Telescope Science Insitute, 3700 San Martin Drive,
  Baltimore, MD 21218, USA \\  
  $^{5}$ D\'epartement de Physique, Universit\'e de Montr\'eal, CP
  6128 Succ. C-V, Montr\'eal, QC, H3C 3J7, Canada }

\begin{document}
\date{}
\pagerange{\pageref{firstpage}--\pageref{lastpage}} \pubyear{}

\maketitle

\label{firstpage}

\begin{abstract}
  Deep narrow-band HST imaging of the iconic spiral galaxy M101 has
  revealed over a thousand new Wolf Rayet (WR) candidates.  We report
  spectrographic confirmation of 10 He\,{\sc ii}-emission line sources
  hosting 15 WR stars. We find WR stars present at both sub-- and
  super--solar metalicities with WC stars favouring more metal-rich
  regions compared to WN stars. We investigate the association of WR
  stars with H\,{\sc ii} regions using archival HST imaging and
  conclude that the majority of WR stars are in or associated with
  H\,{\sc ii} regions. Of the 10 emission lines sources, only
    one appears to be unassociated with a star-forming region. Our
  spectroscopic survey provides confidence that our narrow-band
  photometric candidates are in fact bonafide WR stars, which will
  allow us to characterise the progenitors of any core-collapse
  supernovae that erupt in the future in M101.
\end{abstract}


\section{Introduction}
Wolf-Rayet (WR) stars are the descendants of massive O stars. They
display powerful stellar winds, resulting in unique, broad
emission-line spectra which allow us to detect WR stars in both Local
Group \citep{Moffat1983, Massey1998, Neugent2011} and more distant
star-forming galaxies \citep{Conti1990, Hadfield2005}. The strong
stellar winds strip the outer layers, revealing the nuclear by-products
of central core burning in the photosphere of the star. The products of
hydrogen burning via the CNO cycle are seen at the surface in
nitrogen-rich WN stars while the products of helium burning are
revealed in more evolved carbon-rich WC stars and even more evolved
oxygen-rich WO stars.

Stellar evolutionary theory suggests that massive stars end their
lives as core-collapse supernovae, providing chemical enrichment
within the interstellar medium (ISM). Indeed, Red Supergiants (RSGs)
between 8--16M$_{\odot}$ have been directly observed to produce
hydrogen--rich Type II-P ccSNe \citep{Mattila2010,
  Smartt2009}. Theoretical models have also supported WN and WC stars
as progenitors of H--poor Type Ib and H-- and He--poor Type Ic ccSNe,
respectively, since such elements have been stripped from the star via
strong stellar winds \citep{Groh2013}. 

Since most O and WR stars are in massive binaries, one expects the
ccSN of the initially more massive star to lead to a Black Hole (BH)
$+$ O star , followed by a BH$+$WR star binary. Examples of such
systems include Cyg X-1 \citep{Gies2003} as well as IC10 X-1
\citep{Prestwich2007, Silverman2008}, NGC300 X-1 \citep{Crowther2010}
and M33 X-7 \citep{Orosz2007}. However such systems are rare and
normally the asymmetric SN explosion combined with the orbital motion
in the original binary will lead to two runaway stars, one of which
will evolve as a single WR star.

One problem with this massive star evolution scenario is the growing
lack of direct detections of Type Ib/c progenitors, which has called
into question the WR-ccSNe connection \citep{Smartt2009}. For example,
pre-explosion images at the location of Type Ic SN 2002ap in M74
failed to reveal a progenitor down to M$_{B}$\,=\,--4.2\,mag. A binary
scenario was, instead, favoured by \citet{Crockett2007}. In addition,
\citep{Eldridge2013} suggest that \textit{all} Type Ib/c progenitors
are the result of low--mass helium cores produced during binary
evolution, proposing that the WR stars may produce black holes with
which no visible component is associated. However, \citet{Smith2011}
show that the inclusion of both binary and single evolutionary paths
is optimum for reproducing the observed ccSNe rate from a standard
initial mass function whilst \citet{Cao2013} report that the
photometric properties of the progenitor of Type Ib SN iPTF13bvn are
consistent with those of a WR star. Similarly, analysis of the light
curve and ejecta mass of the Type Ibn SN OGLE-2014-SN-131 suggests a
massive WR progenitor, although no direct detection has been made
\citep{Karamehmetoglu2017}.

We need a catalogue of $\sim$ 20,000 WR stars to definitively
demonstrate that WR stars are, or are not, the progenitors of Type Ibc
SNe. This is because the mean lifetime of a WR star is $\sim$300,000
years so assuming that each star has an average of $\sim$150,000 years
left as a WR star, one WR star from a sample of 20,000 should explode
as a Ibc SNe within 7 years, although one within 50 years would
  be a more conservative estimate. Conversely, the demonstration that
Ibc SNe come from objects that are fainter in He\,{\sc ii} than the
faintest known WR stars, such as those in the metal-poor Small
  Magellanic Cloud, would indicate that most WR stars do not end
their lives as supernovae.

M101 is an ideal galaxy with which to study the content and associated
environments of WR stars. At a distance of 6.4Mpc \citep{Shappee11} we
are still able to resolve large clusters, and the face-on orientation
is favourable to an acceptable level of extinction, typically between
0.25--1\,mag \citep{Cedres2002}. Based on H$\alpha$ imaging,
\citet{Kennicutt1998} determined a Star Formation Rate
(SFR)$\sim$1.7\,M$_{\odot}$yr$^{-1}$. This is a lower limit as they
note that coverage is not complete, which is consistent with the upper
value of SFR$\sim$3.3\,M$_{\odot}$yr$^{-1}$ found by
\citet{Jarrett2013} from UV and IR observations. Based on a Milky Way
SFR$\sim$2\,M$_{\odot}$yr$^{-1}$ \citep{Chomiuk2011} and a predicted
Galactic WR population $\sim$1900 \citep{Rosslowe2015} we expect
$\sim$3000 WR stars in M101.

In addition, M101 has a strong metallicity gradient, extending from a
super-solar metallicity of log(O/H)$+$12\,=\,$+$8.9 in the central
regions to log(O/H)$+$12\,=\,$+$7.5 in the outer regions
\citep{Rosa1994,Bresolin2007,Cedres2004}, allowing us to test
N(WR)/N(O) and N(WN)/N(WC) ratios as a function of ambient metallicity
predicted from stellar evolutionary models \citep{Eldridge2006,
  Meynet2005}.

Currently the only published spectrum of a WR star in M101 is that of
the WR counterpart to ULX-1 which is identified as a WN8 star by
\citet{Liu2013}. In this paper we present new Gemini/GMOS spectroscopy
of 10 WR candidates identified from F469N narrow-band imaging using
WFC3 on HST. In Section \ref{observations} we describe details of our
observations, followed by our data reduction techniques in Section
\ref{reductions}. Our results are presented in Section \ref{results},
along with H$\alpha$ analysis in Section \ref{hii}. Discussion of our
non-detections is presented in Section \ref{nondetections} followed by
a summary in Section \ref{discussion}.

\section{Observations}
\label{observations}

M101 was observed in Cycle 17 by the Hubble Space Telescope, Wide
Field Camera 3 (HST/WFC3), under program ID 11635
(PI. Shara). Eighteen pointings, each with a 2.7$\times$2.7~arcmin
field of view, were obtained using the narrow-band F469N filter (tuned
to He\,{\sc ii}$\lambda$4684\,\AA ) to identify Wolf-Rayet (WR)
candidates. A detailed account of the image analysis and first results
is presented in \citet{Shara2013}. At a distance of 6.4\,Mpc
\citep{Shappee11} our M101 imaging has a spatial scale of
$\sim$ 1.24pc\,pixel$^{-1}$ based on HST/WFC3 0.04\,arcsec\,pixel$^{-1}$.

Full details of our imaging survey and initial results are
  presented in \citet{Shara2013}. In summary, we found 25 F469N
bright objects as WR candidates and 71 candidate Red Supergiant (RSG)
stars in one of our 18 HST fields.  We noted their distribution,
namely that the WR stars were much more concentrated in the young
star-forming complex NGC 5462 compared to the RSG.  This is all
predicated on the assumption that the F469N bright objects
\textit{really} are WR stars. This demands follow-up spectroscopy of
the WR candidates, which will also allow us to distinguish between WN
and WC stars. This is also a test of stellar evolutionary theory which
predicts more WC stars in metal-rich regions of the galaxy.

As a first step we obtained observations with the Gemini-North Multi-
Object Spectrograph (GMOS) under program ID GN-2012A-Q-49 (PI:
Bibby). In total, 7 GMOS masks were designed, providing good coverage
of the central region of M101. By observing additional
spectrophotometric standard stars we were also able to
flux calibrate the spectra to ascertain whether each WR candidate is
in fact multiple WR stars, as is commonly seen in previous studies
e.g. in NGC 2403 \citep{Drissen1999} and NGC 7793 (Bibby \& Crowther
2010).

\subsection{Pre-Imaging}
\label{pre-imaging}

The 5$\times$5\,arcmin GMOS field is approximately twice the size of
the HST/WFC3 field of view and as such each GMOS pointing covers
several HST pointings. To design the MOS masks from the WFC3/F469N
imaging would have required multiple pointings to be combined onto a single
mosaic image. Unfortunately this was not possible as the F469N fields did
not overlap sufficiently. Consequently we opted to use GMOS imaging
to ensure that we could take full advantage of the $\sim$30 slits GMOS can
accommodate.

\begin{figure}
\centering
\includegraphics[width=\columnwidth]{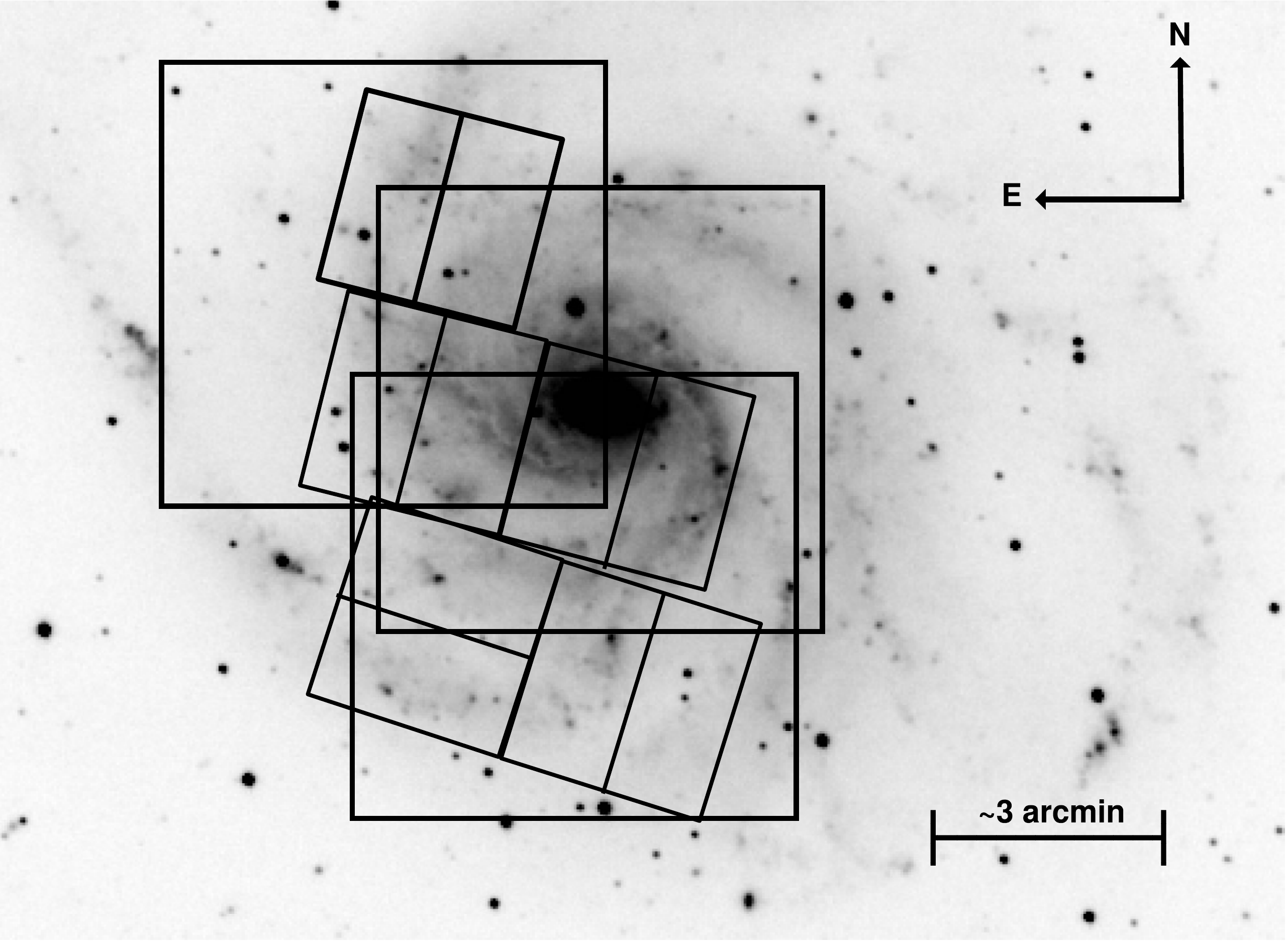}
\caption{An image of M101 taken with the KPNO Schmidt telescope
  trimmed to $\sim$16.5$\times$12 arcmin showing the location of the
  three GMOS pointings, each of 5.5$\times$5.5 arcmin and the
  corresponding WFC3 pointings of 2.7$\times$2.7 arcmin.}
\label{m101_fov}
\end{figure}

The Gemini Science Archive contained g-band imaging obtained on
February 21 2007, under program ID GN-2007A-Q-72 (PI: Chandar) which
covered the central and north-east regions of M101. However, in order
to sample the complete radial distribution of WR stars we obtained
additional pre-imaging on February 16 2012, covering the southern
central region of M101. The coverage of M101 with both HST and Gemini
is shown in Figure \ref{m101_fov}. The archival and new GMOS imaging
had similar spatial resolutions of $\sim$1.0\,arcsec and
$\sim$1.1\,arcsec, respectively.

\subsection{GMOS spectroscopy}

Gemini Multi-Object Spectroscopy of WR candidates in M101 was obtained
during April-June 2012 in seeing conditions ranging between $\sim$0.7
and $\sim$1.0~arcsec. The R150 grism was used, with no blocking filter,
to allow a large wavelength range to include numerous stellar and
nebular diagnostic lines. Standard dithering techniques were used with
a central wavelength of 510nm and 530nm to provide full coverage
across the chip gaps on GMOS in the spectral dimension.

MOS masks were designed using the Gemini mask preparation software and
co-ordinates were transformed from HST/WFC3 imaging to Gemini
pre-imaging. Any gaps in the MOS mask design were filled with slits
placed on H\,{\sc ii} regions to maximise the science output;
results of these regions will be presented in a future paper.

The spectral dispersion obtained $\sim$3.5\,\AA\,pixel$^{-1}$ and the
resolution was derived from nebular lines to be
$\sim$15\,\AA. Exposure times for MOS masks \#1--3 and \# 5--7 were
4$\times$2240 sec, while MOS mask \#4 was observed for 6$\times$2300
seconds as we tried to sample the faintest WR candidates, at the limit
of GMOS's capability. A summary of the spectroscopic observations is
presented in Table \ref{observations_info}.

\begin{table}
\centering
\caption{Observational log for Gemini/GMOS observations of M101
  obtained under program ID GN-2012A-Q49 (PI: Bibby). The number in parenthesis indicates the number of exposures obtained.}
\begin{tabular}{lccccc}

\hline 
Date & MOS &  Exposure &  $\lambda_c$      &  Airmass & Seeing \\
     & Mask & time (s)& (\AA) & & (arcsec)     \\
\hline
28 Feb 2012  & \# 1  & 2240 (2)     & 510       &  1.28   & 0.75 \\
28 Feb 2012  & \# 1  & 2240 (2)    & 530       &  1.22   & 0.87 \\
3 Apr 2012   & \# 2  & 2240 (2)    & 510       &  1.30   & 0.73 \\
12 Apr 2012  & \# 2  & 2240 (2)    & 530       &  1.43   & 0.73 \\
13 Apr 2012  & \# 3  & 2240 (2)     & 510       &  1.29   & 0.93 \\
26 Apr 2012  & \# 3  & 2240 (2)    & 530       &  1.24   & 0.73 \\
26 Apr 2012  & \# 4  & 2300 (3)    & 510       &  1.25   & 0.73 \\
20 May 2012  & \# 4  & 2300 (2)   & 530       &  1.34   & 0.71 \\
21 May 2012  & \# 4  & 2300 (1)    & 530       &  1.34   & 0.80\\
21 May 2012  & \# 5  & 2240 (2)    & 510       &  1.22   & 0.80 \\
21 May 2012  & \# 5  & 2240 (1)    & 530       &  1.23   & 0.80\\
29 Jun 2012 & \#5    & 2240 (1)    & 530       &  1.26    & 0.79 \\
27 Jun 2012  & \# 6  & 2240 (2)    & 510       &  1.29   & 0.59\\
29 Jun 2012  & \# 6  & 2240 (1)    & 530       &  1.22   & 0.79 \\
29 Jun 2012  & \# 7  & 2240 (2)    & 510       &  1.36   & 0.79 \\
30 Jun 2012  & \# 7  & 2240  (2)   & 530       &  1.23   & 0.75  \\
\hline

\end{tabular}
\label{observations_info}
\end{table}

\section{Data Reduction \& Calibration}
\label{reductions}

Spectroscopic data reduction was performed using standard Gemini
\textsc{iraf} \footnote{IRAF is distributed by the National Optical
  Astronomy Observatories, which are operated by the Association of
  Universities for Research in Astronomy, Inc., under cooperative
  agreement with the National Science Foundation.}  reduction packages
including \textsc{gprepare} and \textsc{gsreduce}. Wavelength
calibration was performed from observations of a CuAr lamp using the
same instrumental setup and MOS mask as the science
observations. 

In order for flux calibration to be possible we obtained observations
of the spectrophotometric standard HZ44 \citep{Oke1990} on the same
night as MOS mask \#1. The second order contamination from the R150
grism prevented the Gemini \textsc{iraf} software from being used so
flux calibration was achieved using the \textsc{starlink} package
\textsc{figaro} instead \citep{Shortridge2004}. The wavelength range was trimmed to
4000--7000 \AA\, to remove the majority of the second order contamination
and both the observed and tabulated standards were fit with a spline
function and then divided by each other to produce a calibration
curve.

Since our observing program only allowed for one standard star
observation we cross-calibrated the other MOS masks using a common
object and determining a calibration factor. The exception was MOS
mask \#6 and \#7 which covered the most southernly region of M101 and
which did not contain the calibration object so an additional object
was used. Unfortunately this object was too faint to extract with the
Gemini packages so cross-calibration could not be performed in this
way, however the calibration factor for the other MOS masks were
relatively consistent so the average was used.

To achieve an absolute flux calibration we have to account for slit
losses since not all of the WR light will enter the slit. This was
achieved by comparing the empirical photometric magnitude of each
source to its spectroscopic magnitude calculated by convolving the
observed spectrum with the filter bandpass. For masks \#1,2,5,6 we
found the fraction of light in the slit to be 0.86, and 0.76 for mask
\#4. No WR sources were extracted from Mask \#7 so no slit losses were
determined. The majority of the synthetic magnitudes revealed a
brighter source than in the HST imaging, suggesting that the spectra was
contaminated by additional sources. This is not surprising, or unexpected
since the GMOS slit width is significantly larger than the HST/WFC3
resolution.

\section{Results}
\label{results}

The GMOS spectra were analysed using the Emission Line Fitting routine
(ELF) within \textsc{dipso} to measure line fluxes of both stellar and
nebular emission lines\citep{Howarth2004}.

\subsection{Nebular Analysis}
\label{neb_analysis}

Where possible we made local estimates of extinction using Balmer line
ratios, assuming Case B recombination for an optically thick nebula
and a Standard Galactic Extinction Law. We used the H$\alpha$/H$\beta$\,=\,2.86
line ratio for all nebular sources apart from \#48 and \#112 where the
H$\beta$ line unfortunately fell within the chip gap due to the source
being unable to be extracted in one or more of the exposures. In both cases,
the H$\gamma$$\lambda$4343 emission line was present so we used the
H$\alpha$/H$\gamma$\,=\,0.164 line ratio \citep{Osterbrock2006} to
determine extinction. The line fluxes and resulting E(B-V) values for
each nebular source are presented in Table \ref{neb_fluxes}. We note
that where there were no nebular lines present we adopt a value of
E(B-V)\,=\,0.44\,mag from \citet{Lee2009}, which is consistent
with our average of E(B--V)\,=\,0.42$\pm$0.07\,mag.

For estimates of the metallicity of the nebular regions we rely on the
strong line method of \citet{Pettini2004} using the [N\,{\sc
    ii}]/H$\alpha$ ratio as a proxy for metallicity since the [O\,{\sc
    ii}]$\lambda$3727 line was not detected in our spectra. Given the
proximity of [N\,{\sc ii}]\,$\lambda$6583 to the H$\alpha$\,$\lambda$6568
emission line, the narrow-band filter includes a contribution from
both, and hence we need to correct for this contribution in our
measurement of L(H$\alpha$). Where nebular lines are present in our
GMOS spectra the [N\,{\sc ii}]/H$\alpha$ ratio has been determined
directly whilst for others a ratio of [N\,{\sc ii}]/H$\alpha$\,=\,0.54
has been taken from \citet{Kennicutt2008}. This is slightly higher
than our average [N\,{\sc ii}]/H$\alpha$ ratio of 0.37$\pm$0.07 albeit
from a small sample size. 

Using the N2 method from \citet{Pettini2004} to determine the
metallicity of those H\,{\sc ii} regions hosting WR stars we find a
range of log(O/H)$+$12\,=\,8.41--8.80 with errors of
$\pm$0.4\,dex. Using the O3N2 indicator we find a similar range of
log(O/H)$+$12\,=\,8.56--8.92 ($\pm$0.25 dex) with regions in general
agreement by $\pm$0.15\,dex. The derived metallicity for Source \#1024
differs by $\sim$0.3\,dex between methods but is still in agreement
within errors. Overall we find an average value of
log(O/H)$+$12\,=\,8.66$\pm$0.24\,dex for the metallicity of H\,{\sc
  ii} regions in M101. This is consistent with the metallicities
derived in \citet{Bresolin2007} and \citet{Rosa1994} but is slightly
higher than we would expect from the metallicity gradient of
8.769($\pm$0.06)--0.90($\pm$0.08)(R/R$_{0}$) found by \citet{Cedres2004}. One
  explanation may be that all of our regions lie within the central
  regions of M101 with R/R$_{25}$ of 0.04 to 0.22, based on an
  inclination of 18 degrees, PA\,=\,45 degrees \citep{Kenney1991} and
  a distance of 6.4Mpc. In contrast, the majority of the 90 regions used by
  \citet{Cedres2004} are within R/R$_{0}$\,=\,0.30.

\begin{table}
\caption{Nebular analysis of H\,{\sc ii} regions hosting WR
  stars. Fluxes and Intensities are given relative to
  H$\alpha$\,=\,100 where H$\alpha$ is in units of
  $\times$10$^{-16}$ergs$^{-1}$ cm$^{-2}$ and extinctions are
  determined using Balmer line ratios of H$\alpha$/H$\beta$\,=\,2.86
  or H$\gamma$/H$\alpha$\,=\,0.165 \citep{Osterbrock2006}.  The
  metallicity of the region is determined using both the N2 and O3N2
  methods of \citet{Pilyugin2005}. c/g indicates that the line was in
  the chip gap so no flux was measured} \centering

\begin{tabular}{l@{\hspace{1.5mm}}l@{\hspace{1.5mm}}c@{\hspace{2mm}}c@{\hspace{2mm}}c@{\hspace{2mm}}c@{\hspace{2mm}}c@{\hspace{2mm}}c@{\hspace{2mm}}c@{\hspace{1mm}}}
\hline 
 &  & \multicolumn{7}{c}{Source ID} \\
ID &    $\lambda$(\AA)        & 48   & 1012 & 1016  & 49   & 112  & 1024  & 2053  \\
\hline

\hline
F(H$\delta$)          & 4100    & 7.71 & --     & --      & --   & --   &  --    & --  \\
I(H$\delta$)          & 4100    & 9.67  & --     & --      & --   & --   &  --    & -- \\

F(H$\gamma$)       & 4343    & 13.3 & 16.5 & 6.47 & --   & 7.19  & 10.9  & 5.57 \\
I(H$\gamma$)       & 4343    & 16.3  & 18.8 & 17.9 & --   & 15.9  & 21.7  & 24.0 \\

F(H$\beta$)           & 4861    &  c/g & 32.0 & 19.0 & 30.8  & c/g  & 21.7 & 12.7 \\
I(H$\beta$)           & 4861    &  c/g  & 35.1 &  35.2 & 33.9 & c/g  & 35.3 & 9.03 \\

F([O\,{\sc iii}])       & 4959    & 22.1 & --    & --    & 9.74 & 3.03 & 0.37 & 6.73 \\
I([O\,{\sc iii}])        & 4959    & 25.5 & --    & --    & 10.4 & 5.07 & 0.51 & 21.2 \\

F([O\,{\sc iii}])       & 5007    & 53.8 & 1.68 & 3.96 & 49.5 & 8.94 & 2.80 & 7.71 \\
I([O\,{\sc iii}])       & 5007     & 61.7 & 1.59 & 7.16 & 54.7 & 14.8 & 4.21 & 23.1 \\

F(H$\alpha$)         & 6563    & 100  & 100  & 100    & 100 & 100   & 100 & 100 \\
I(H$\alpha$)          & 6563    & 100  & 100  & 100    & 100 & 100   & 100 & 100 \\

F([N\,{\sc ii}])        & 6583    & 31.2 & 34.8 & 64.6  & 66.6 & 30.0 & 22.6 & 14.7 \\
I([N\,{\sc ii}])        & 6583    & 31.2  & 34.7 & 63.3  & 66.7 & 29.9 & 22.5 & 14.2 \\

\hline
F(H$\alpha$)       & 6563    & 42.0 & 71.3 & 236.6 & 3.80 & 143.2 & 66.7 & 224.3 \\
I(H$\alpha$)        & 6563    & 57.7 & 86.5 & 929.1 & 4.98 & 475.7 & 189.4 & 2058.0 \\
\hline

E(B-V)                  &            & 0.14 & 0.09 & 0.59  & 0.12 & 0.53  & 0.46 & 0.98  \\
$[$NII$]$/H$\alpha$ &      & 0.31 & 0.35 & 0.63  & 0.67 & 0.30  & 0.22 & 0.14  \\
log(O/H)$+$12$^{1}$ &       & 8.61 & 8.64 & 8.79  & 8.80 & 8.60  & 8.53 & 8.41  \\
O\,{\sc iii}/H$\beta$  &      & --   & 0.40 & 0.20  & 1.61 & --    & 0.12 & 0.60  \\
log(O/H)$+$12$^{2}$  &       & --   & 8.74 & 8.92  & 8.64 & --    & 8.85 & 8.56  \\
log(O/H)$+$12$^{3}$  &       & 8.61 & 8.69 & 8.85  & 8.72 & 8.60  & 8.69 & 8.49  \\ 
R/R$_{25}$                   &       & 0.18 & 0.11 & 0.13 & 0.16 & 0.16 & 0.08 & 0.22 \\
log(O/H)$+$12$^{4}$  &      & 8.60 & 8.66 & 8.64 & 8.62 & 8.62 & 8.67 & 8.56 \\

\hline
\multicolumn{9}{l}{$^{1}$ from N2, $^{2}$ from O3N2, $^{3}$ Mean
  log(O/H)$+$12, $^{4}$ log(O/H)$+$12} \\
\multicolumn{9}{l}{based on gradient in \citet{Cedres2004} } \\
\hline

\end{tabular}
\label{neb_fluxes}
\end{table}

\subsection{Stellar Analysis}

Wolf-Rayet candidates can be identified from narrow-band He\,{\sc ii}
imaging and photometry but they can only be confirmed as bonafide WR
stars from spectroscopy. The photometric properties of each WR
candidate presented in this paper are taken from HST imaging (see
\citet{Shara2013} for details) and are presented in Table
\ref{wr_photometry}. Spectroscopy allows us to identify Nitrogen-rich
(WN) stars predominantly from their He\,{\sc ii}\,$\lambda$4686 emission
line while carbon-rich (WC) stars are seen to be dominated by C\,{\sc
  iii}\,$\lambda$4650 and C\,{\sc iv}\,$\lambda$5808. Rarer oxygen-rich
(WO) stars are identified by O\,{\sc iv}\,$\lambda$3811-34 emission
lines; however, these lines lie outside the spectral range of our
observations.

We have spectroscopically confirmed 10 WR sources within M101 as
indicated in Figure \ref{m101_fov_wr}. For completeness we present more detailed
finding charts in Appendix \ref{finding_charts_all}. As for the nebular analysis we
used the \textsc{dipso} emission line fitting routine ELF to
measure the flux from the stellar source, results of which are
presented in Table \ref{wr_spec_tab}. Typical errors for the flux
measurements were higher than expected at $\sim$$\pm$20\% for strong
lines such as He\,{\sc ii} and C\,{\sc iv}. This is likely due to the
fact that the 1$\arcsec$ slit covers $\sim$30pc in physical size so
there are more surrounding massive stars to contribute to the
continuum and dilute the WR emission lines, producing a weak excess.


\begin{figure}
\centering
\includegraphics[trim=0 40 0 0, clip, width=\columnwidth]{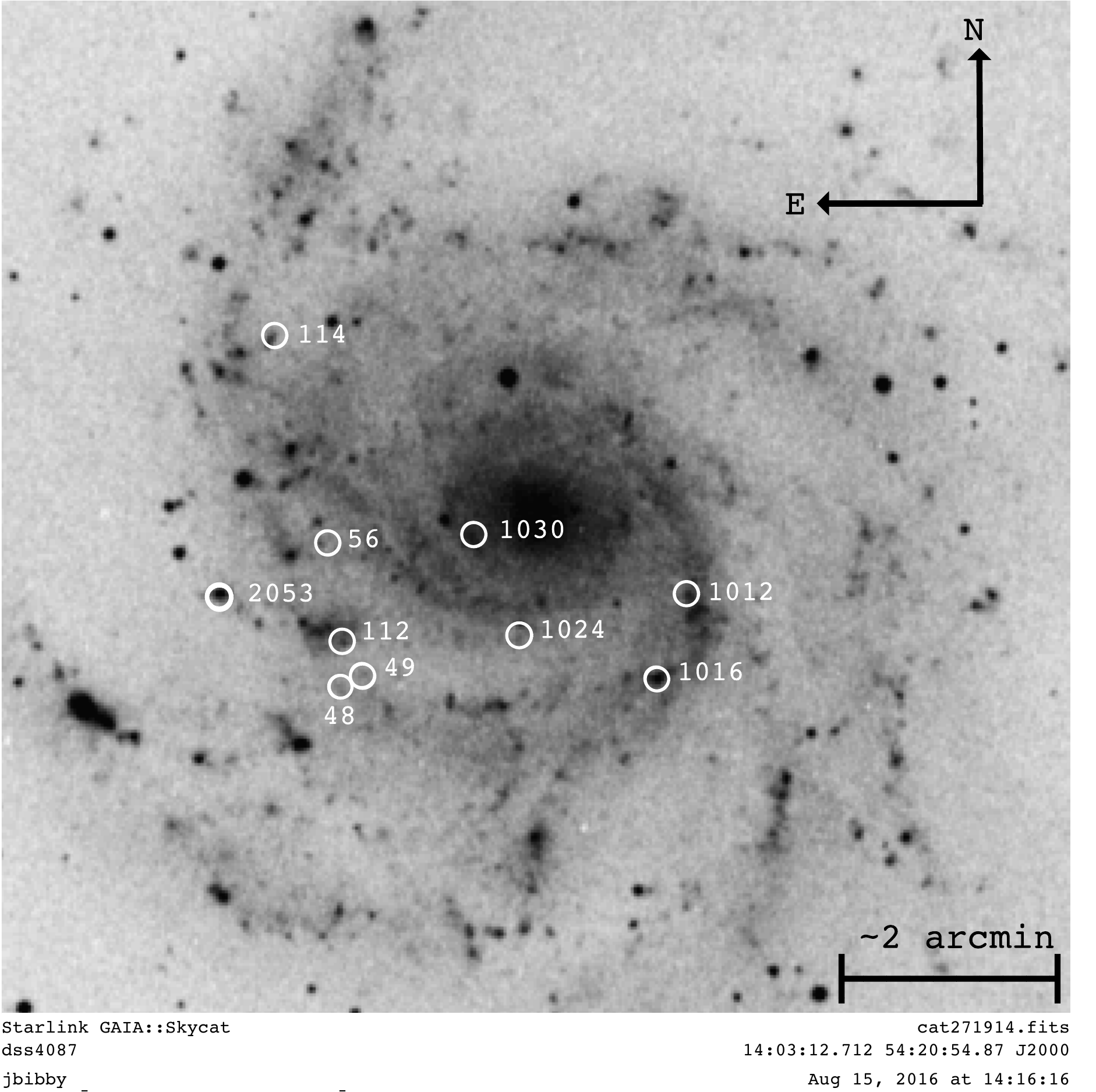}
\caption{An archival DSS image of M101 showing the location of the
  spectrographically confirmed WR stars along with their ID's. }
\label{m101_fov_wr}
\end{figure}

Previous studies use the line ratios of \citet{Smith1996} and
\citet{Smith1990} to classify WN and WC subtypes, respectively along
with the LMC line luminosities from \citet{Crowther2006} to estimate
the number of WN and WC stars within each source. However, M101 is
known to be a metal-rich galaxy \citep{Bresolin2007} and we
consistently find our WR sources in solar and super-solar regions (see
Section \ref{neb_analysis}) so instead we use Galactic templates from
observations of WR stars in \citet{Rosslowe2015} to estimate the
number of WR stars in M101. The templates are adjusted to a Galactic distance of
  1kpc and assume an average extinction for each subtype. The
  properties of each star used in each template are summarised in
  Appendix \ref{galactic}. We note that for super-solar metallicity
  regions, such as \#1016, the WR emission is expected to be stronger
  therefore the number of WR stars stated is an upper limit.

\begin{figure}
\centering 
\subfigure[Source \#112]{\includegraphics[width=0.7\columnwidth, angle=-90]{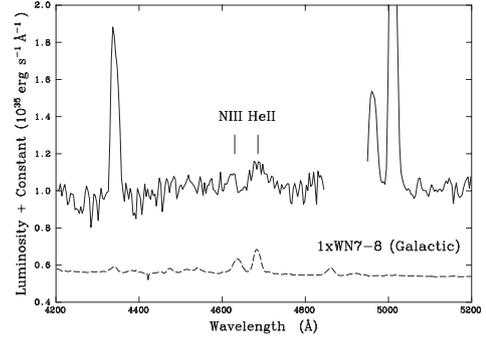}}
\subfigure[Source \#1012]{\includegraphics[width=0.7\columnwidth, angle=-90]{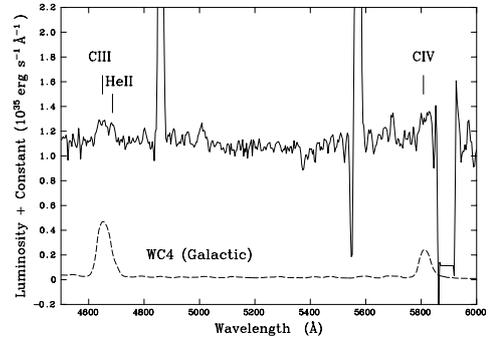}}
\subfigure[Source \#56]{\includegraphics[width=0.7\columnwidth, angle=-90]{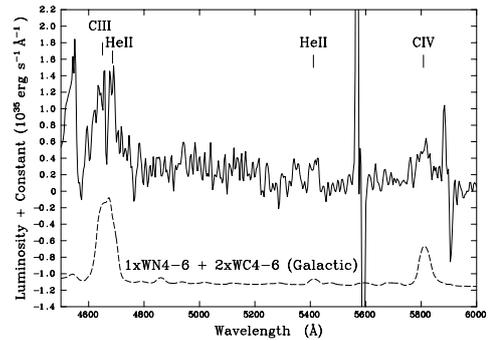}}

\caption{Normalised, extinction and distance corrected spectra of WR
  sources in M101 along with Galactic WR template spectra from
  \citet{Rosslowe2015}. The units shown are Luminosity per unit
  wavelength. Source a) shows a single WN7-8 star, b) a
  single WC4 star and c) a composite of one WN4-6 and two WC4-6
  stars. The template spectra are offset from the observed spectra for
  clarity. We note the feature just short of 4600\AA in Source \#56 is
due to the chip gap extrapolation with the Gemini packages and is not real.}

\label{spectra_source}
\end{figure}

Within our 10 WR sources we have identified 11 WC stars and 4 WN
stars.  Figure \ref{spectra_source} shows an example of our spectra
along with the Galactic template stars whilst additional spectra are
presented in Appendix \ref{WR_spectra_templates}. We find mid- and
late-type WN stars \citep{Smith1996} in our M101 survey but no
early-type WN stars, which is consistent with the online WR Galactic
catalogue hosted by Crowther \footnote{P.Crowther hosts an up-to-date
 Galactic WR catalogue at
 http://pacrowther.staff.shef.ac.uk/WRcat/index.php} (and references therein), which
shows that over 90\% of the WR stars classified in the Milky Way to have
subtype later than WN5. For our WC stars we see both WC4-6 and WC7-8
subtypes, indicating a trend towards later types but not as clear as
for the Galactic sample. Again, small number statistics prevent any firm conclusion.

Interestingly, we see our WN stars located in regions of sub-solar
metallicity between log(O/H)$+$12\,=\,8.49--8.61, whereas our WC stars
favour slightly more metal-rich regions of
log(O/H)$+$12\,=\,8.69--8.85. This is consistent with stellar
evolutionary theory, which predicts that the WC/WN ratio should increase in metal-rich
regions due to stronger stellar winds and enhanced stripping of the
outer layers \citep{Eldridge2006, Meynet2005}. Since our WR
stars are all within the inner galaxy we are cautious not to make any
direct comparisons but the presence of WN7-8 stars in the most central
regions is consistent with the presence of WN9 stars in the inner
regions of our own Galaxy. Obtaining spectra of
additional WR candidates will allow us to test this theory further.
We note that 3 of our WR sources, including the composite WN$+$WC
source, do not exhibit nebular lines in their spectra so we cannot
determine their local metallicity.

\subsection{Synthetic Magnitudes}
\label{synthetic_mag}

As described in Section \ref{reductions} we used synthetic magnitudes
to determine the slit losses for each MOS mask. When comparing the
magnitudes derived from the ground-based spectra we find that the
synthetic He\,{\sc ii} magnitudes are at least 1 magnitude greater than
the WFC3/F469N magnitudes in all but one source. This suggests that
most of our spectra have multiple stars in the slit which is
unsurprising given the 0.8$\arcsec$ versus 0.1$\arcsec$ resolution
difference. This is further highlighted in the 2$\times$2\,arcmin
images of each source presented in Appendix \ref{post_stamps} where
additional sources can be seen within 1$\arcsec$ (our GMOS slit size) of
each WR candidate with the exception of source \#56. Unsurprisingly,
Source \#56 is the only source where the WR component dominates the
spectra, indicating an He\,{\sc ii} excess of --1.16 mag. The crowding
owing to the 1$\arcsec$ slit width results in a synthetic He\,{\sc ii}
excess of a m(He\,{\sc ii}) - m(Continuum) excess $\leq$--0.15\,mag for
all but one source (\#49). This excess is within typical errors of
ground-based photometry and as such these WR stars would not have been
identified as WR candidates from ground-based imaging such as that
presented in \citet{Bibby2012}.

\subsection{Non Detections}
\label{nondetections}
In total 208 WR candidates were included in the 7 MOS mask designs.
There was an error in the co-ordinate transformation from the HST
imaging to the GMOS pre-imaging so one MOS mask (\#7) yielded no WR
spectra. In addition, we used one MOS mask (\#4) to try to
acquire spectra of the faintest WR stars with
m$_{F435W}$$\leq$24\,mag, however none of these 17 candidates in this
mask were extracted successfully. This reduced the total number of WR
candidates to 159 sources.

Out of these 19\% of spectra revealed nebular lines, 11\% absorption
lines, 7\% WR emission lines and 3\% had the He\,{\sc ii}\,$\lambda$4686
emission line located in a chip gap. We hoped to avoid diagnostic
lines falling into chip gaps by using 2 central wavelengths however
the combined spectra often was not clean enough to identify a WR
emission line confidently. Unfortunately the remaining spectra showed
only noise or could not be extracted. On further inspection, the
majority of our WR candidates had m$_{F435W}$$\leq$24\,mag in the
continuum so it is not surprising that they could not be extracted as
was found with MOS mask \#4.

No WR stars fainter than m$_{F435W}$\,=\,23.64\,mag were detected by
Gemini/GMOS and we note for future observations that this limits us to
the brighter stars in M101. Moreover, the photometry presented in
Table \ref{wr_photometry} indicates that the WR stars had
m$_{F435W}$-m$_{F469N}$ excesses $\geq$0.36\,mag suggesting that our
GMOS survey is favouring the strongest emission-line
stars. Nonetheless the WR stars we have detected are strongly
supportive of our identification of candidates as WR stars. We
expect that a number of the objects we were unabel to detect as WR
strs in these observations will ultimately be shown to be WR stars, as
their luminosity function suggests. To confirm this however, more
sensitive observations will be required.

\onecolumn

\begin{table}
  \caption{Photometric properties of the WR stars spectrographically
    confirmed in M101. The RA and DEC positions are taken from the WFC3
    F469N image.}
\centering
\begin{tabular}{l@{\hspace{1.9mm}}c@{\hspace{1.9mm}}c@{\hspace{1.9mm}}c@{\hspace{1.9mm}}c@{\hspace{1.9mm}}c@{\hspace{1.9mm}}c@{\hspace{2mm}}c@{\hspace{1.9mm}}c@{\hspace{1.9mm}}c@{\hspace{1.9mm}}c@{\hspace{1.9mm}}c@{\hspace{1.9mm}}c@{\hspace{1.9mm}}}
\hline
ID & RA & Dec & R/R$_{25}$ & m$_{f469N}$ & err & m$_{F435W}$ & err & m$_{F555W}$ & err & m$_{F814W}$ & err & Mask \#  \\
\hline
 56 & 14:03:25.915 & 54:20:39.01    & 0.14 & 22.59 & 0.03 & 23.08 & 0.02 & 23.94 & 0.02 & 25.23 & 0.03 & 1  \\

1030 & 14:03:16.533 & 54:20:44.45   & 0.04 & 22.92 & 0.11 & 23.64 & 0.46 & 24.67 & 0.05 & -- & -- & 1   \\

48 & 14:03:25:286 & 54:19:17.87     & 0.18 &  22.61 & 0.03 & 23.56 & 0.03 & 24.30 & 0.03 & 25.33 & 0.05 & 1   \\

1012 & 14:03:02.861 & 54:20:11.61   & 0.11 &  22.16 & 0.02 & 22.62 & 0.02 & 23.59 & 0.03 & 24.60 & 0.05 & 2   \\

1016 & 14:03:04.711 & 54:19:24.14 & 0.13 &  22.40 & 0.08 & 23.20 & 0.05 & 24.12 & 0.05 & 24.68 & 0.08 & 2  \\

49 & 14:03:23.604 & 54:19:24.71 & 0.16  & 22.43 & 0.04 & 23.49 & 0.01 & 24.33 & 0.02 & 25.65 & 0.03 & 4  \\

112 & 14:03:24.893 & 54:19:43.16 & 0.16 &  22.18 & 0.09 & -- & -- & 22.71 & 0.04 & 23.96 & 0.10 & 5   \\
114 & 14:03:29.640 & 54:22:35.91 & 0.21 & 20.70 & 0.04 & 20.24 & 0.06 & 20.60 & 0.04 & 21.68 & 0.05 & 5   \\

1024& 14:03:13.561 & 54:19:47.72 & 0.08 & 22.67 & 0.08 & 23.03 & 0.02 & 23.81 & 0.03 & 24.74 & 0.05  & 6   \\

2053 & 14:03:32.816 & 54:20:07.79 & 0.22 & 24.83 & 0.10 & -- &-- &-- &-- &--  &-- & 6  \\
\hline

\end{tabular}
\label{wr_photometry}

\end{table}

\begin{table}
  \caption{Observed fluxes (F$_{\lambda}$) and corrected luminosities
    (L$_{\lambda}$) for WR stars in M101 confirmed with Gemini/GMOS
    observations. Errors on the line fluxes are shown in
    parenthesis. Sources are corrected for slit loss and dereddened
    using the extinction determined from the nebular emission in the
    spectrum. Where no nebular emission is present a value of
    E(B-V)\,=\,0.44 is taken from \citet{Lee2009} and luminosities are
    based on a distance of 6.4\,Mpc from \citet{Shappee11}. WN and WC
    numbers are derived from Galactic WR templates from
    \citet{Rosslowe2015}.}  \centering
\begin{tabular}{{l@{\hspace{2mm}}c@{\hspace{2mm}}c@{\hspace{2mm}}c@{\hspace{2mm}}c@{\hspace{2mm}}c@{\hspace{2mm}}c@{\hspace{2mm}}c@{\hspace{2mm}}c@{\hspace{2mm}}c@{\hspace{2mm}}c@{\hspace{2mm}}c@{\hspace{2mm}}c@{\hspace{2mm}}}}
\hline
&   & \multicolumn{6}{c}{F$_{\lambda}$ ($\times$10$^{-16}$ erg s$^{-1}$ cm$^{-2}$)} & & \multicolumn{2}{c}{L$_{\lambda}$ ($\times$10$^{36}$ erg s$^{-1}$)}  & &  \\
ID & R/R$_{25}$ &  F(N\,{\sc iii}) &  F(C\,{\sc iii}) & F(He\,{\sc
  ii}) & F(He\,{\sc ii}) & F(C\,{\sc iii}) & F(C\,{\sc iv}) & E(B-V) & L(He\,{\sc ii}) & L(C\,{\sc iv}) & N(WN) & N(WC) \\

   &  & $\lambda$4612-30 & $\lambda$4650 & $\lambda$4686 & $\lambda$5411 & $\lambda$5696 & $\lambda$5808 & &$\lambda$4686 & $\lambda$5808  &  \\
\hline 
56    & 0.14 &   --    &  1.27(0.14) & 1.35(0.14)  & 0.54(0.12) & --    & 1.64(0.14)     & 0.44   & 3.19    &   2.56   & 1$\times$WN4-6 & 2$\times$WC4-6 \\  
1030 & 0.04 &   --     &  1.10(0.17) & 1.22(0.23)  &  --        & 0.92(0.26)   & 0.76(0.19) & 0.44  & 2.84   & 1.23 & --       & 2$\times$WC7-8 \\
48    &  0.18 &   --   &   --        & 3.17(0.40)  &  --        & --    & --             &  0.14   & 2.62    &   --     & 1$\times$WN4-6 &   --           \\
1012 & 0.11 &  --      &  3.83(0.54) & 2.58(0.47)  &  --        & --    & 3.69(1.18)     &  0.09   & 1.68    &   2.24  & --      & 1$\times$WC4-6 \\
1016 & 0.13  &  --     &  11.6(1.68) & 6.13(1.56)  & --         & 6.27(0.68)   & 2.38(0.48) & 0.59 & 13.8    &   12.0   & --       & 1$\times$WC7-8\\
49    & 0.16 &   --    &  5.07(0.27) &   --        &   --       & --           & 1.88(0.12) & 0.12 & --      &   1.33   & --             & 3$\times$WC4-6 \\
112  & 0.16 &   --     &   --        & 1.93(0.40)  &   --       & --           & --         &  0.53   & 5.25 &   --     & 1$\times$WN7-8 &   --           \\
114  & 0.21 &   --     &  1.45(0.33) &   --        &  --        & --           & 1.23(0.32) &  0.44   & --      &   2.49   & --     & 1$\times$WC4-6 \\
1024 &  0.08 &   --    &  1.86(0.30) &   --        &  --        & 1.00(0.15)   & 0.24(0.11) &  0.46   & --      &   0.28   & --     & 1$\times$WC7-8 \\
2053 & 0.22 & 6.3(0.11)&   --        & 6.59(0.16)  &  --        & --           & --         &  0.98   & 90.7    &   --   & 1$\times$WN7-8 & --              \\
\hline

\end{tabular}
\label{wr_spec_tab}
\end{table}

\twocolumn

\section{WR Stars in H\,{\sc ii} regions}
\label{hii}

H\,{\sc ii} regions are formed and powered by the ionising
  radiation produced by massive O stars. The number of O stars and
  amount of ionising radiation influences the size of the H\,{\sc ii}
  region. Classical H\,{\sc ii} regions contain only a handful of O
  stars and are typically of order 10pc in diameter. Giant H\,{\sc ii}
  regions are of order $\sim$100pc in size, hosting $\sim$100 O7\,{\sc v}
  stars. However, giant H\,{\sc ii} regions often host
  separate star-forming regions of different sizes and ages so there
  is not a direct size-luminosity correlation. Since only O stars are
  capable of producing such regions, one would expect O stars, WR
  stars and resulting core-collapse supernovae to be associated with
  such H\,{\sc ii} regions. However, in the Milky Way only 27\% of WR
  stars are seen to be associated with such regions
  \citep{Crowther2015a} which is most likely due to the fact that they
  are ejected or that the star-forming region is unbound so dissolves
  quickly \citep{Crowther2013}. It is also possible that the H\,{\sc
    ii} region fades quicker than the average lifetime of the WR stars
  or that the cluster itself may still be embedded due to its young
  age \citep{Gvaramadze2012}; rarely these stars might also form in
  isolation.

In addition to investigating the association with H\,{\sc ii}
  regions we can also estimate the O star population itself. We have
used flux calibrated KPNO/JAG H$\alpha$ images from
\citet{Knapen2004}, corrected for extinction and [N\,{\sc ii}]
contamination, to determine the H$\alpha$ luminosity for each
region. Using the relation of \citet{Kennicutt1998} we can determine
the number of ionising photons from the strength of the H$\alpha$
emission which in turn allows us to estimate the number of O stars
present, assuming 10$^{49}$\,photons\,s$^{-1}$ for an O7V star
\citep{Vacca1994}. For six of our WR sources we can compare
  the WR/O7V ratio as a function of metallicity to predictions from
  evolutionary models (such as \citet{Eldridge2006}) but with so few
  points it is not possible to make any helpful comparison. Also, we
  do not account for the contribution of WR stars to the ionising
  photons \citep{Vacca1992}.  For example, in source \#49 the 3 WC
  stars present would be expected to contribute significantly towards
  the ionising flux so the WR/O\,=\,0.75 ratio is a lower limit since
  the number of O stars is not required to be as high to account for
  the measured flux. Our full survey of M101 will allow us to
  investigate this in more detail. 

We have identified whether a WR star is directly in, on the edge of an
H\,{\sc ii} region or is truly isolated and have identified the region
from \citet{Hodge1990} and use their nomenclature in this
work. Aperture photometry was performed in \textsc{gaia} and each
aperture size was determined based on a 95\% luminosity cut with background subtraction performed by using an additional
  aperture of the same radius on an 'empty' region local to the
  H\,{\sc ii} region. The corresponding aperture radii, luminosity
measurements and resulting O star population for each source are
presented in Table \ref{HII_properties}. The error on the
  final number of O stars is $\sim$10\% which results from the choice
  of aperture size although this does not account for the error
  associated with the ionising flux from WR stars; the errors on the
  flux measurements themselves are of order 1\%.

Seven of our 10 sources exhibit H$\alpha$ emission in their spectra,
from which we conclude that 70\% of WR stars in M101 are found in
H\,{\sc ii} regions (based on small number statistics of course). On
inspection of archival KPNO imaging \citep{Knapen2004} we also
conclude that 70\% of WR stars are in H\,{\sc ii} regions, however
there are discrepancies between the imaging and spectroscopic data.

Source \#112 is directly located in an H\,{\sc ii} region in both
archival KPNO and HST narrow-band H$\alpha$ imaging and, as one would
expect, shows nebular emission lines in its spectrum. Conversely, the
spectrum of source \#114 does not show nebular emission but does show
stellar absorption features typical of a main-sequence A star along
with WR emission lines. Ground-based KPNO imaging suggests that the
source is in an H\,{\sc ii} region, however the superior spatial
resolution of HST/ACS imaging reveals that the source is actually
$\sim$2\,arcsec NW of the H\,{\sc ii} region. We note that the
resolution of the KPNO imaging is $\sim$2.5\,arcsec.

\begin{figure}

\subfigure[KPNO/JAG,
H$\alpha$]{\includegraphics[width=0.48\columnwidth]{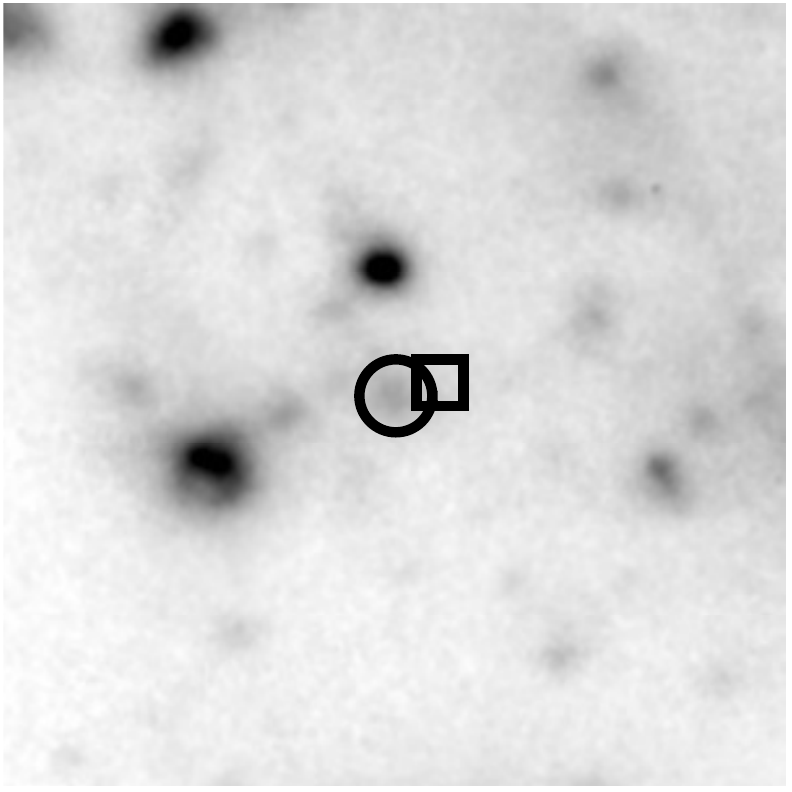}}
\subfigure[Gemini/GMOS, g'-band]{\includegraphics[width=0.48\columnwidth]{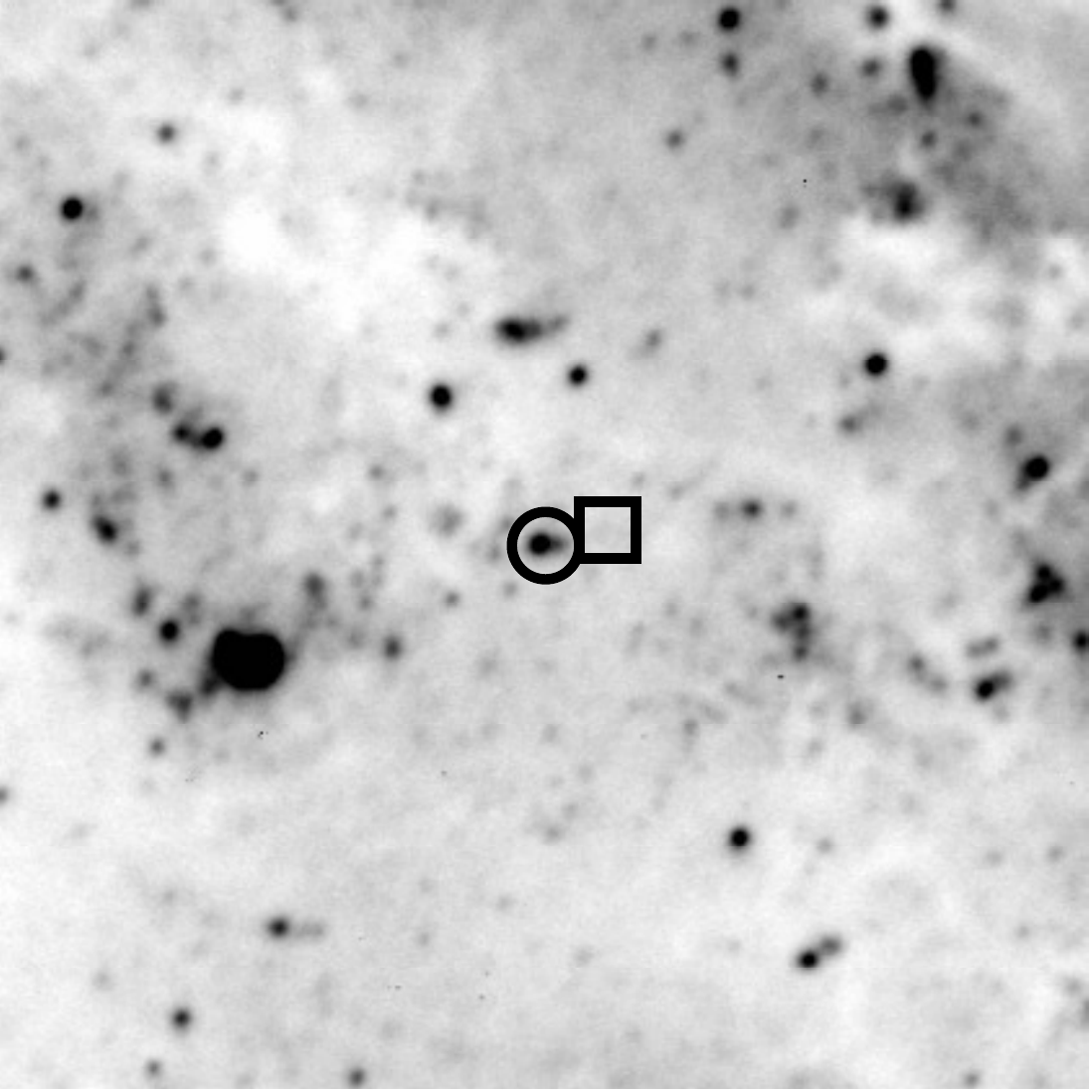}}

\subfigure[HST/WFC3,
F469N]{\includegraphics[width=0.48\columnwidth]{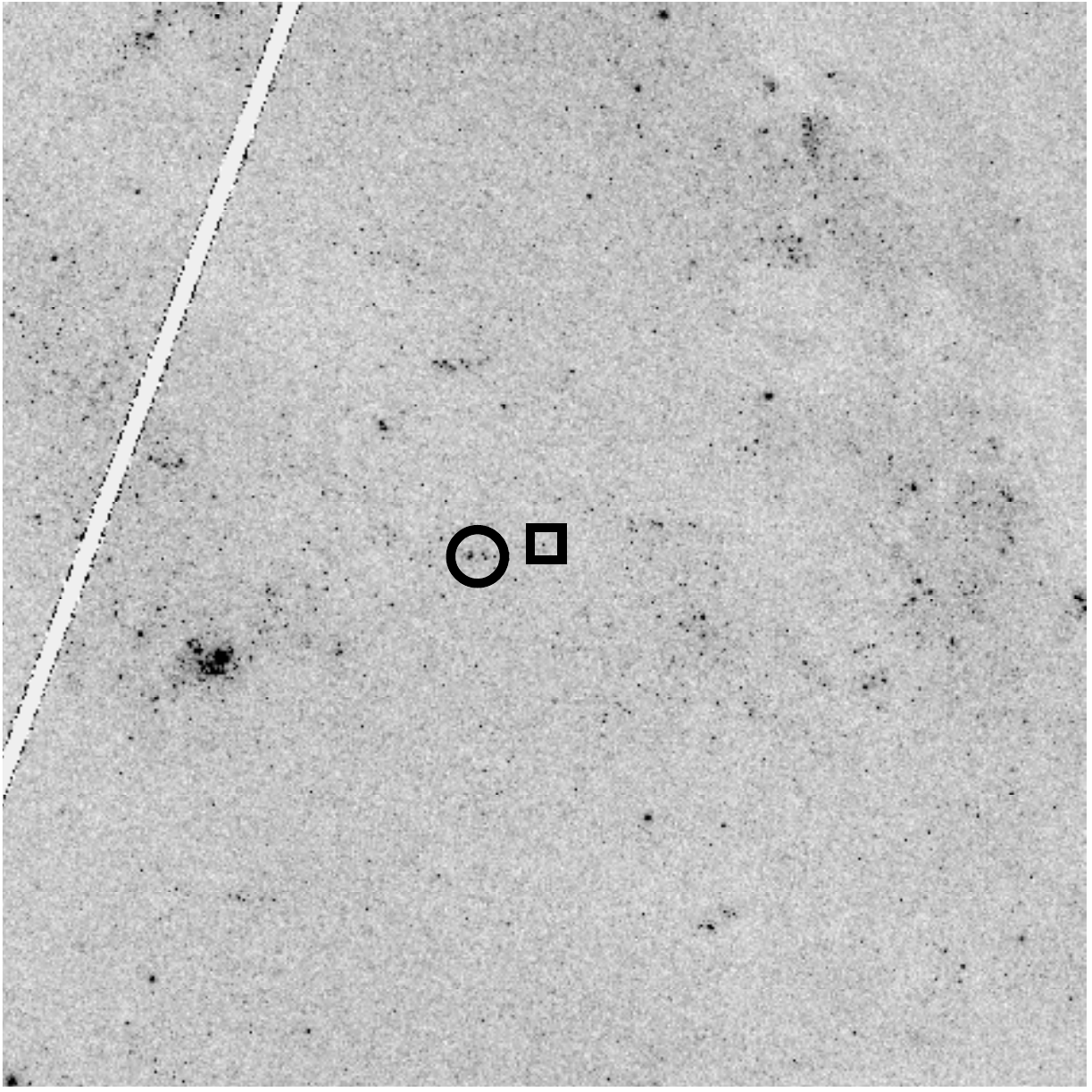}}
\subfigure[HST/WFC3, F469N]{\includegraphics[width=0.48\columnwidth]{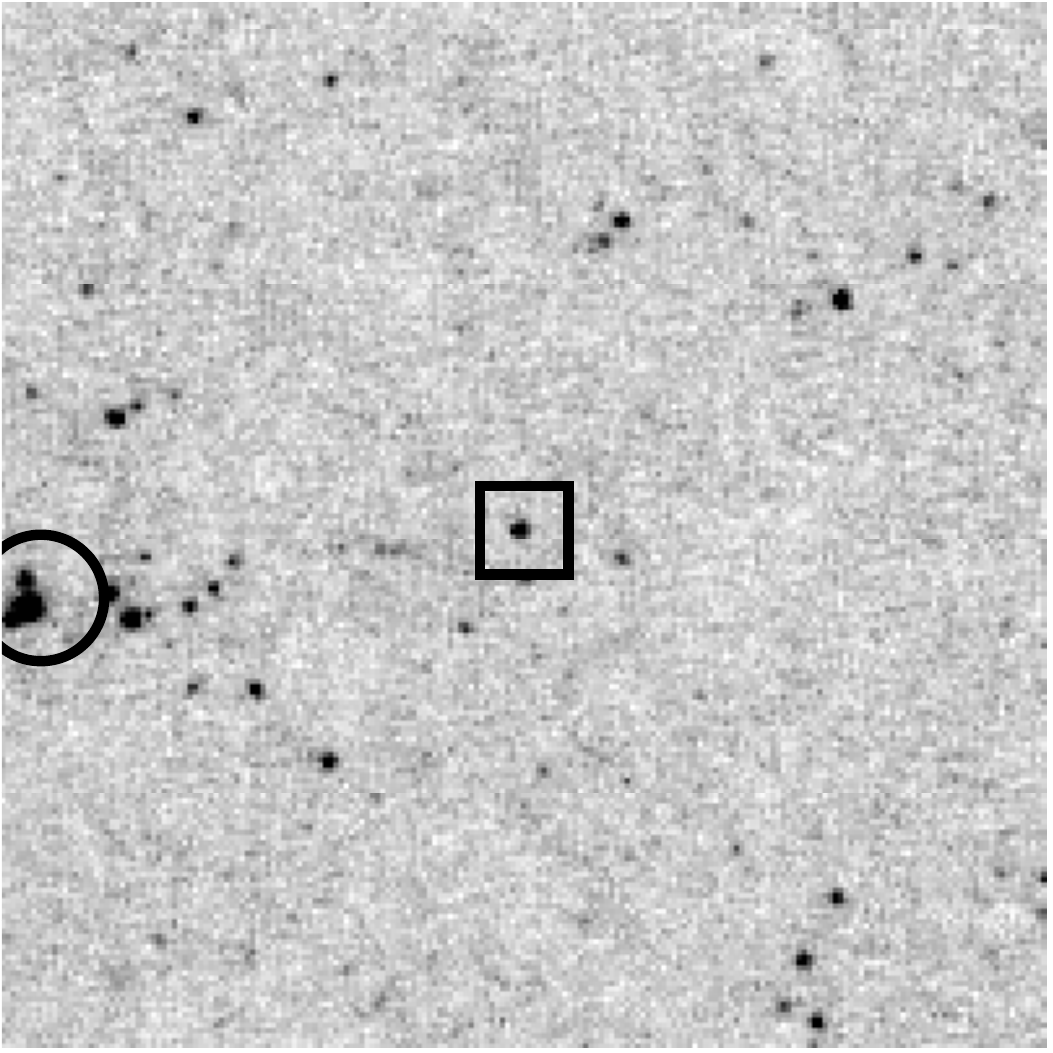}}

\caption{Postage stamp images of $\sim$1~arcmin showing the location
  of WR source \#56 (square) and associated H\,{\sc ii} region \#937
  (circle) as observed with different telescopes (a-c). Image (d) is a
  close up on (c) clearly identifying the WR star. The orientation of
  the images is North up and East left.}
\label{HII_regions_image}

\end{figure}

The lack of nebular emission in Source \#56 is consistent with no
H$\alpha$ detection in the KPNO imaging (Figure
\ref{HII_regions_image}). Since the spectrum of source \#56 reveals
the presence of both WN and WC stars it is likely that the H\,{\sc ii}
region is beyond the detection limits of the KPNO imaging. In
addition, archival HST F658N imaging does not reveal the presence of
an H\,{\sc ii} region, suggesting that the gas from the star cluster
has been expelled. The projected distance of Source \#56 from the
  closest cluster is $\sim$150\,pc which is consistent with the distance a
  runaway WR star can travel in its lifetime \citep{Eldridge2011}. The
  presence of multiple WR stars in the GMOS spectrum makes this
  scenario unlikely.

The spectrum of Source \#1030 has a chip gap at the location of the
H$\alpha$$\lambda$6563 emission, but the star but is located on the
edge of an H\,{\sc ii} region in the KPNO image, which the HST imaging
reveals to be a stream connecting two H\,{\sc ii} regions. The most
likely explanation is that the spatially extended H$\alpha$ emission
is too weak to be detected in the GMOS spectra. Interestingly, source
\#48 does show H$\alpha$ emission in its spectrum, however the KPNO
imaging does not reveal any H\,{\sc ii} region. This would suggest
that there is a faint, underlying H\,{\sc ii} region beyond the
detection limits of the KPNO data which is confirmed to be the case
from archival HST F658N imaging.
 
In addition to \#112, a further three of our WR sources, \#1012, 1016, and
1024 present a nebular spectrum and are located in H\,{\sc ii} regions
seen in both KPNO and HST imaging. Although sources \#49 and \#2053 exhibit
H$\alpha$ emission in the GMOS spectroscopy and KPNO imaging the
resolution of HST reveals that they are in fact on the edge of the
H\,{\sc ii} region, suggesting that the spectroscopic detection is of
the H\,{\sc ii} region itself. The location of each WR source relative
to the H$\alpha$ emission is presented in Appendix \ref{post_stamps}.

In summary, from KPNO H$\alpha$ imaging we conclude that 70\% of WR
stars in our sample are in H\,{\sc ii} regions, 10\% are associated
with H\,{\sc ii} regions and 20\% are not associated with H\,{\sc ii}
regions. However, from the improved spatial resolution of HST our
conclusions are 50\% in, 40\% associated with and 10\% not in H\,{\sc
  ii} regions. The high fraction of WR stars being associated with
H\,{\sc ii} regions is not consistent with the low fraction (27\%) of
WR stars associated with star-forming regions presented by
\citet{Crowther2015a}. Whilst we do not claim these results to be
statistically robust due to small number statistics, the difference
between the conclusions based on the two sets of imaging, particularly
between those directly in or on the edge of a H\,{\sc ii} region,
demonstrates how the spatial resolution and sensitivity of imaging can
lead to different interpretations of stellar
environment. Moreover, the spectroscopic results presented
  here are limited to the brighter WR stars which are more likely to
  be found in bright H\,{\sc ii} regions. We expect a full
  spectroscopic survey including the fainter WR candidates to reduce
  the fraction of WR stars seen in H\,{\sc ii} regions.

\onecolumn
\begin{table}
\caption{H$\alpha$$+$[N\,{\sc ii}] flux measurements of the H\,{\sc ii} regions
  confirmed to host WR stars. We note that the radius of the H\,{\sc ii}
  region is largely influenced by the spatial resolution of the images
  and is quoted for information only. The H$\alpha$ flux
  is corrected for distance, extinction and [N\,{\sc ii}] contribution
  and the H$\alpha$ luminosity is determined using the calibration of \citet{Knapen2004} and the N(07\,{\sc v}) stars
  using 10$^{49}$ photons/s \citep{Vacca1994}.}
\centering
\begin{tabular}{l@{\hspace{2mm}}c@{\hspace{2mm}}c@{\hspace{2mm}}c@{\hspace{2mm}}c@{\hspace{2mm}}c@{\hspace{2mm}}c@{\hspace{2mm}}c@{\hspace{2mm}}c@{\hspace{2mm}}c@{\hspace{2mm}}c@{\hspace{2mm}}c@{\hspace{2mm}}c@{\hspace{2mm}}c@{\hspace{2mm}}c@{\hspace{2mm}}c@{\hspace{2mm}}c@{\hspace{2mm}}c@{\hspace{2mm}}c@{\hspace{2mm}}c@{\hspace{2mm}}}
\hline
WR & Hodge  & RA$^{1}$ & Dec$^{1}$ & Radius    & F(H$\alpha$$+$[N\,{\sc  ii}]) & E(B-V) &[N\,{\sc ii}]/ & L(H$\alpha$) & log Q$_{0}$ & N(O7\,{\sc v}) & N(WR)/& Spectra & KPNO & HST  \\

 ID  & ID        & (h:m:s)      & ($^{\circ}$:':'') & (arcsec) & (erg s$^{-1}$ cm$^{-2}$) & & H$\alpha$  & (erg s$^{-1}$)  & (s $^{-1}$) & & N(O7\,{\sc v}) &  & & \\

\hline
56     &               & & & & & & & & &    &                                                                                                                                             & No & No & No \\
1030 & 733        & 14:03:10.786 & 54:20:18.34 & 3.32 &  4.29$\times$10$^{-14}$    & 0.44 & 0.54     & 3.71$\times$10$^{38}$  & 50.44  & 28  & 0.07  & No & Edge & Edge \\ 
48     &               & & & & & & & & &    &                                                                                                                                              & Yes & No & Yes \\ 
1012 & 470        & 14:03:07.184 & 54:22:21.02 & 4.22 &  7.69$\times$10$^{-14}$    & 0.09 & 0.35     & 3.79$\times$10$^{38}$  & 50.17  & 15  & 0.14   & Yes & Yes & Yes \\
1016 & 505,507 & 14:03:01.800 & 54:22:03.87 & 6.27 &  3.08$\times$10$^{-13}$    & 0.59 & 0.63     & 3.39$\times$10$^{39}$  & 51.40  & 250 & 0.004 & Yes & Yes & Yes \\
49     & 872        & 14:03:02.108 & 54:19:16.07 & 2.34 &  5.11$\times$10$^{-15}$    & 0.12 & 0.67     & 5.37$\times$10$^{37}$  & 49.60  & 4   & 0.75     & Yes & Yes & Edge \\
112   & 901        & 14:03:04.219 & 54:19:05.39 & 3.86 &  7.49$\times$10$^{-14}$    & 0.53 & 0.30     & 1.30$\times$10$^{39}$  & 50.98  & 96  & 0.01    & Yes & Yes & Yes \\ 
114   & 998        & 14:03:23.969 & 54:18:24.57 & 3.37 &  5.56$\times$10$^{-14}$    & 0.44 & 0.54     & 1.93$\times$10$^{39}$  & 51.16  & 145 & 0.007 & No & Yes & Edge \\
1024 & 671        & 14:03:04.531 & 54:20:45.32 & 2.90 &  1.63$\times$10$^{-14}$    & 0.46 & 0.22     & 2.39$\times$10$^{38}$  & 50.25  & 18  & 0.06     & Yes & Yes & Yes \\
2053 & 1044      & 14:03:07.460 & 54:17:55.79 & 6.87 &  3.88$\times$10$^{-13}$     & 0.98 & 0.14     & 3.28$\times$10$^{40}$  & 52.39  & 2454  & 0.0004 & Yes & Yes & Edge  \\
\hline

\multicolumn{10}{l}{\footnotesize{$^{1}$Co-ordinates of the H\,{\sc
      ii} region are taken from the KPNO H$\alpha$
  image in \citet{Knapen2004}}} \\
\hline
\end{tabular}
\label{HII_properties}
\end{table}
\twocolumn

\section{Summary}
\label{discussion}

We spectrographically confirmed the detection of 15 WR stars within 10
He\,{\sc ii} emission sources in M101. From comparison with Galactic templates
we conclude that these regions host 4 WN and 11 WC stars, indicating
that our detections are biased towards WC stars. This is not
surprising given the higher He\,{\sc ii} excess for WC stars compared to WN
stars. Moreover, all of our regions are metal-rich so we expect a
WC/WN ratio $\geq$1 based on the evolutionary models from
\citet{Eldridge2006}, which is consistent with our findings though our
sample is limited by small number statistics.

We see no distinctive division between the locations of WC and WN
stars, though again, our sample is small. The successful detection
of both WC and WN stars demonstrates that our imaging technique is
sensitive to both, including the coolest WC and WN stars, WC7-8 and
WN7-8, respectively. 

The main challenge to spectroscopically confirming WR stars in
  M101 is the required line to continuum contrast, since the He\,{\sc
    ii} excess, or line-to-continuum ratio, is our primary diagnostic
  for identifying WR candidates. This can be hindered by intrinsically
  weak WR emission (e.g. in low metallicity environments) hence high
  S/N observations are required in future to confirm these candidates
  as bonafide WR stars. A companion star or surrounding massive stars
  within the same slit will also increase the continuum and dilute the
  WR emission line making the confirmation of a WR more difficult;
  high resolution observations are required to overcome this as well
  as careful mapping of sources from HST imaging to ground-based
  spectra.

\section*{Acknowledgments}

This research is based on spectroscopic observations obtained at the
Gemini Observatory, which is operated by the Association of
Universities for Research in Astronomy, Inc., under a cooperative
agreement with the NSF on behalf of the Gemini partnership: the
National Science Foundation (United States), the National Research
Council (Canada), CONICYT (Chile), the Australian Research Council
(Australia), Minist\'{e}rio da Ci\^{e}ncia, Tecnologia e
Inova\c{c}\~{a}o (Brazil) and Ministerio de Ciencia, Tecnolog\'{i}a e
Innovaci\'{o}n Productiva (Argentina). Photometric results presented
here are based on NASA/ESA Hubble Space Telescope observations
obtained at the Space Telescope Science Institute, which is operated
by the Association of Universities for Research in Astronomy
Inc. under NASA contract NAS5-26555. JLB, MMS and MW acknowledge the
interest and generous support of Hilary and Ethel Lipsitz. AFJM is
grateful to NSERC (Canada) and FQRNT (Quebec) for financial
assistance.

\appendix
\onecolumn

\section{Location of WR sources in relation to H\,{\sc ii} regions}
\label{post_stamps}

Postage stamp images of $\sim$1 arcmin showing the location of the
H\,{\sc ii} region closest to the WR source. Circles indicate the
location of the H\,{\sc ii} region in (a) - (c) and of the WR star in
(d). Where the WR star is located outside the H\,{\sc ii} region a
square has been used to identify it's location.

\clearpage

\centering

\begin{figure}

\subfigure[KPNO/JAG,
H$\alpha$]{\includegraphics[width=0.24\columnwidth]{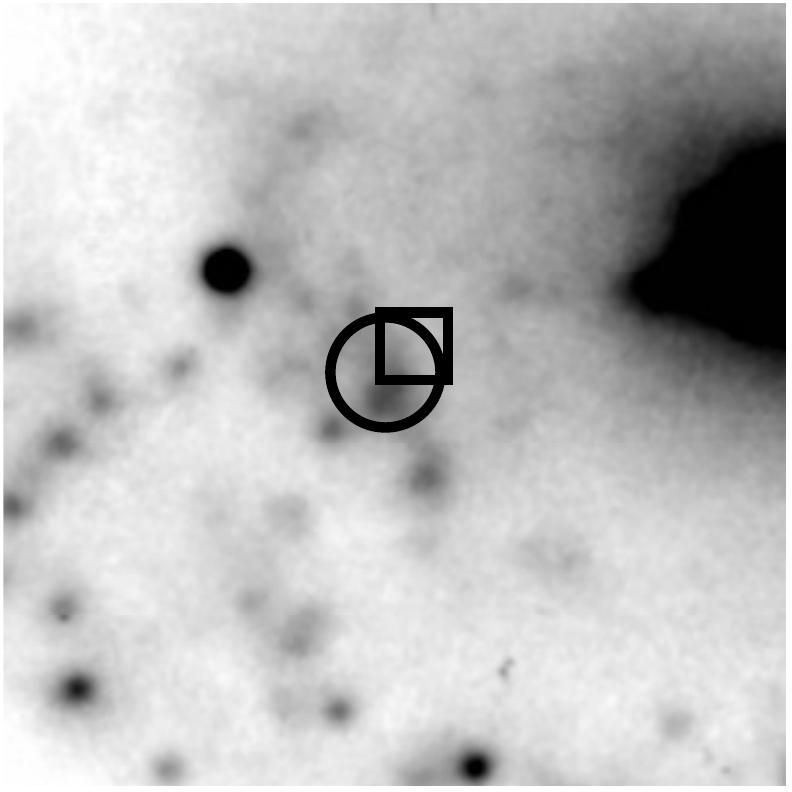}}
\subfigure[Gemini/GMOS,
g'-band]{\includegraphics[width=0.24\columnwidth]{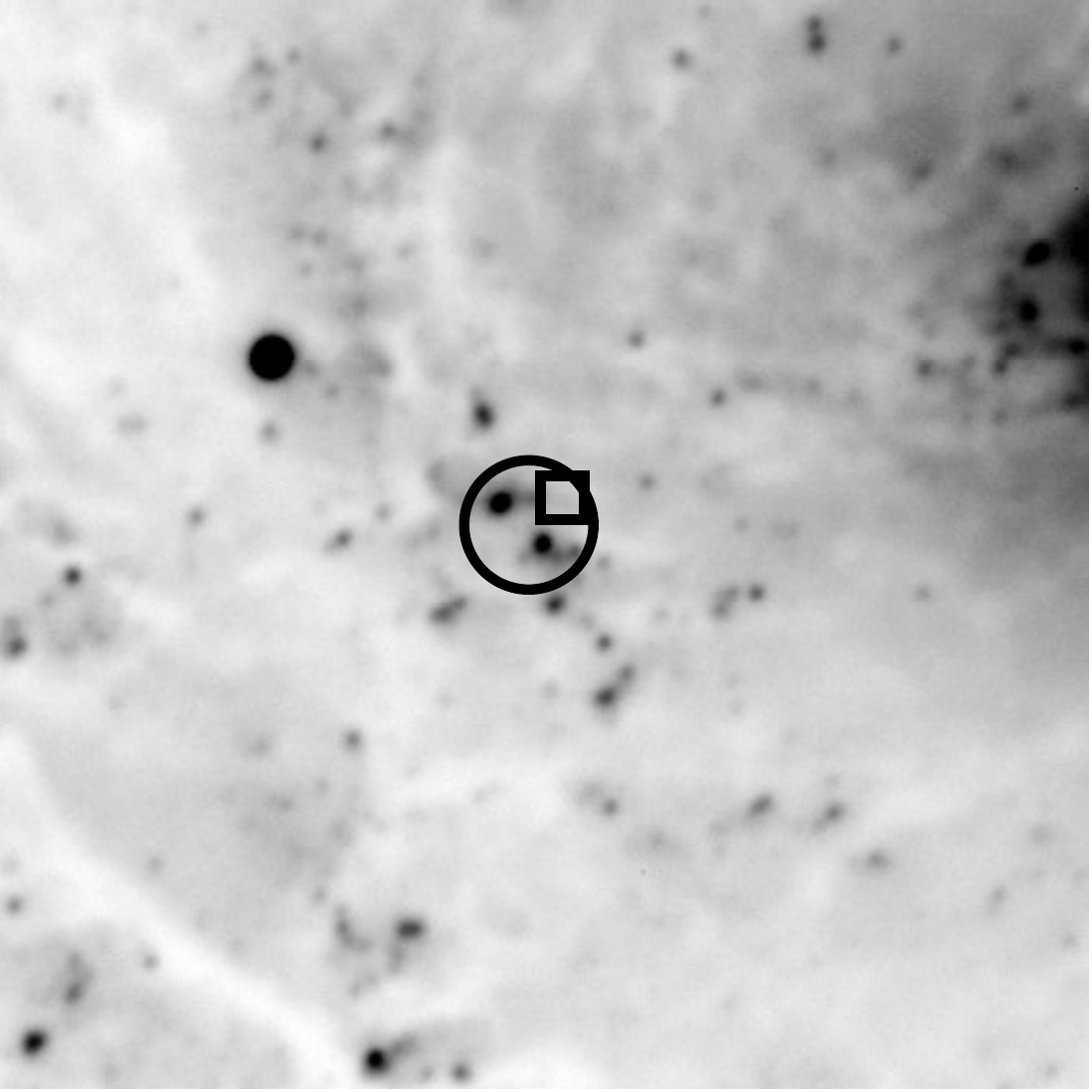}}
\subfigure[HST/WFC3,F469N]{\includegraphics[width=0.24\columnwidth]{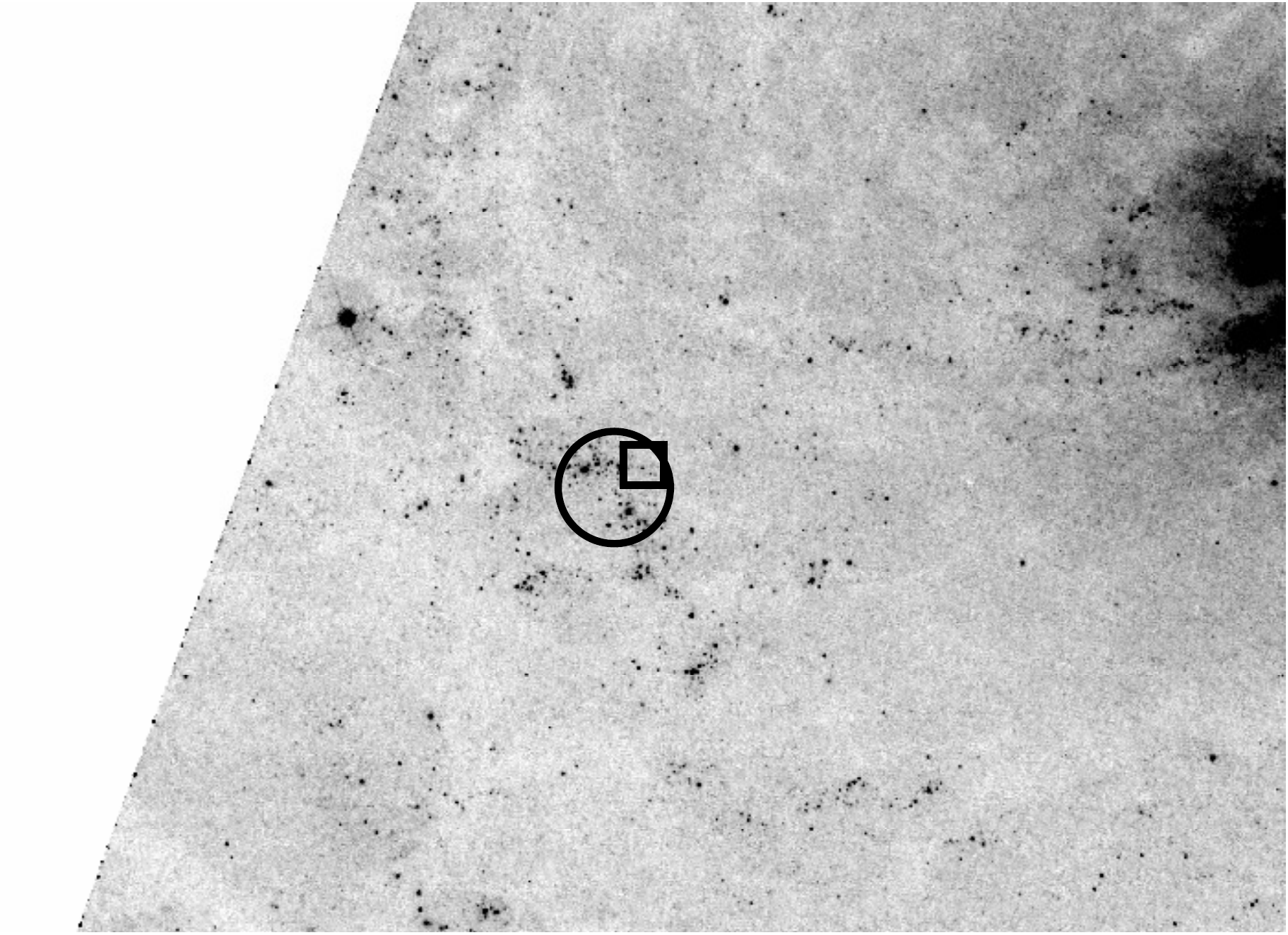}}
\subfigure[HST/WFC3, F469N]{\includegraphics[width=0.24\columnwidth]{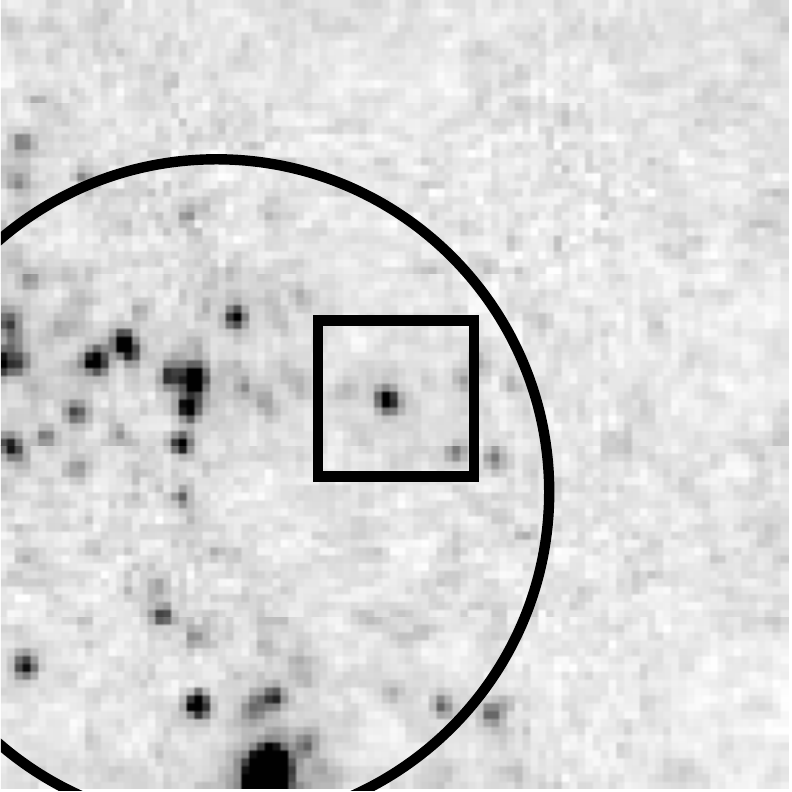}}

\caption{Source \#1030}

\end{figure}

\begin{figure}
\subfigure[]{\includegraphics[width=0.24\columnwidth]{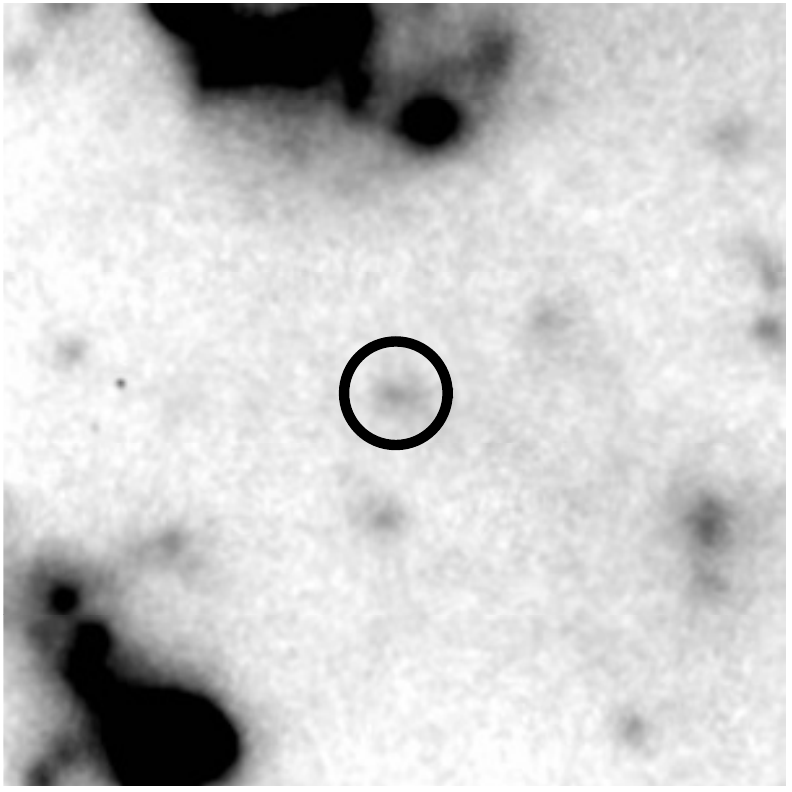}}
\subfigure[]{\includegraphics[width=0.24\columnwidth]{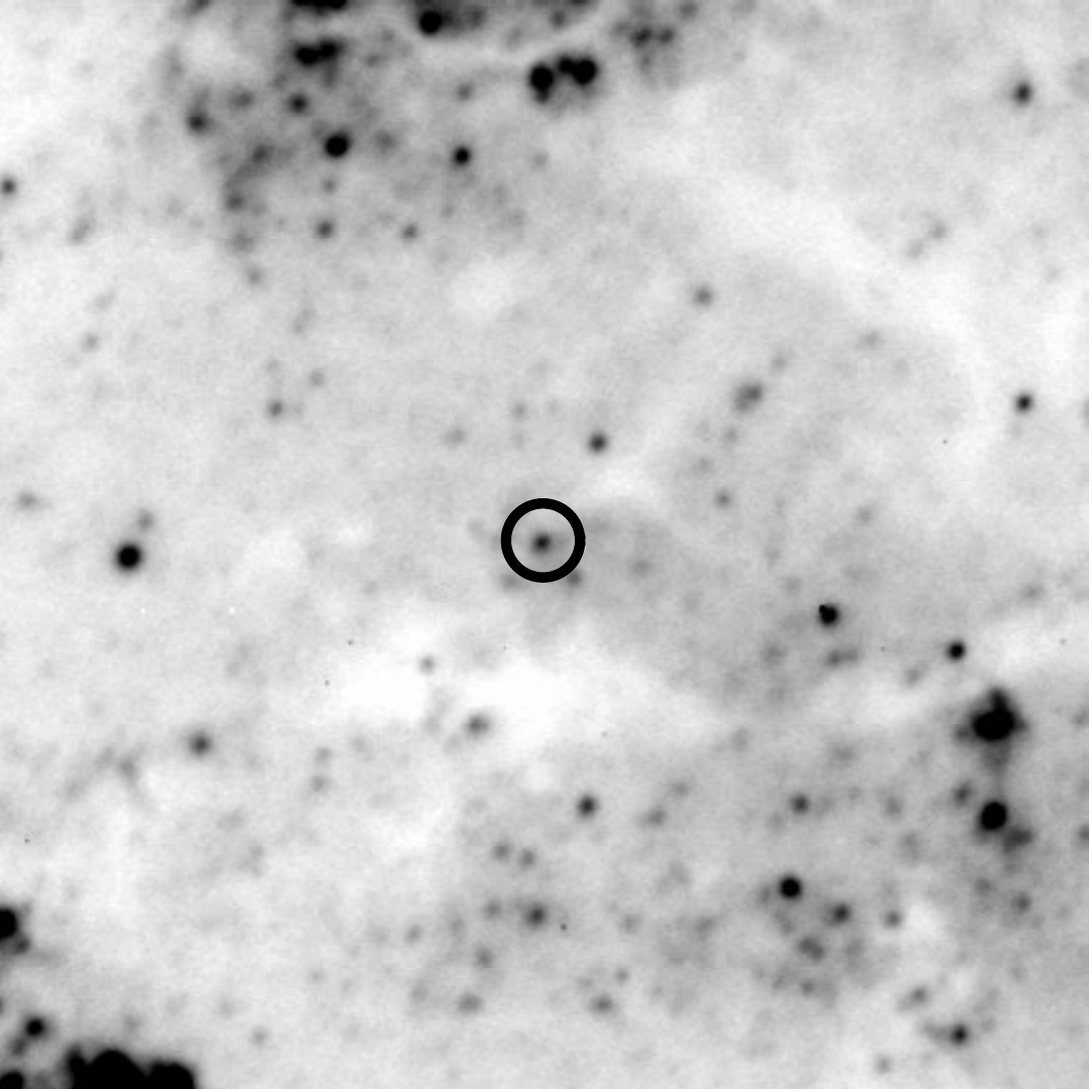}}
\subfigure[]{\includegraphics[width=0.24\columnwidth]{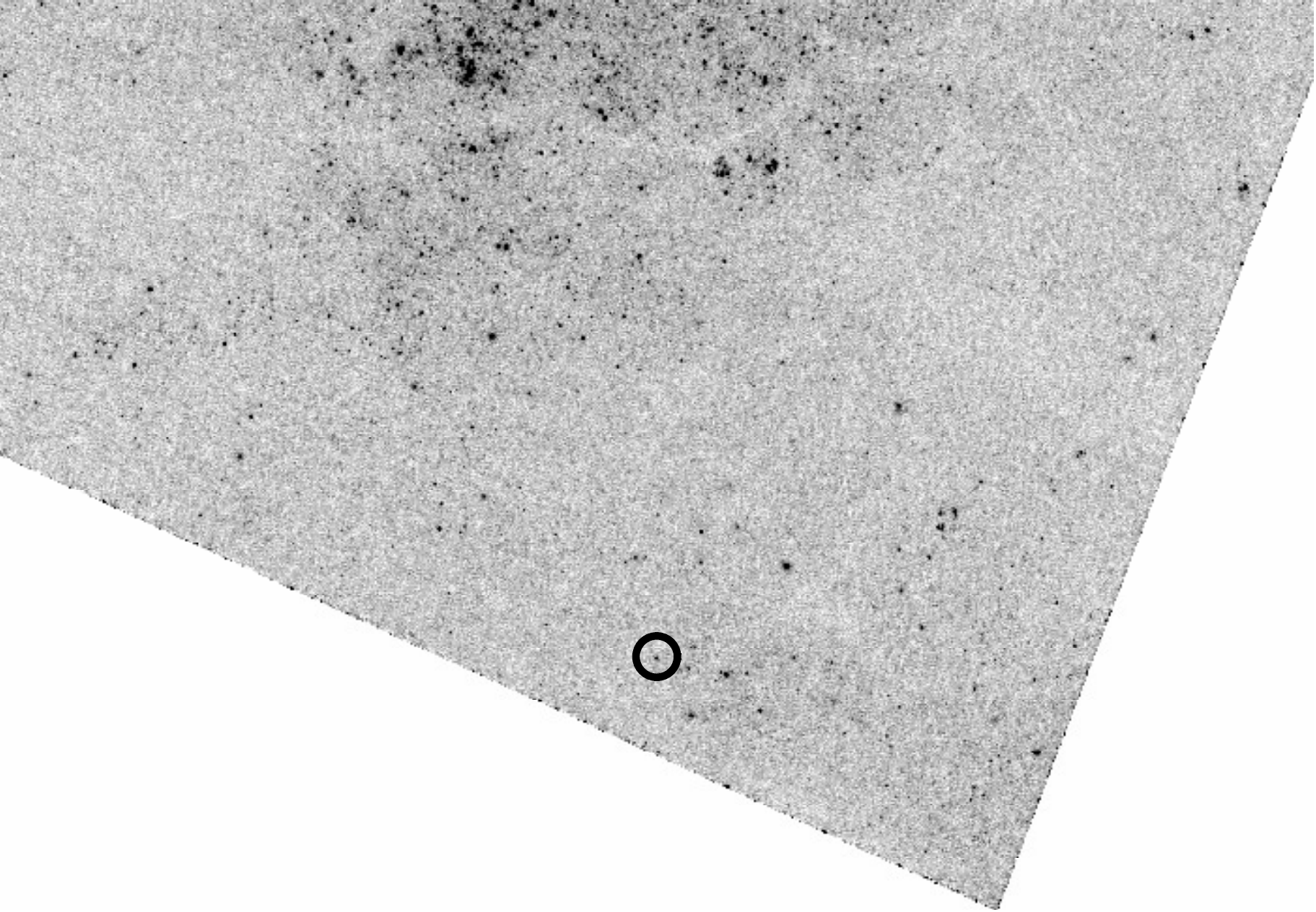}}
\subfigure[]{\includegraphics[width=0.24\columnwidth]{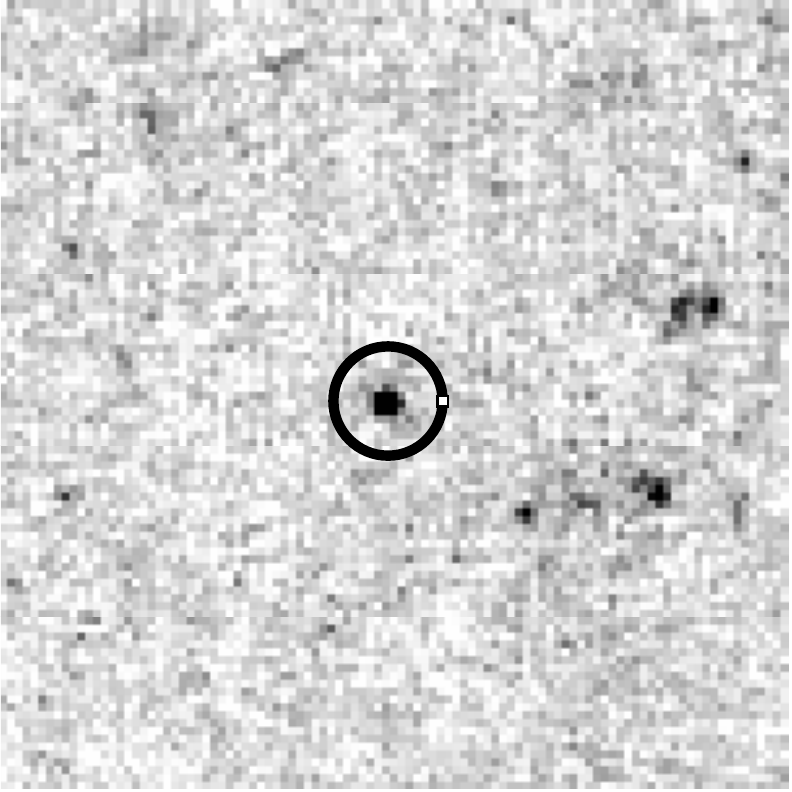}}

\caption{Source \# 48}

\end{figure}

\begin{figure}

\subfigure[]{\includegraphics[width=0.24\columnwidth]{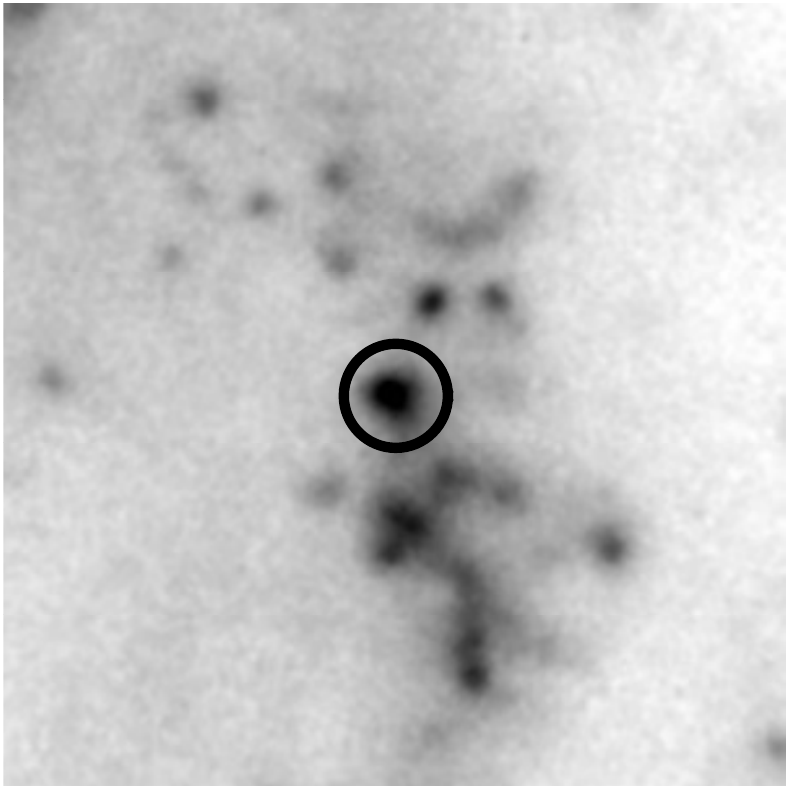}}
\subfigure[]{\includegraphics[width=0.24\columnwidth]{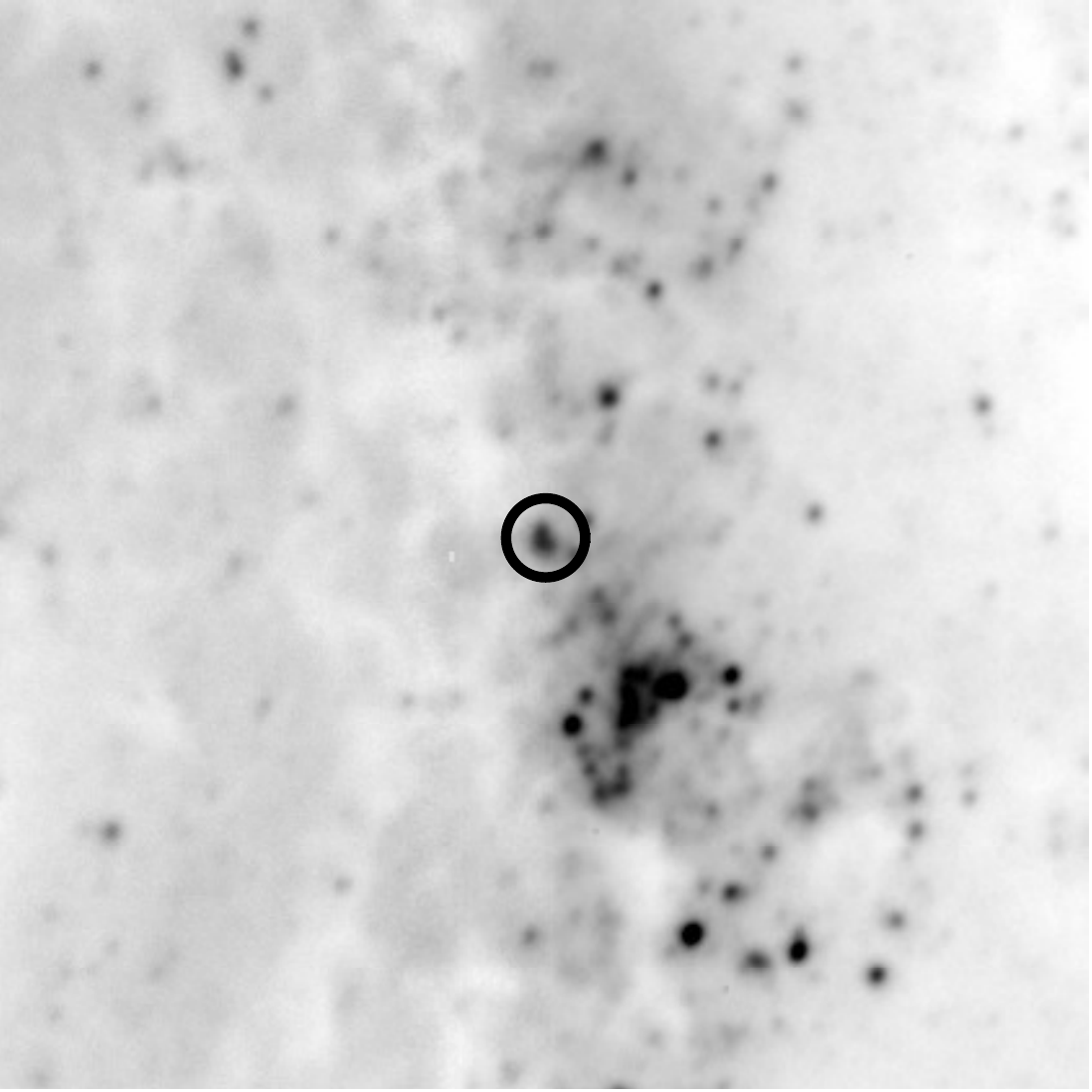}}
\subfigure[]{\includegraphics[width=0.24\columnwidth]{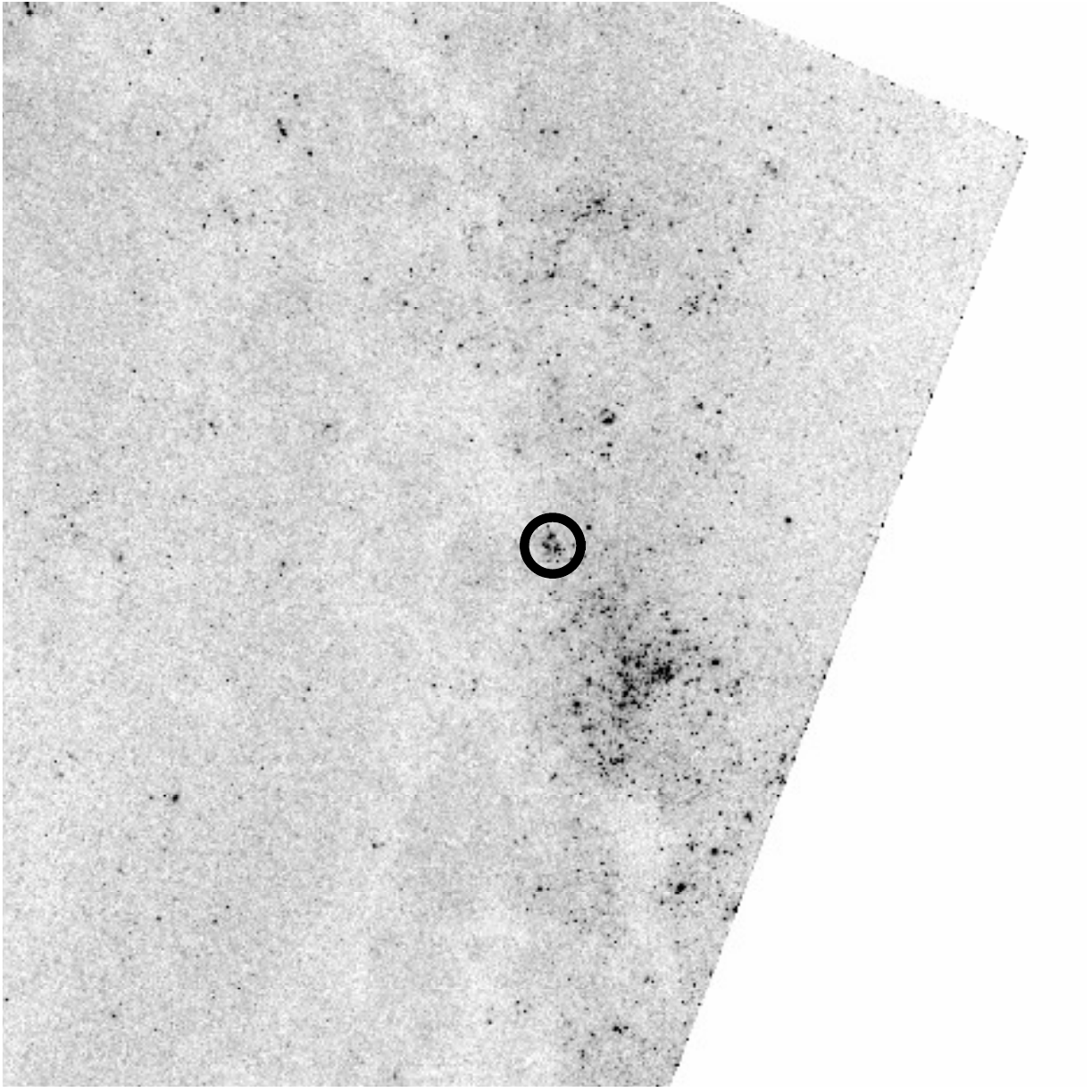}}
\subfigure[]{\includegraphics[width=0.24\columnwidth]{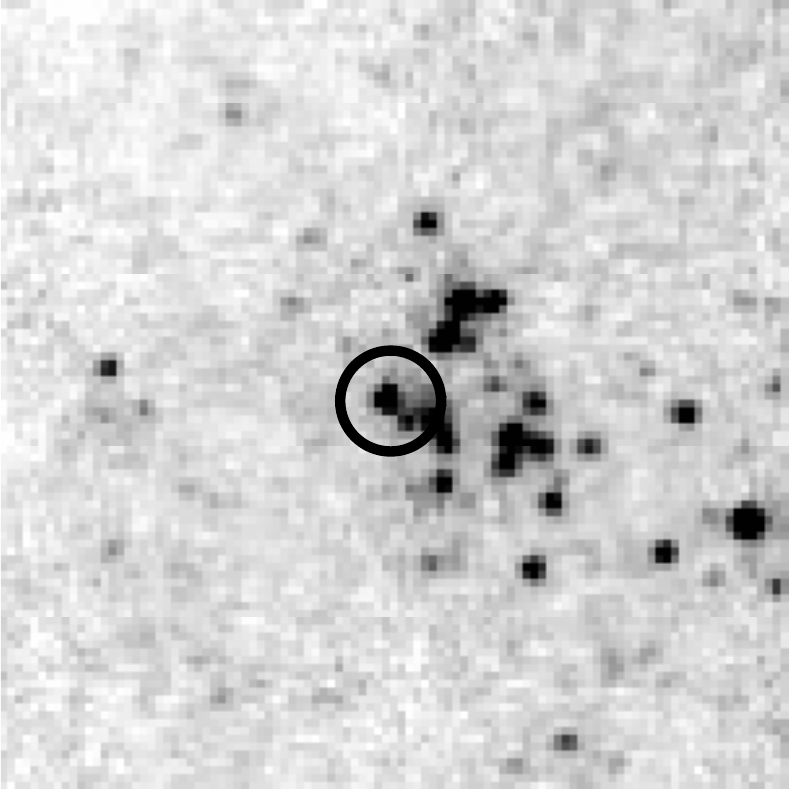}}

\caption{Source \# 1012}

\end{figure}

\begin{figure}

\subfigure[]{\includegraphics[width=0.24\columnwidth]{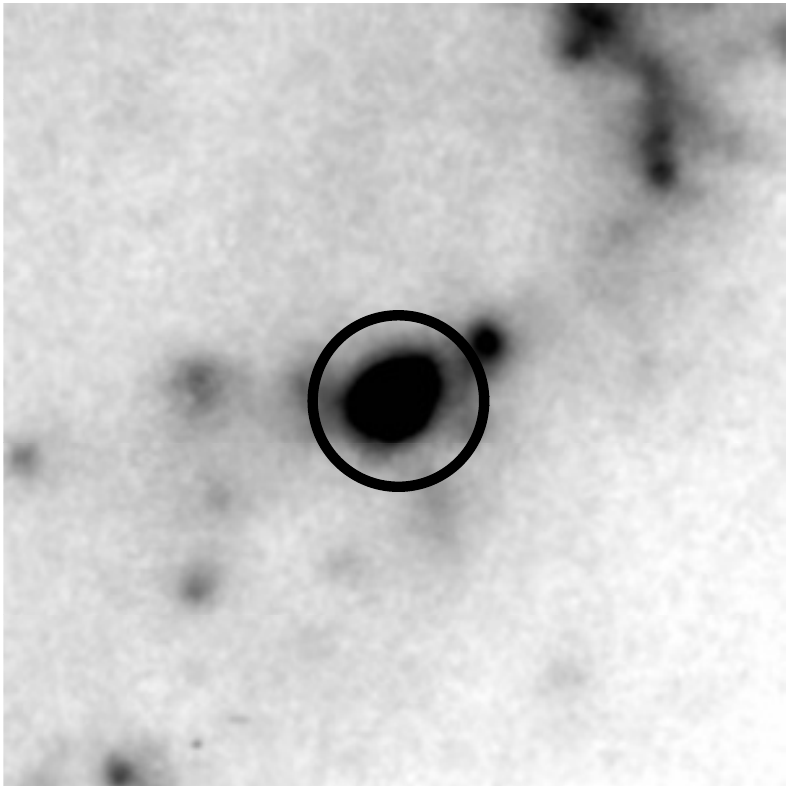}}
\subfigure[]{\includegraphics[width=0.24\columnwidth]{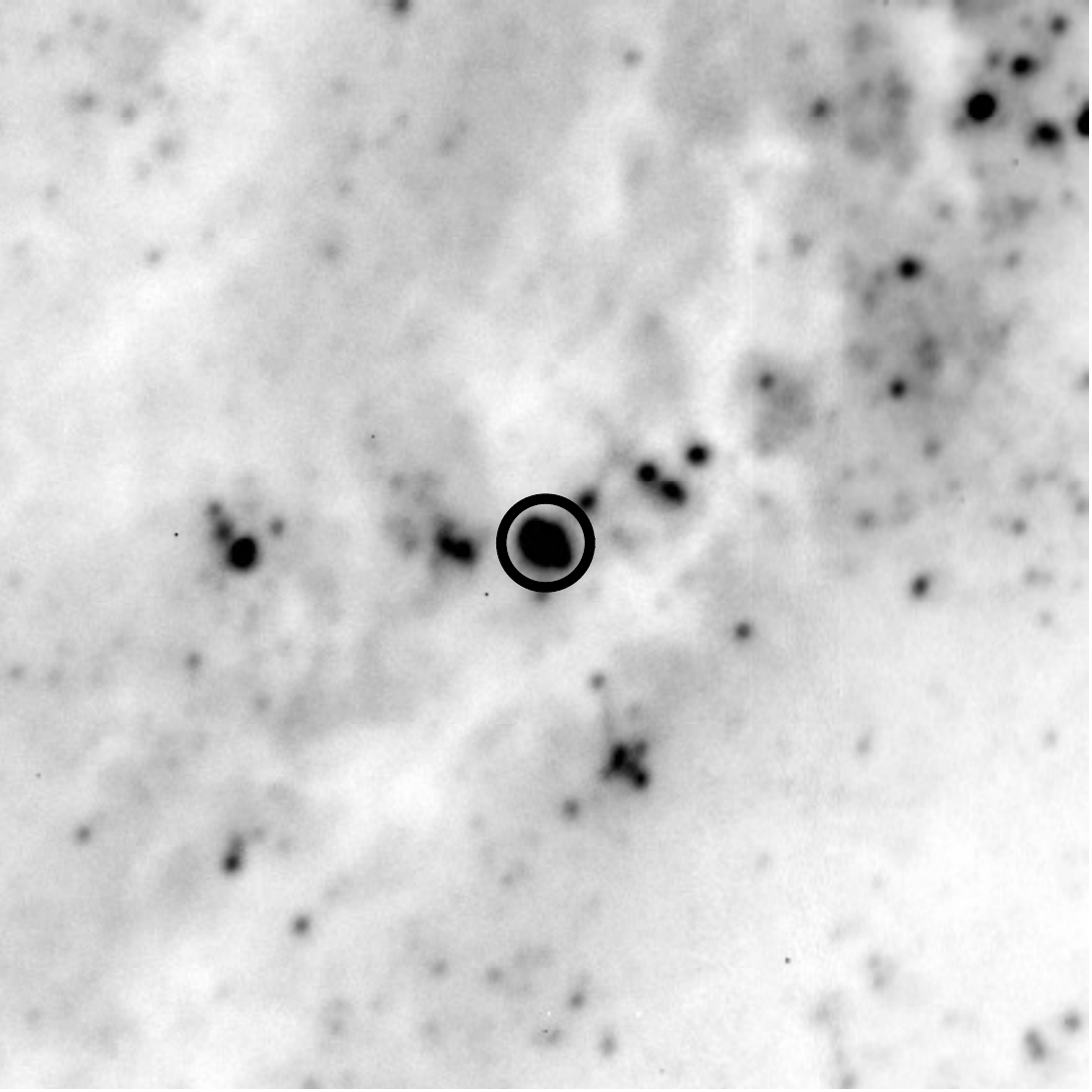}}
\subfigure[]{\includegraphics[width=0.24\columnwidth]{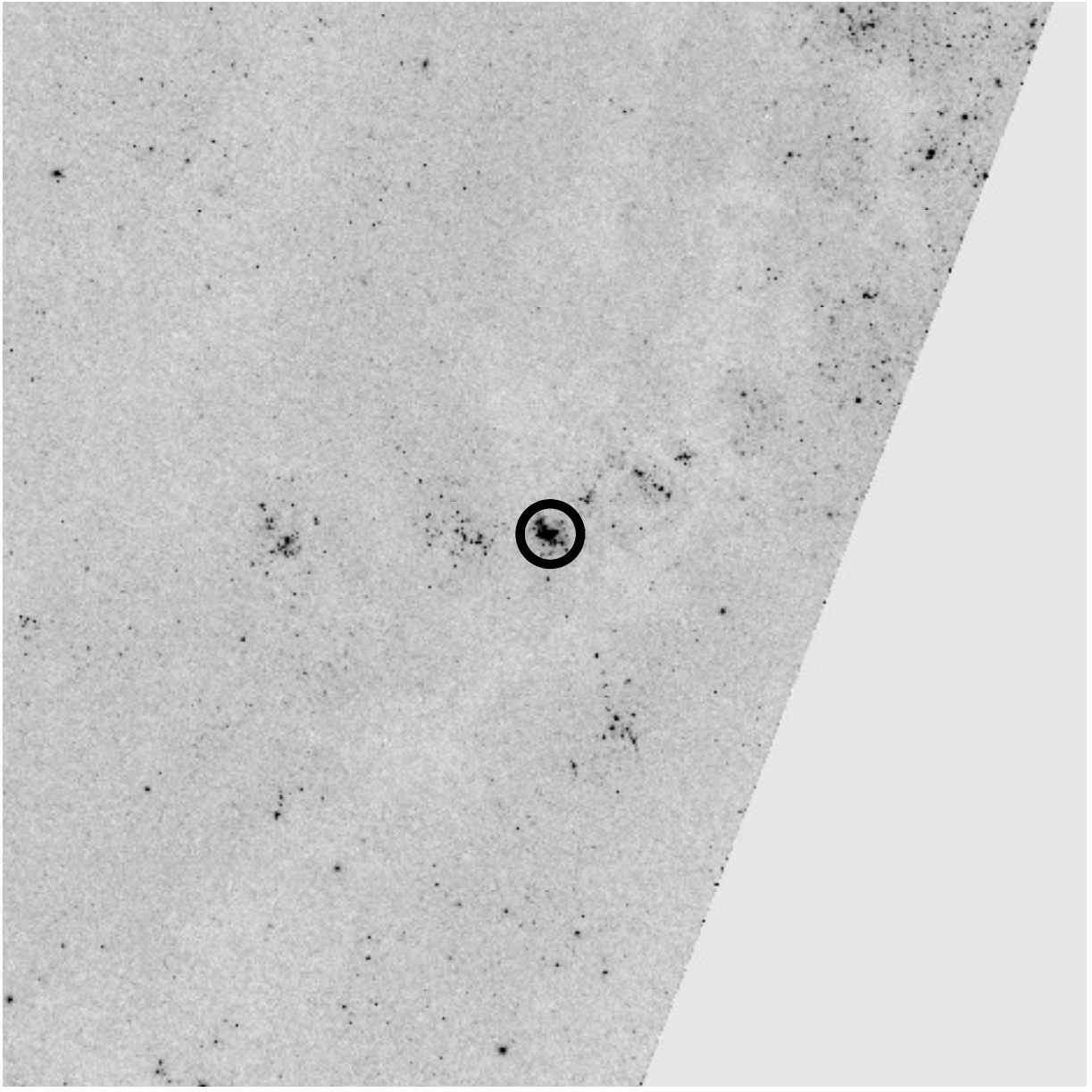}}
\subfigure[]{\includegraphics[width=0.24\columnwidth]{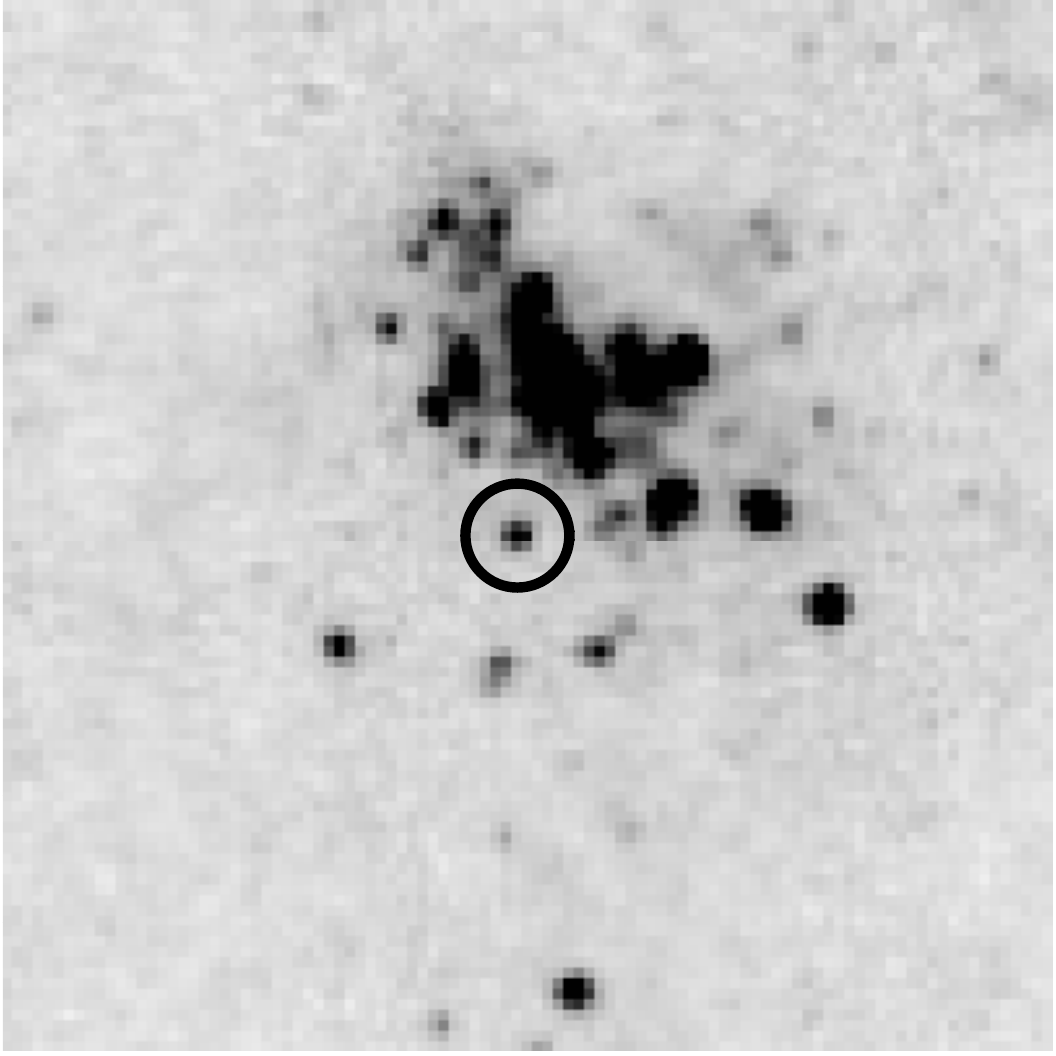}}

\caption{Source \# 1016}

\end{figure}

\begin{figure}

\subfigure[]{\includegraphics[width=0.24\columnwidth]{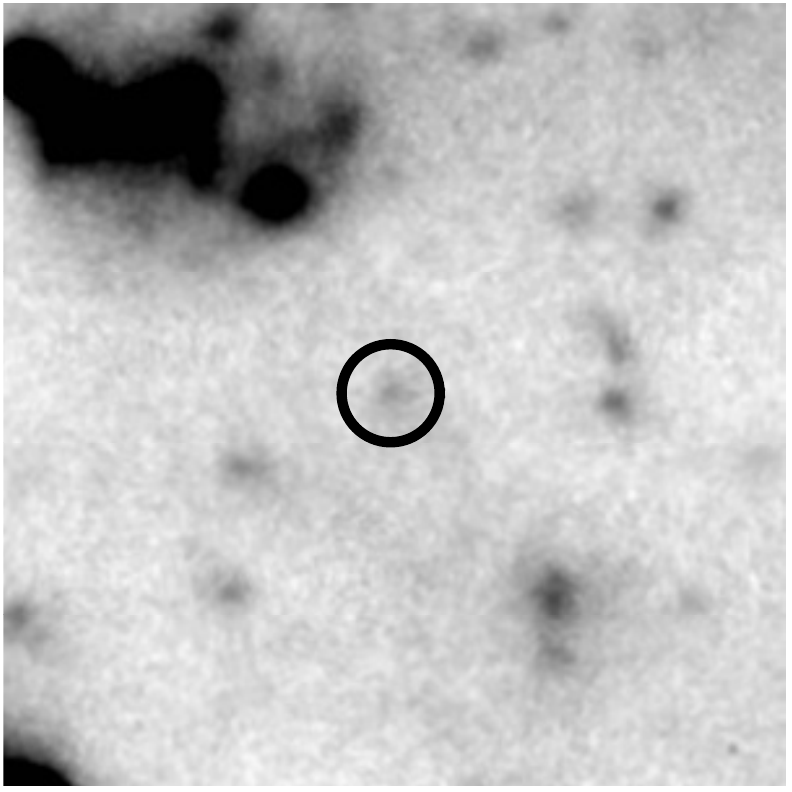}}
\subfigure[]{\includegraphics[width=0.24\columnwidth]{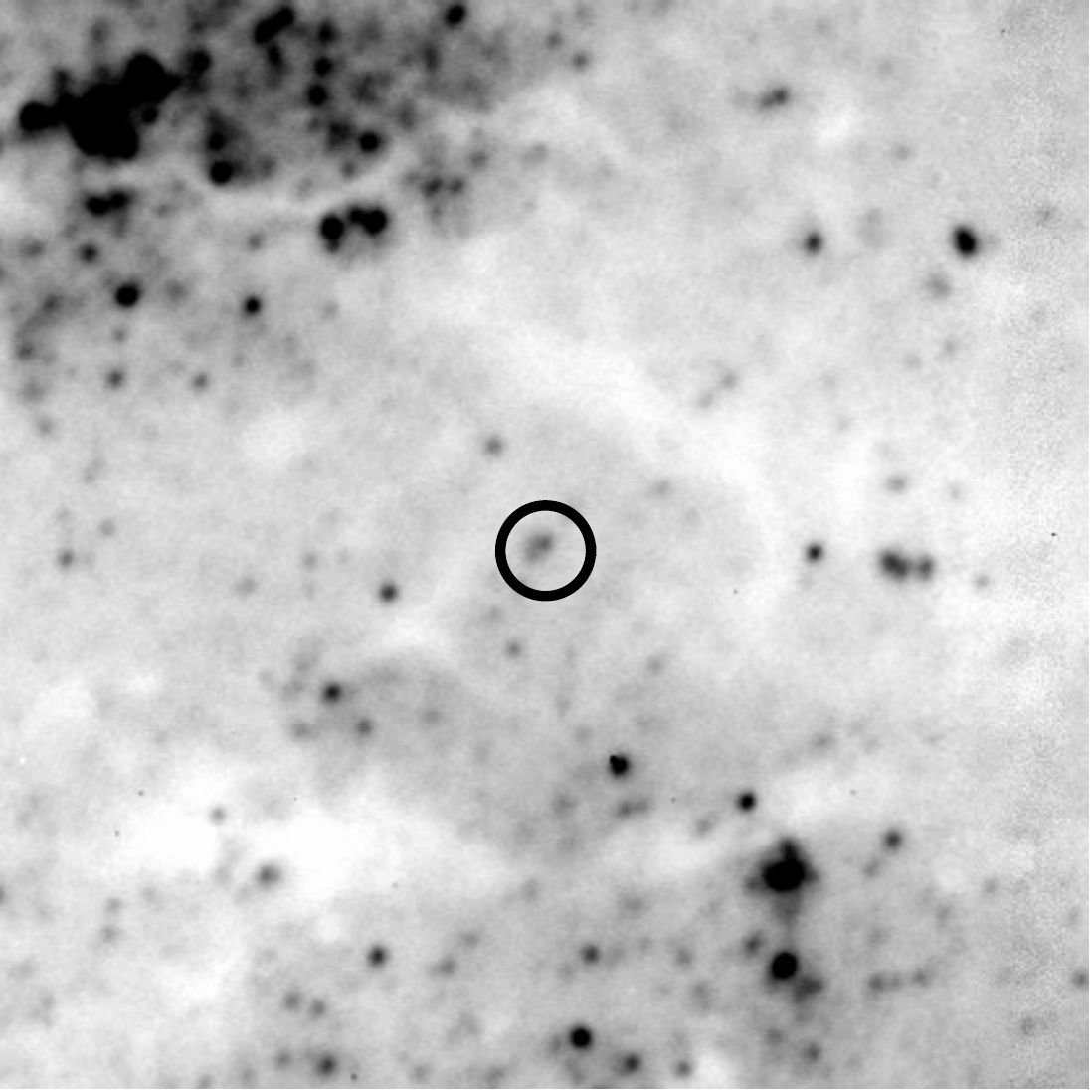}}
\subfigure[]{\includegraphics[width=0.24\columnwidth]{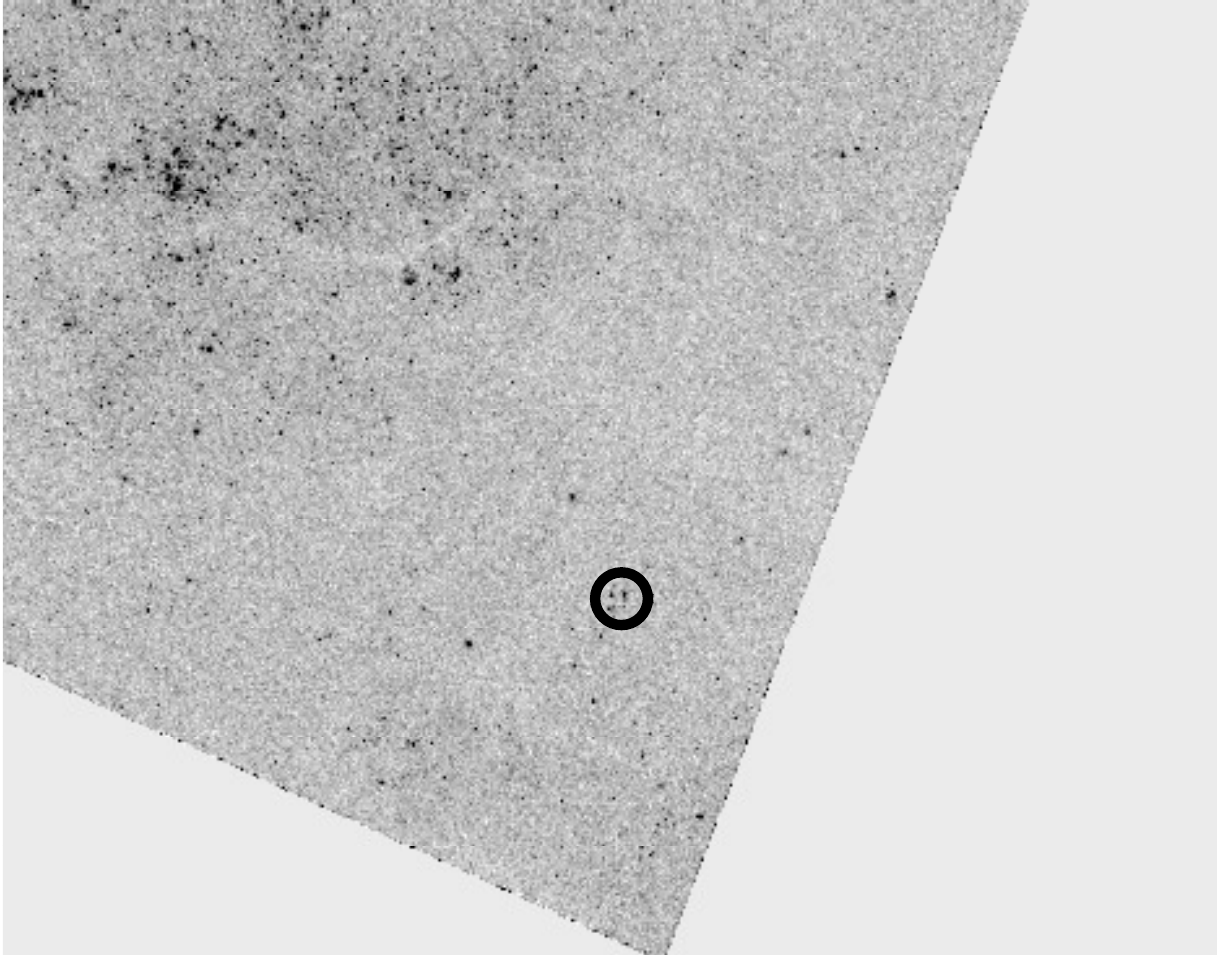}}
\subfigure[]{\includegraphics[width=0.24\columnwidth]{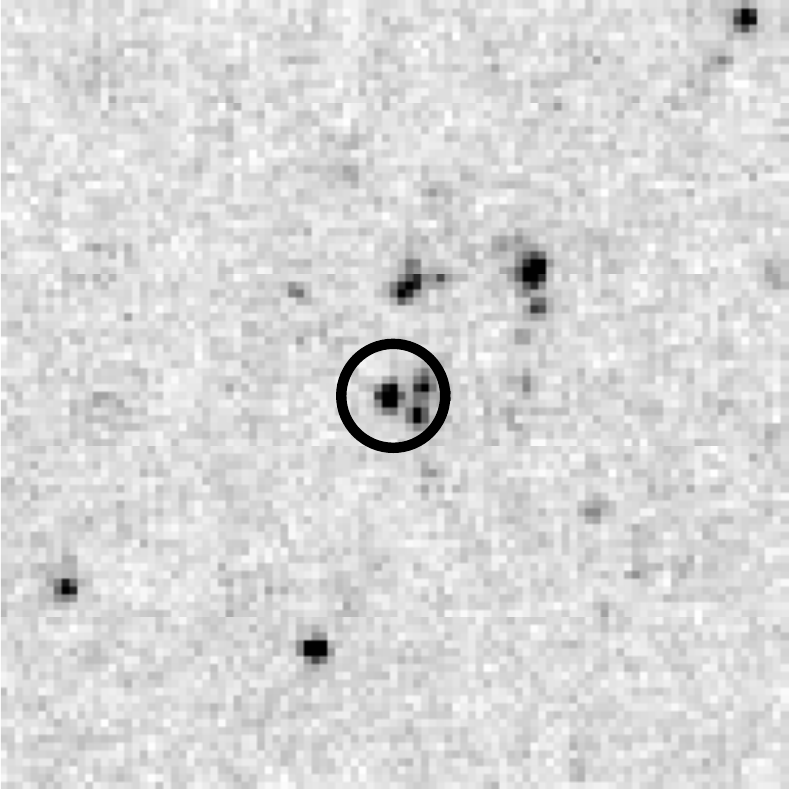}}

\caption{Source \# 49}

\end{figure}

\begin{figure}

\subfigure[]{\includegraphics[width=0.24\columnwidth]{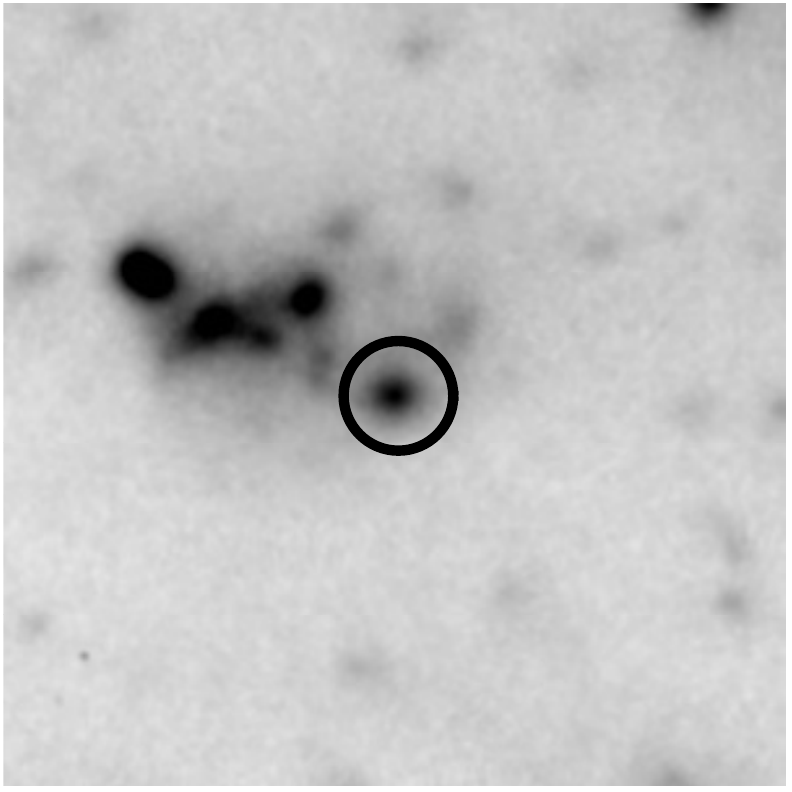}}
\subfigure[]{\includegraphics[width=0.24\columnwidth]{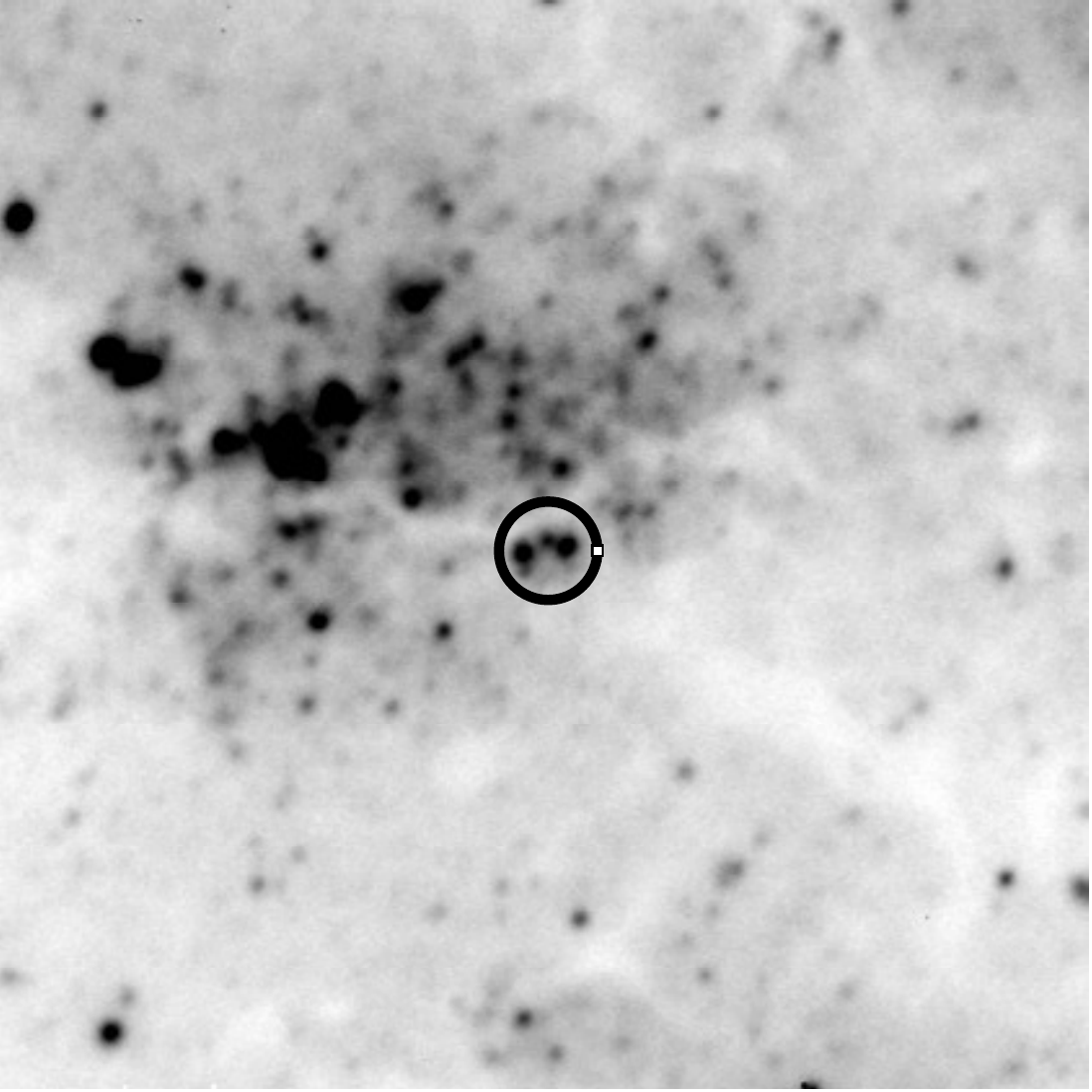}}
\subfigure[]{\includegraphics[width=0.24\columnwidth]{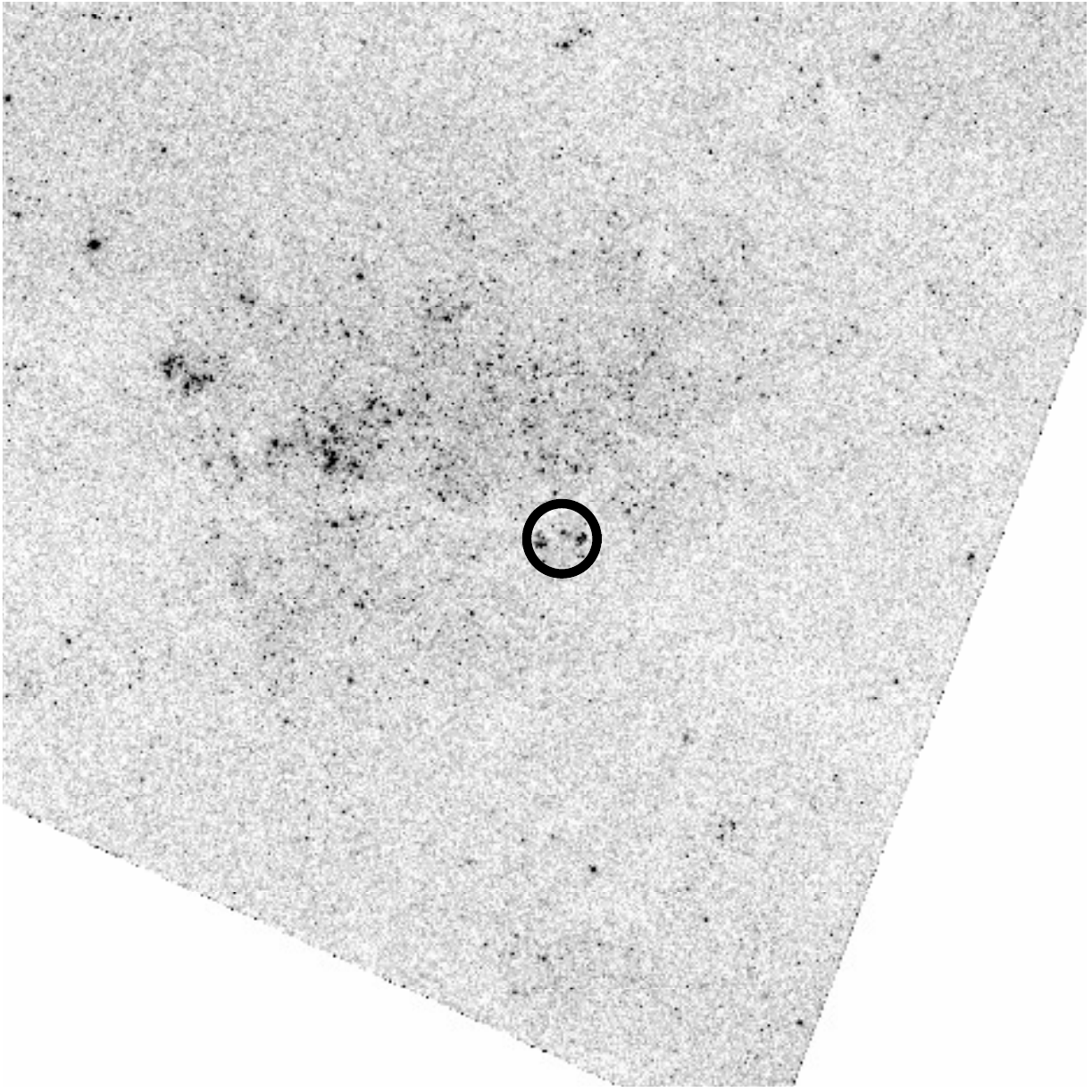}}
\subfigure[]{\includegraphics[width=0.24\columnwidth]{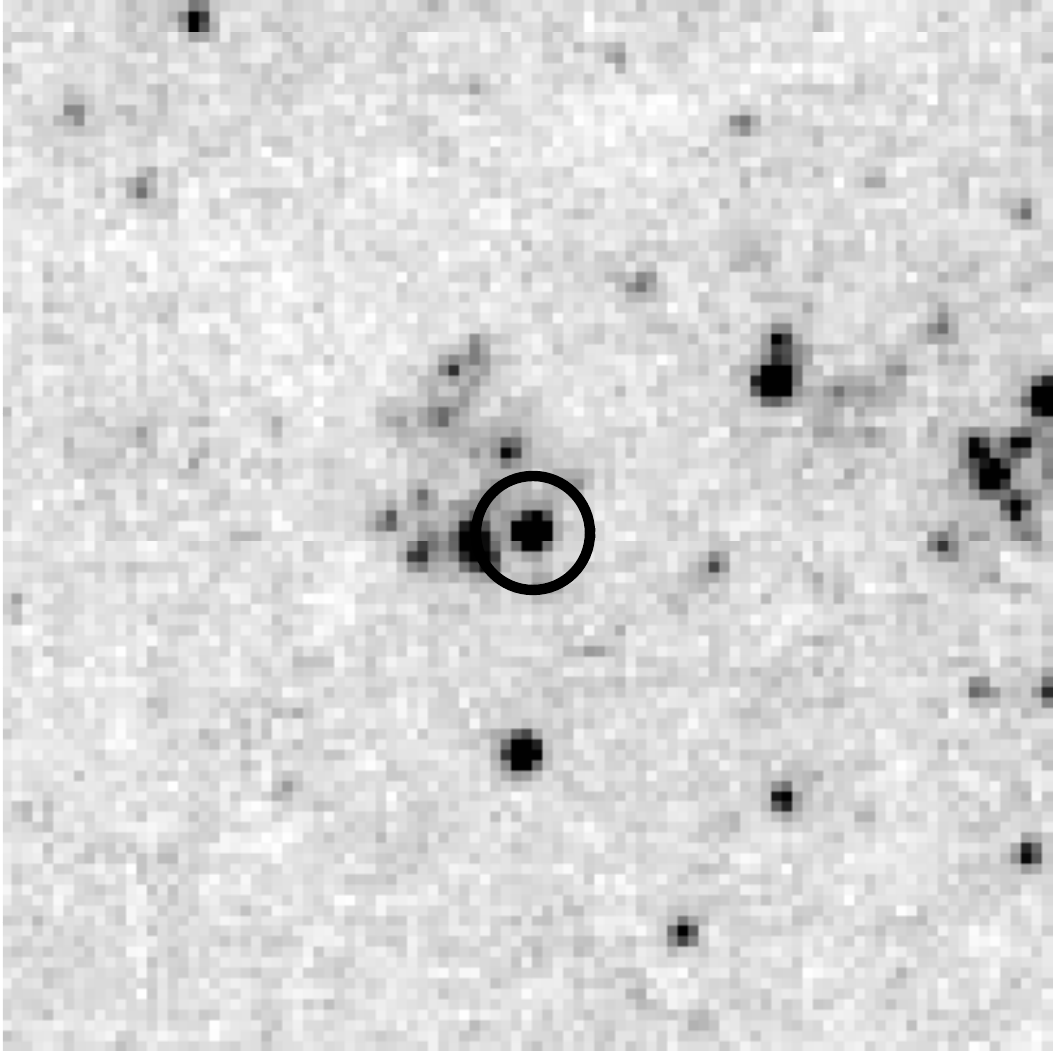}}
\caption{Source \# 112}

\end{figure}

\begin{figure}

\subfigure[]{\includegraphics[width=0.24\columnwidth]{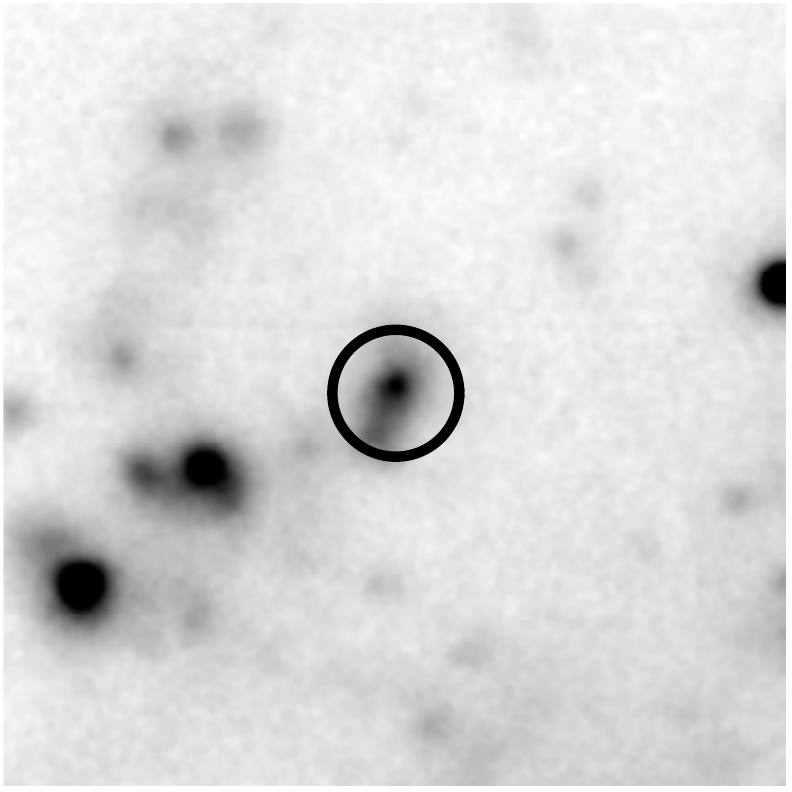}}
\subfigure[]{\includegraphics[width=0.24\columnwidth]{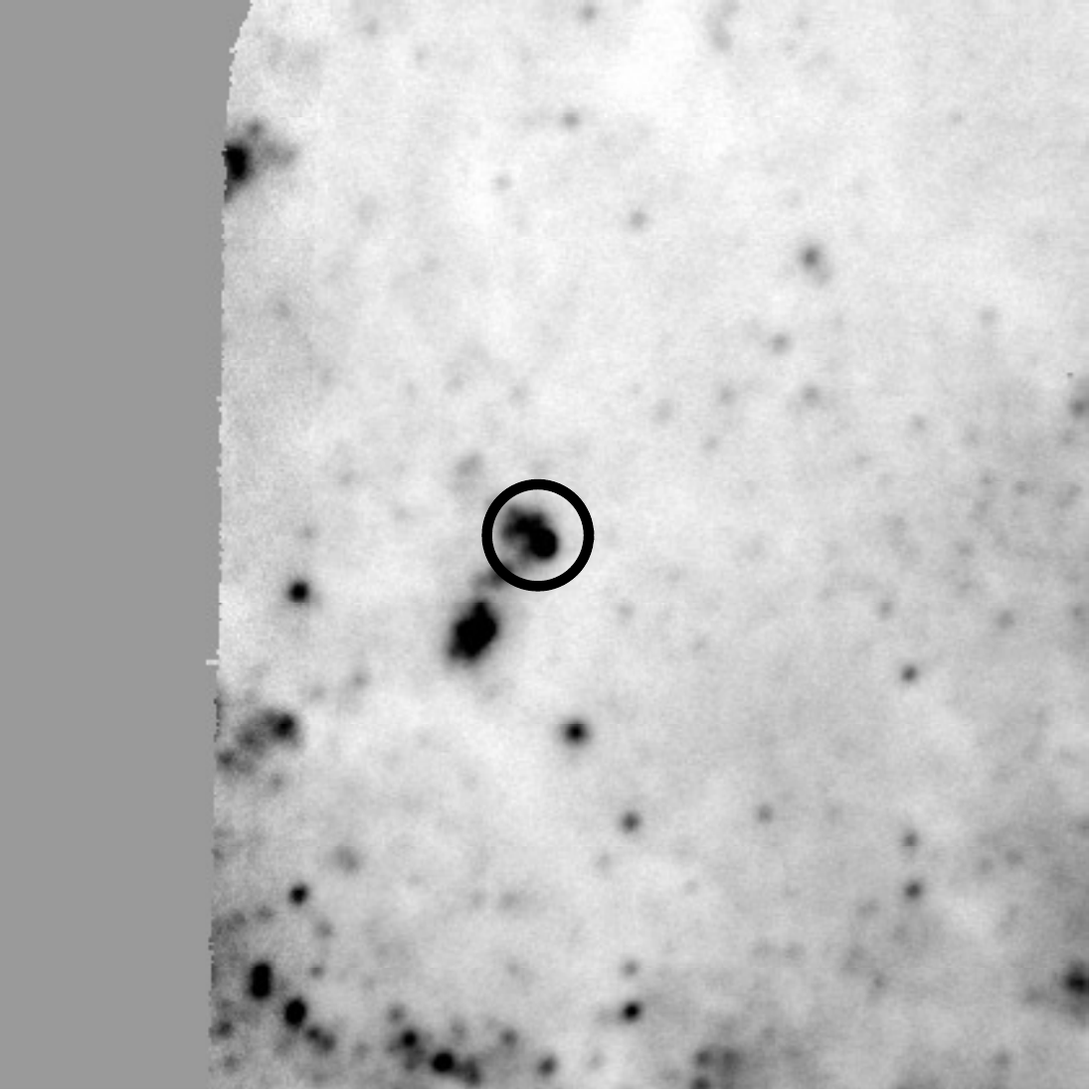}}
\subfigure[]{\includegraphics[width=0.24\columnwidth]{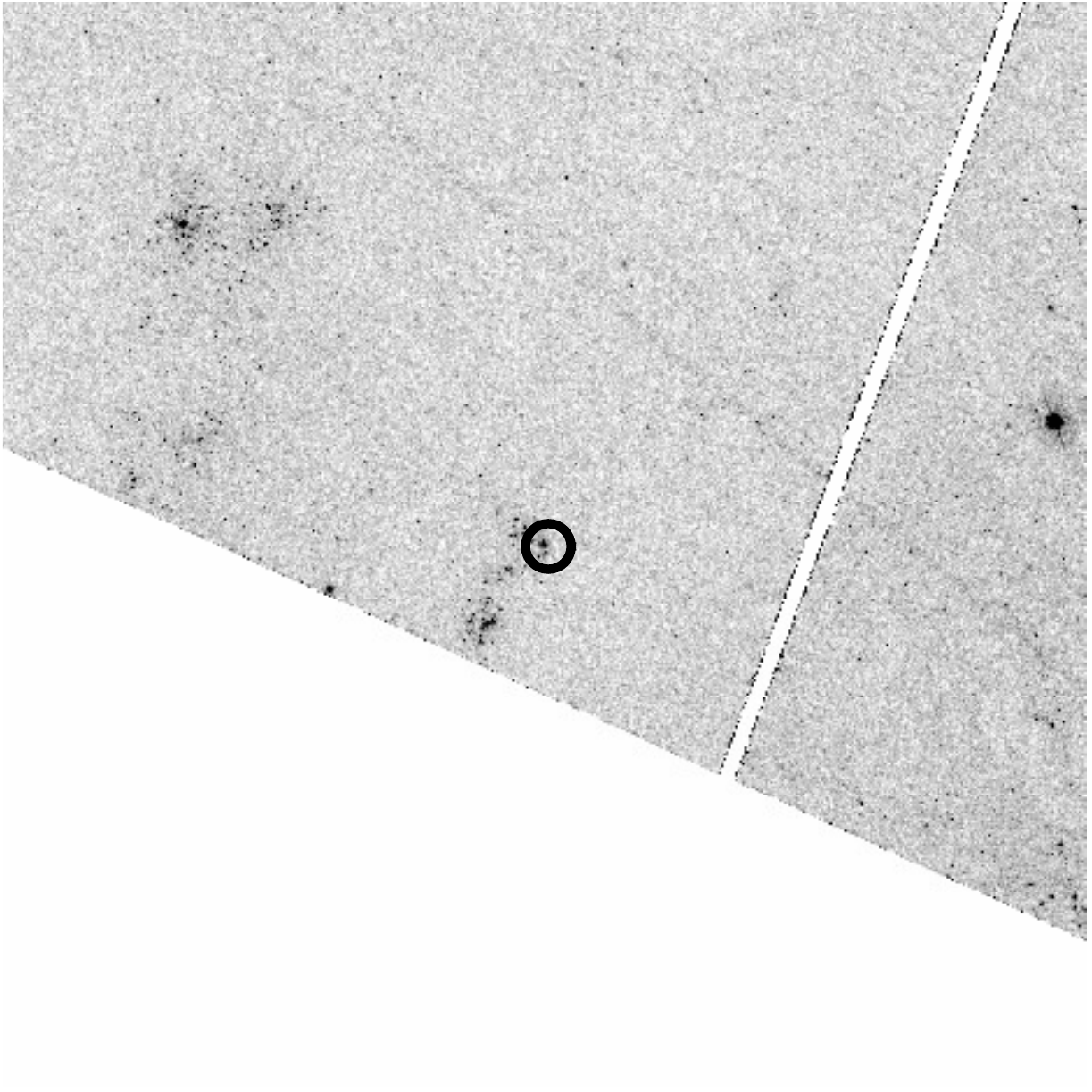}}
\subfigure[]{\includegraphics[width=0.24\columnwidth]{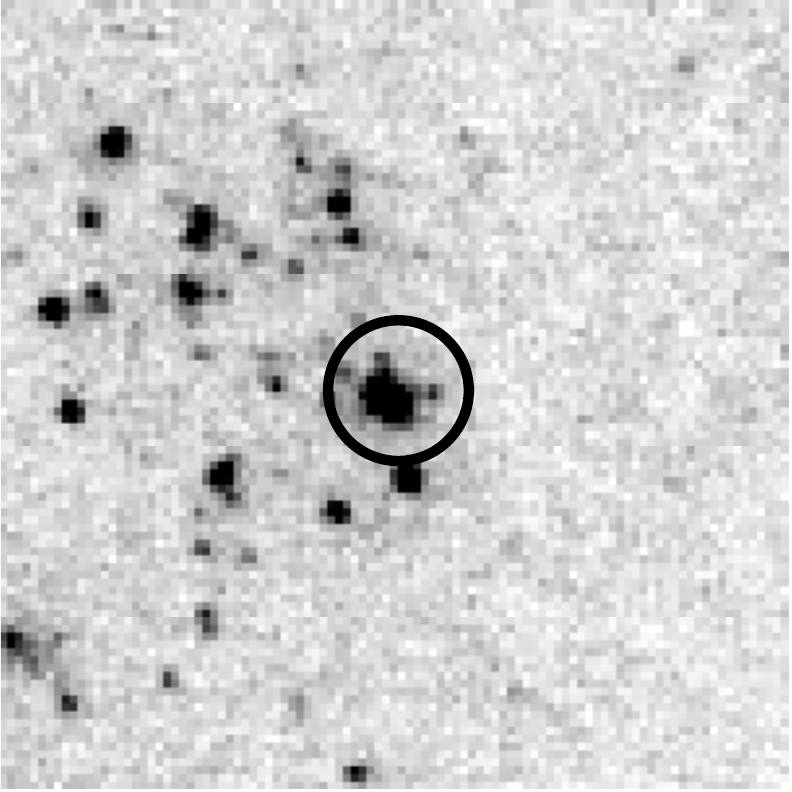}}

\caption{Source \# 114}

\end{figure}

\begin{figure}

\subfigure[]{\includegraphics[width=0.24\columnwidth]{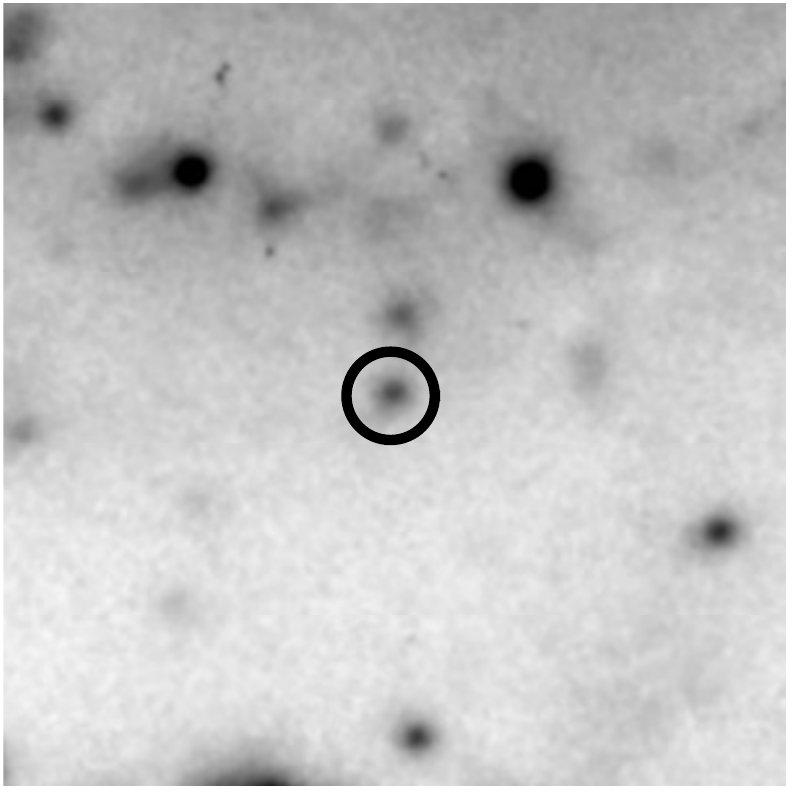}}
\subfigure[]{\includegraphics[width=0.24\columnwidth]{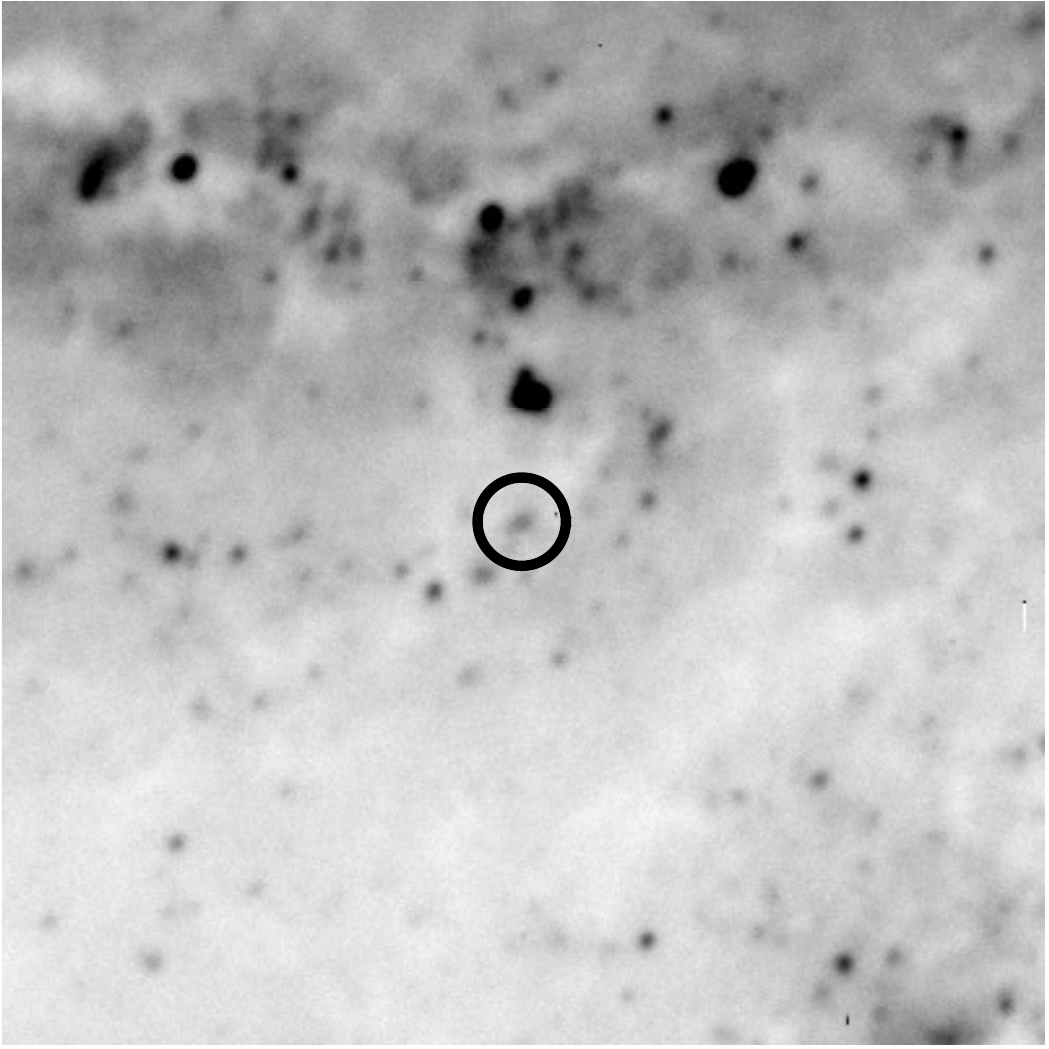}}
\subfigure[]{\includegraphics[width=0.24\columnwidth]{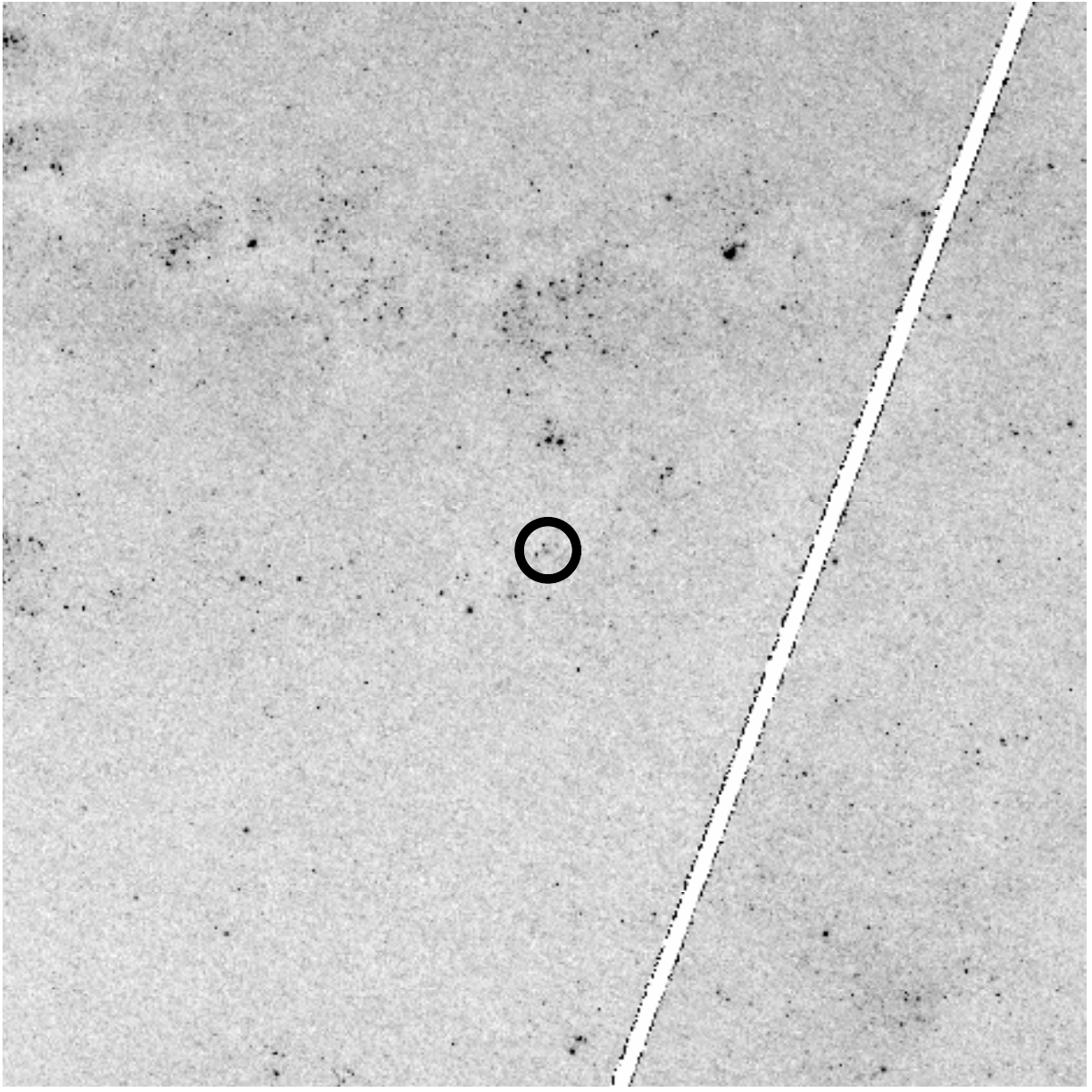}}
\subfigure[]{\includegraphics[width=0.24\columnwidth]{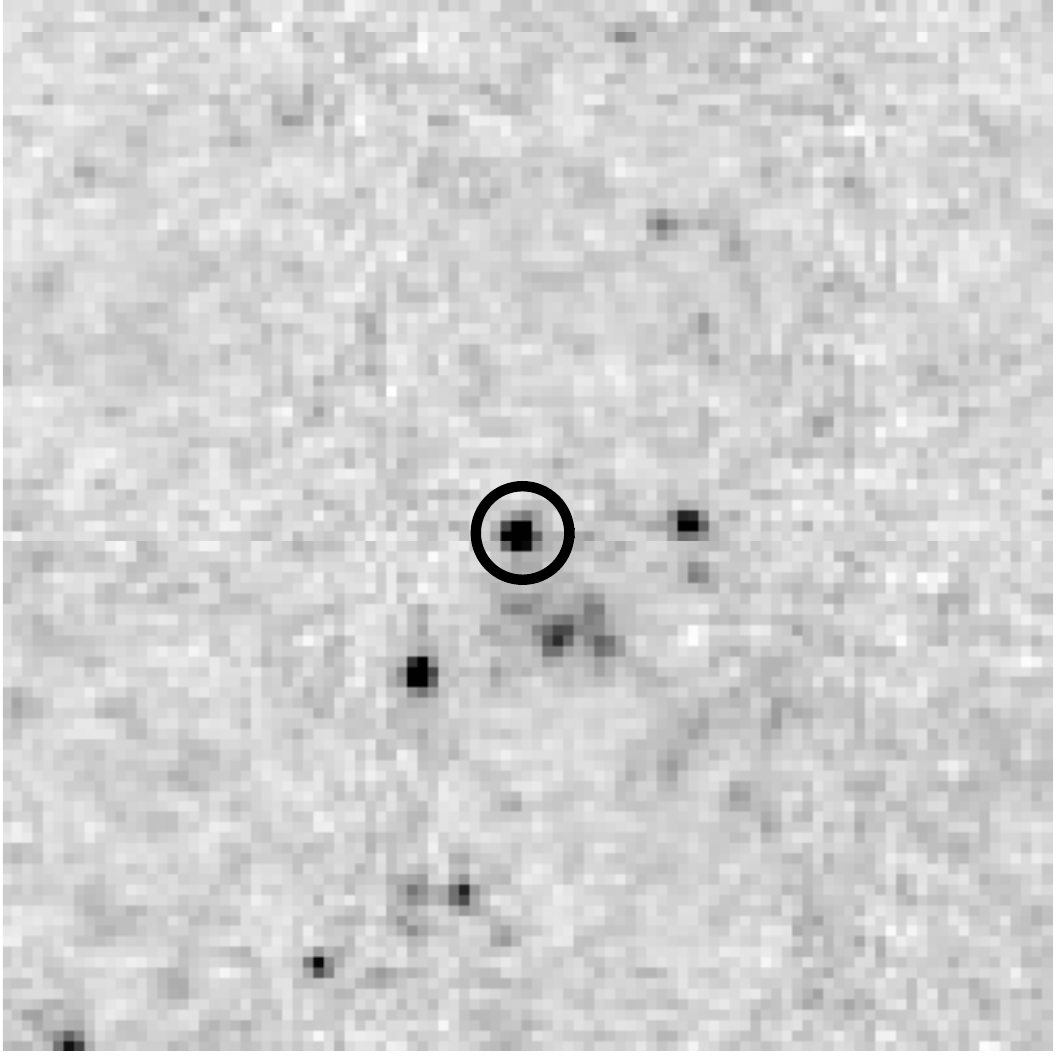}}

\caption{Source \# 1024}

\end{figure}

\begin{figure}

\subfigure[]{\includegraphics[width=0.24\columnwidth]{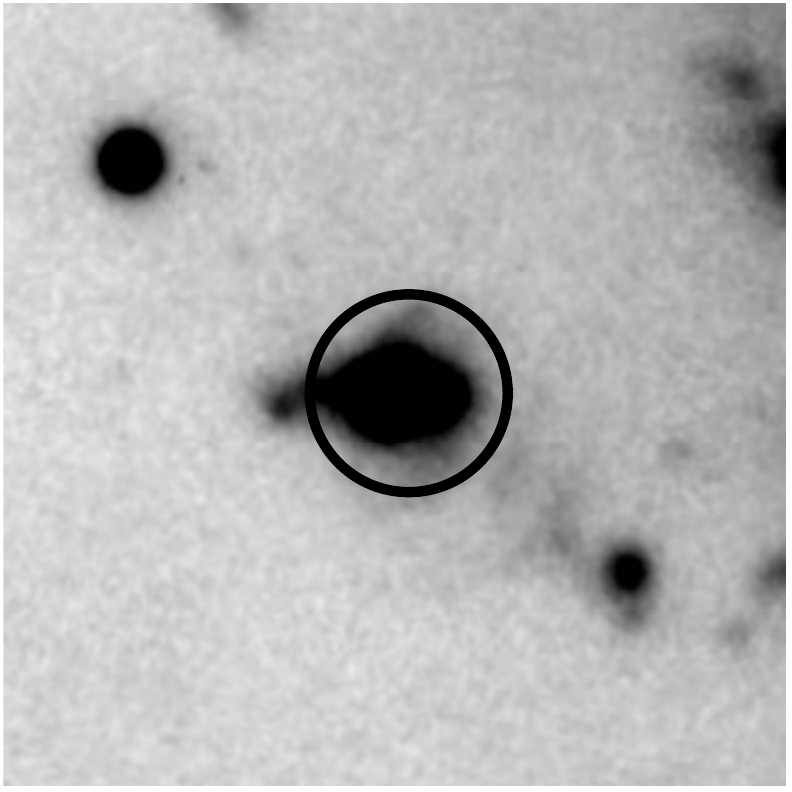}}
\subfigure[]{\includegraphics[width=0.24\columnwidth]{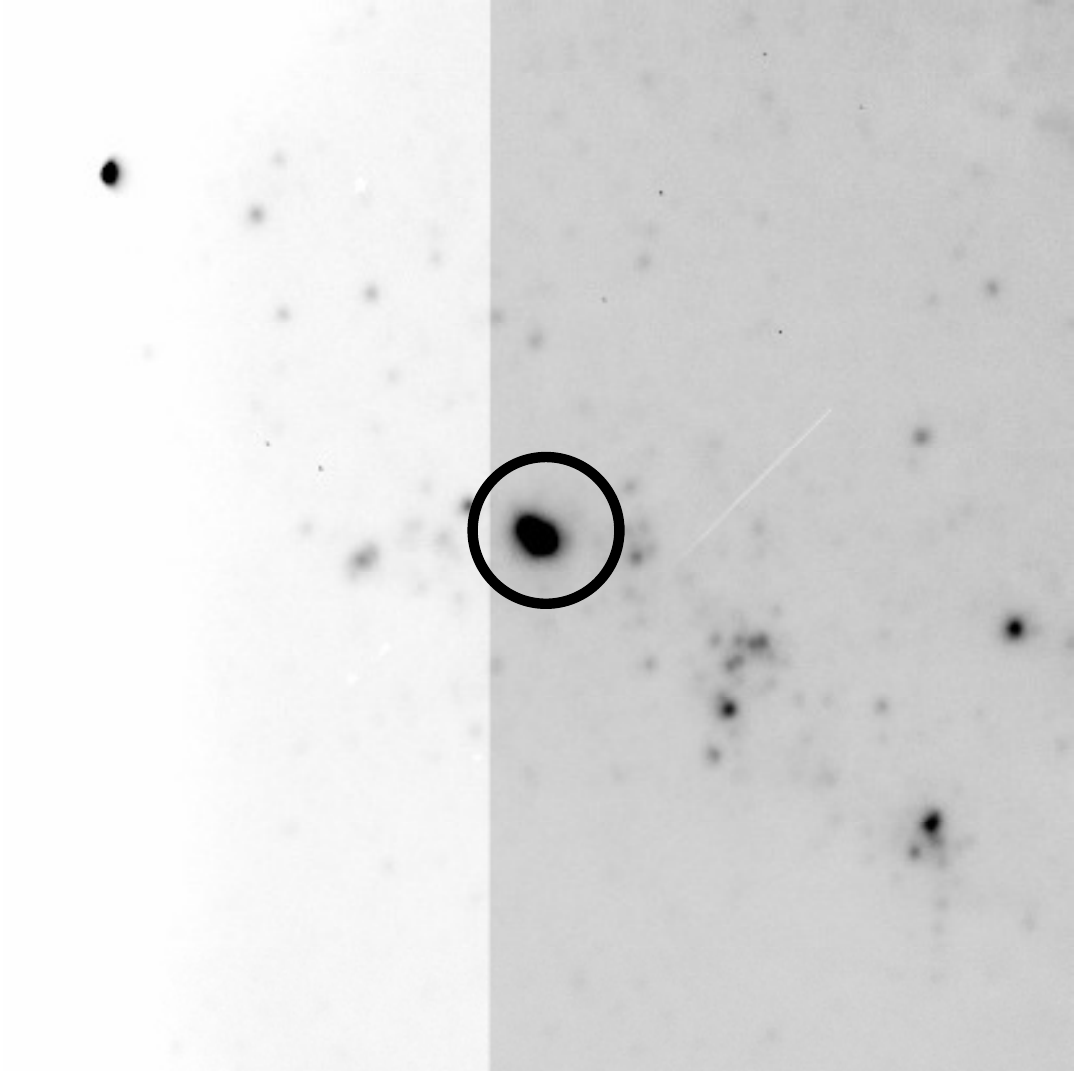}}
\subfigure[]{\includegraphics[width=0.24\columnwidth]{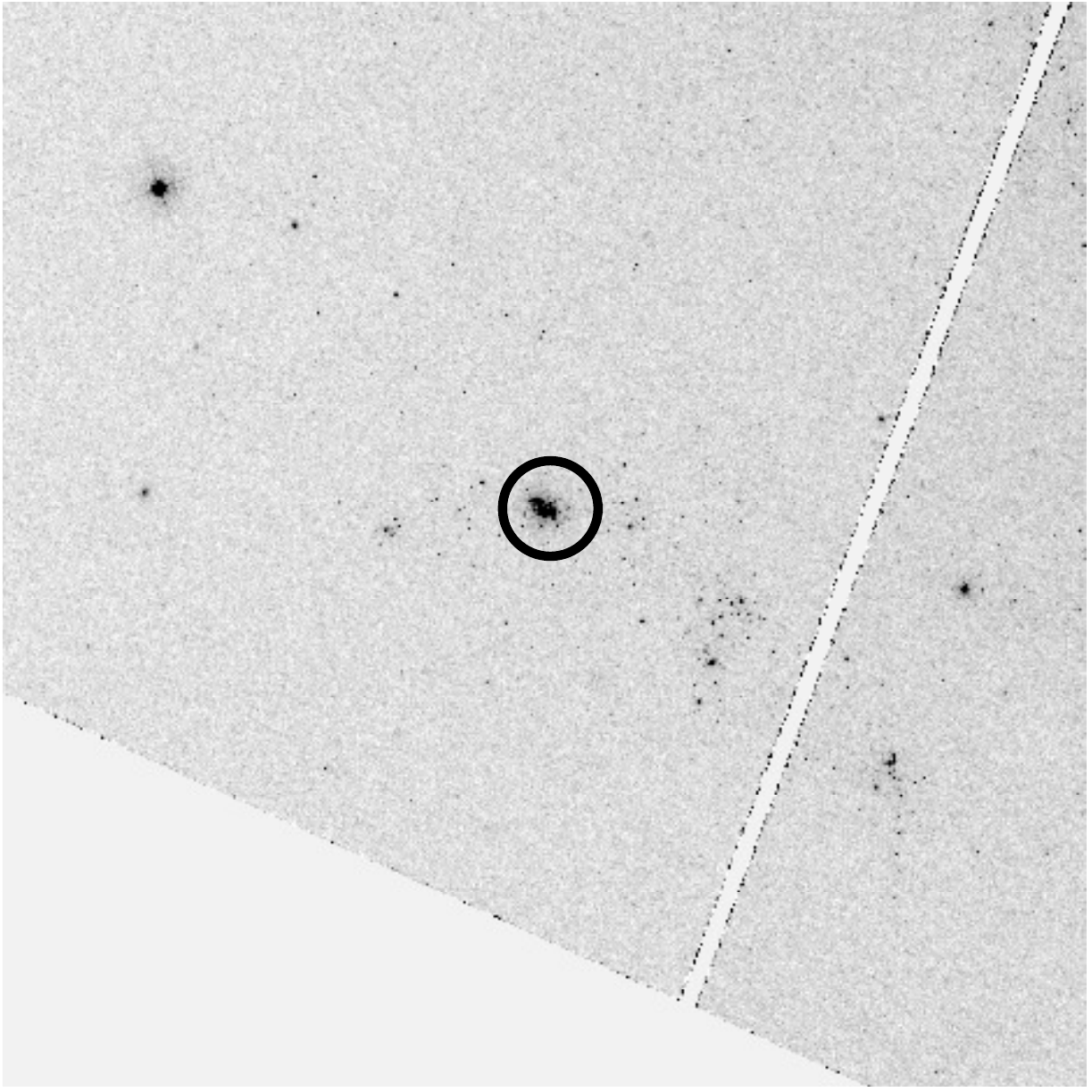}}
\subfigure[]{\includegraphics[width=0.24\columnwidth]{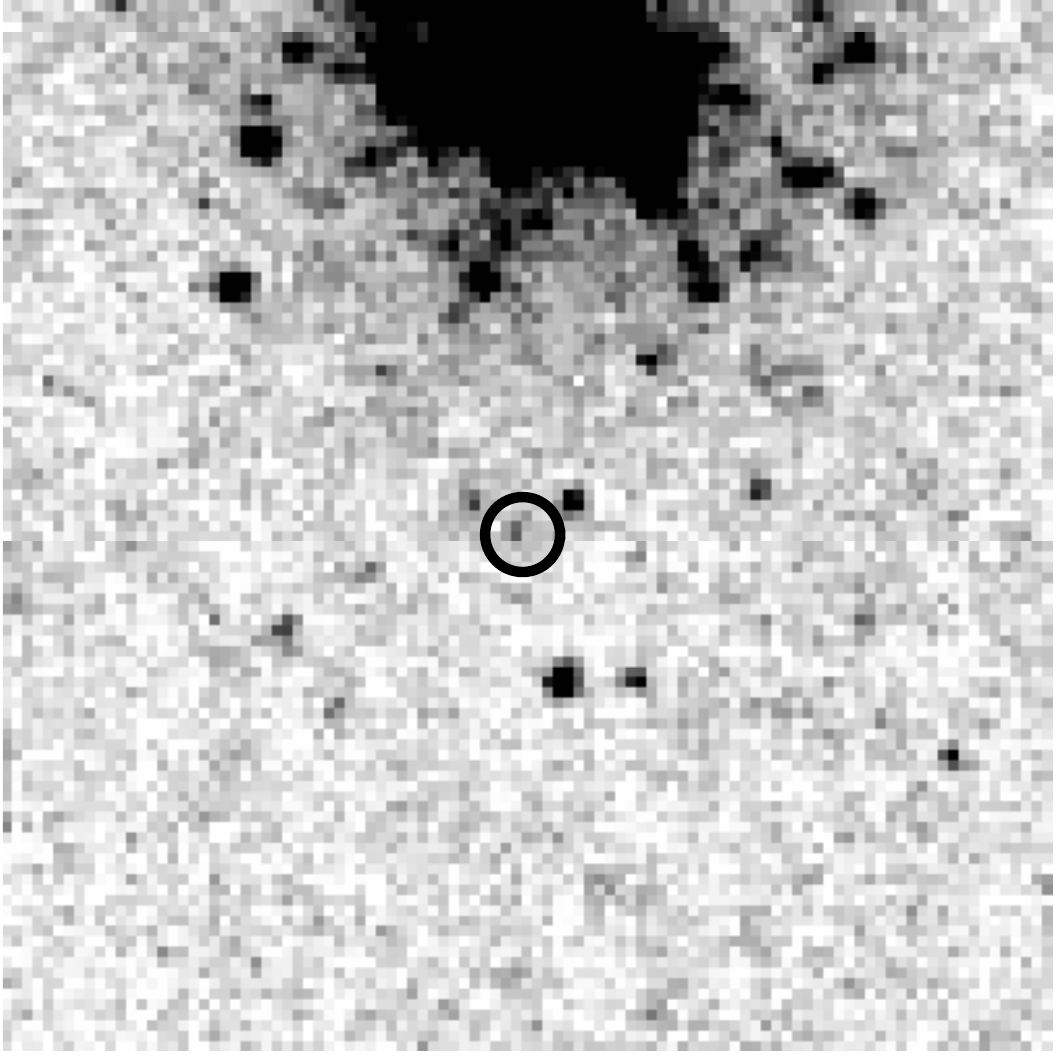}}

\caption{Source \# 2053}

\end{figure}

\clearpage
\section[]{Finding Charts for WR stars}
\label{finding_charts_all}
This appendix contains 2$\times$2\,arcmin stamps of each WR star in
the a) F469N, b) F435W, c) F555W, d) F814W and e) continuum subtracted F469N filters.
\clearpage

\begin{figure}

\subfigure[]{\includegraphics[width=0.19\columnwidth]{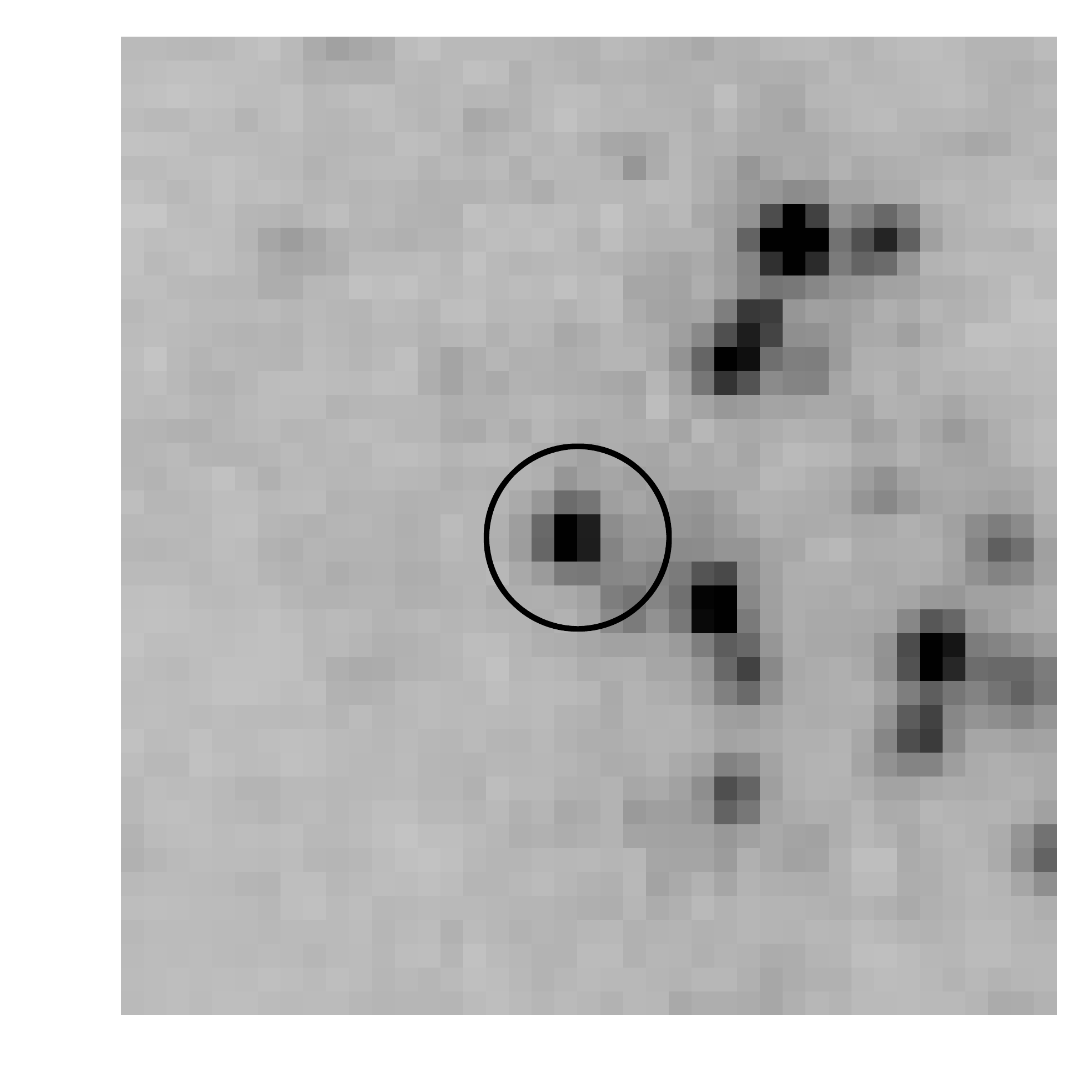}}
\subfigure[]{\includegraphics[width=0.19\columnwidth]{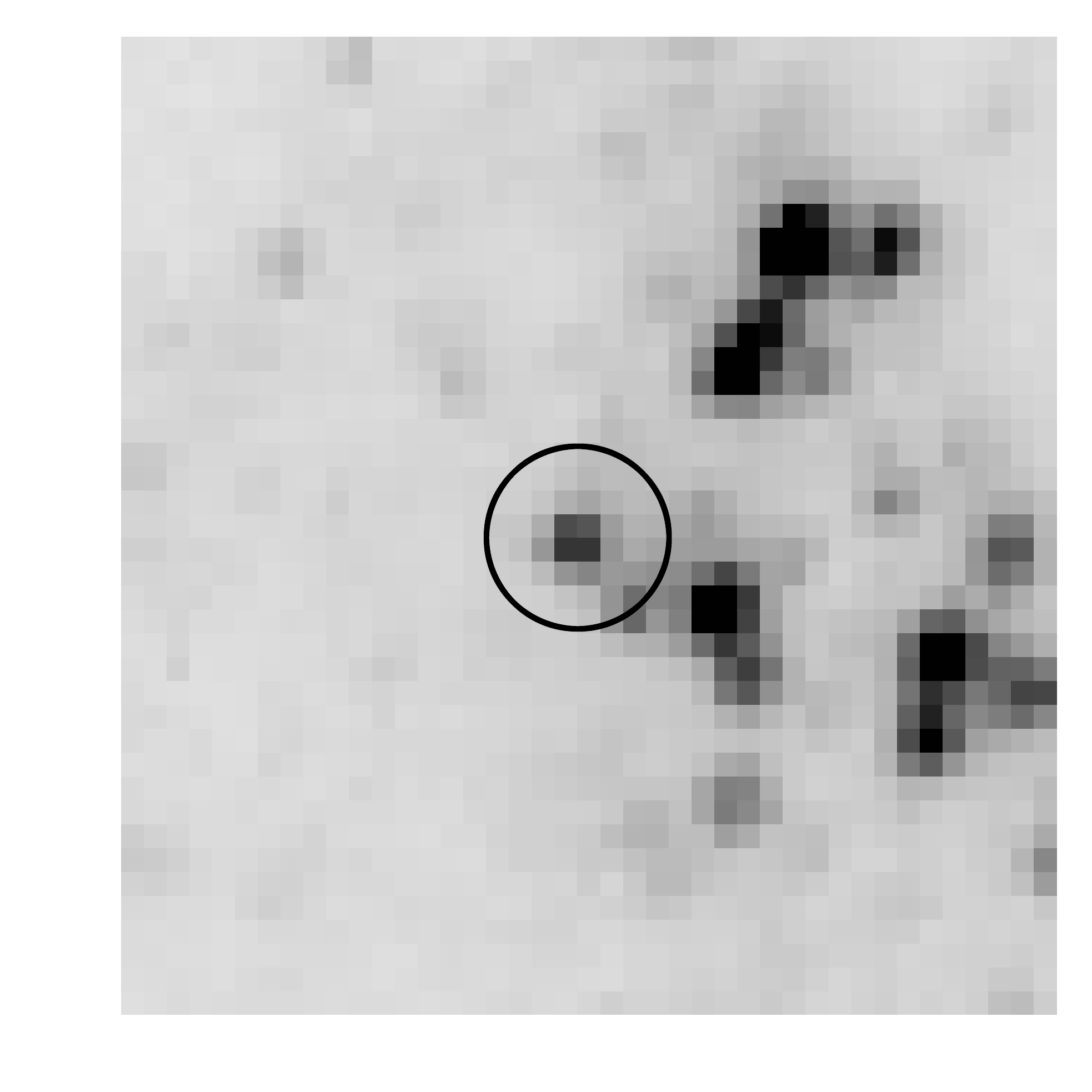}}
\subfigure[]{\includegraphics[width=0.19\columnwidth]{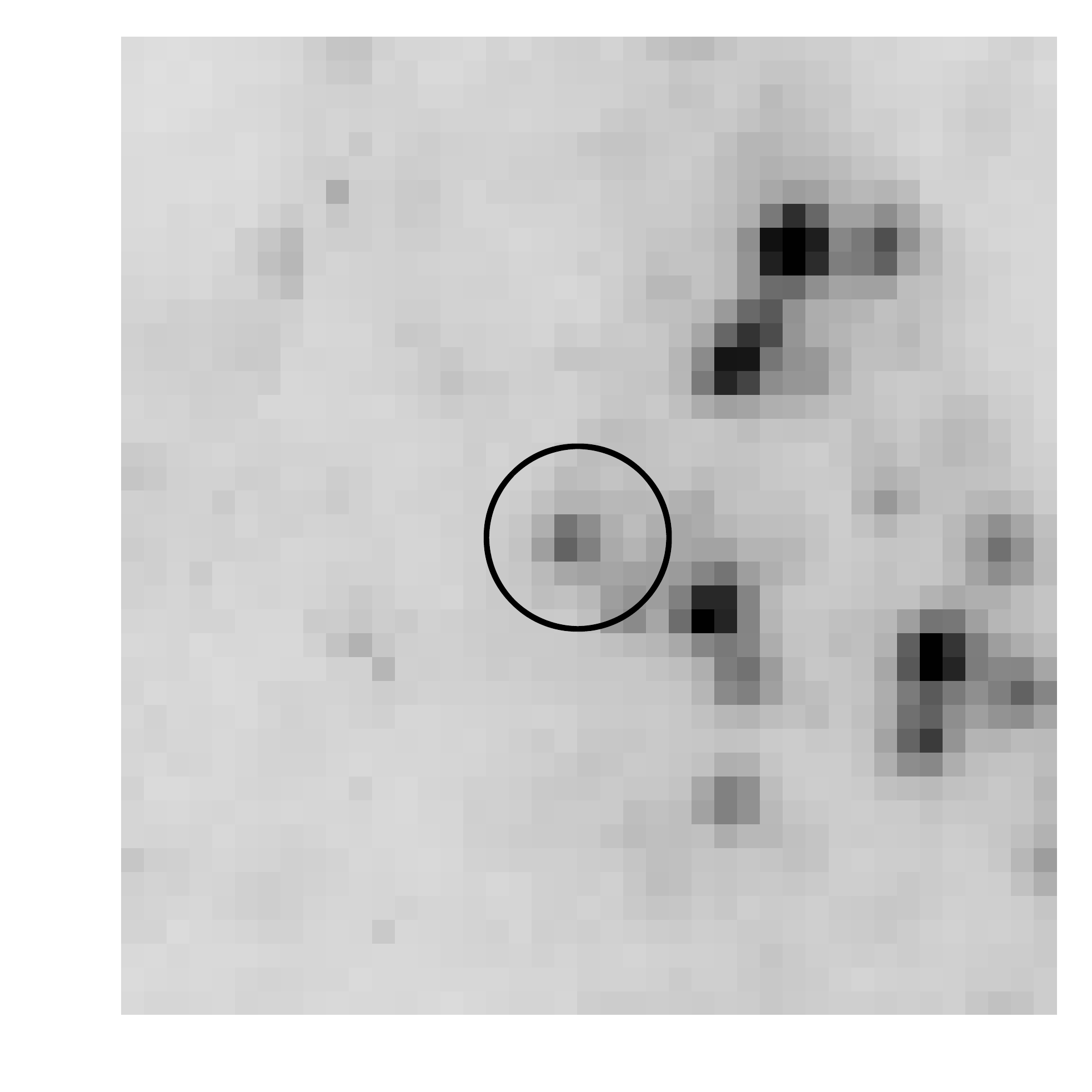}}
\subfigure[]{\includegraphics[width=0.19\columnwidth]{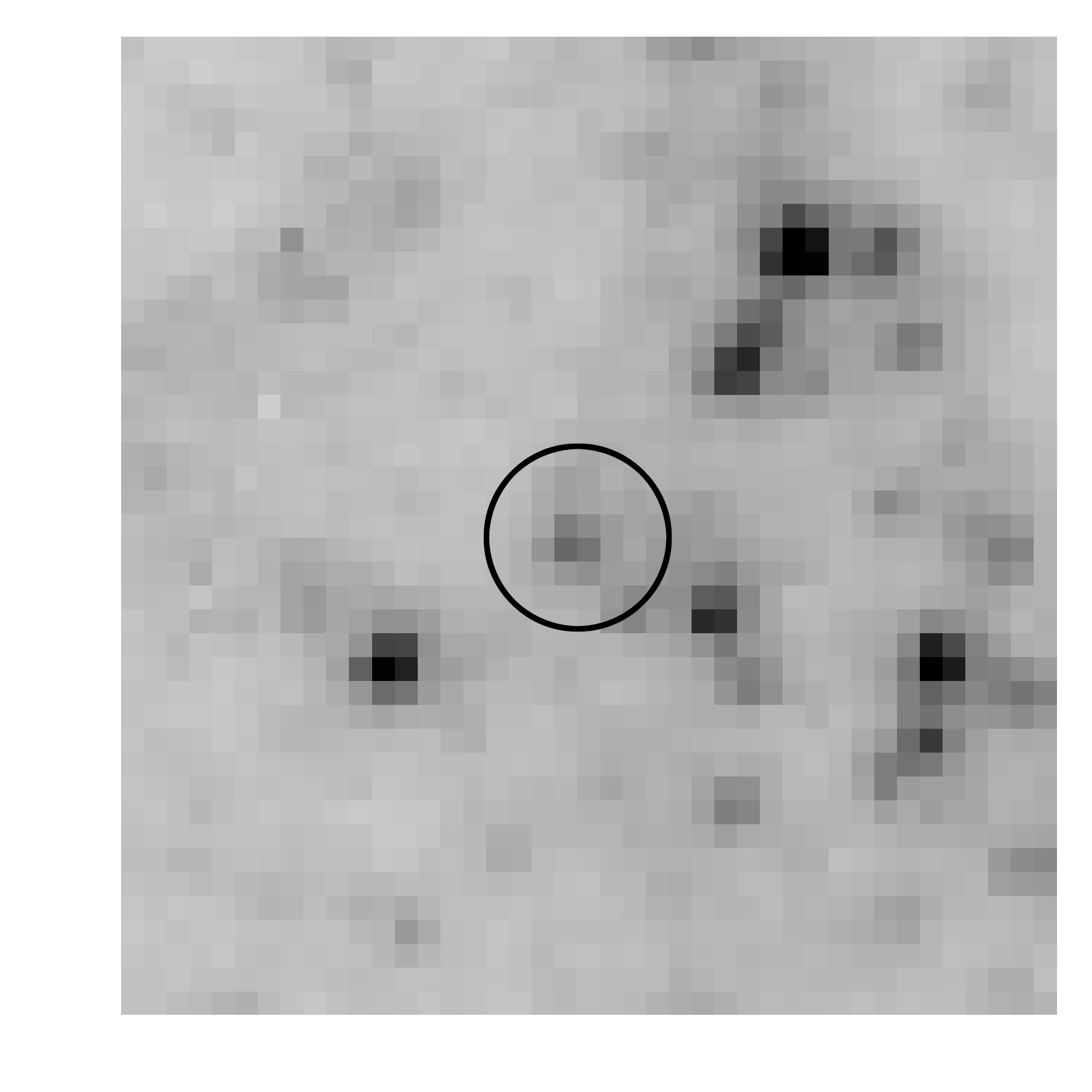}}
\subfigure[]{\includegraphics[width=0.19\columnwidth]{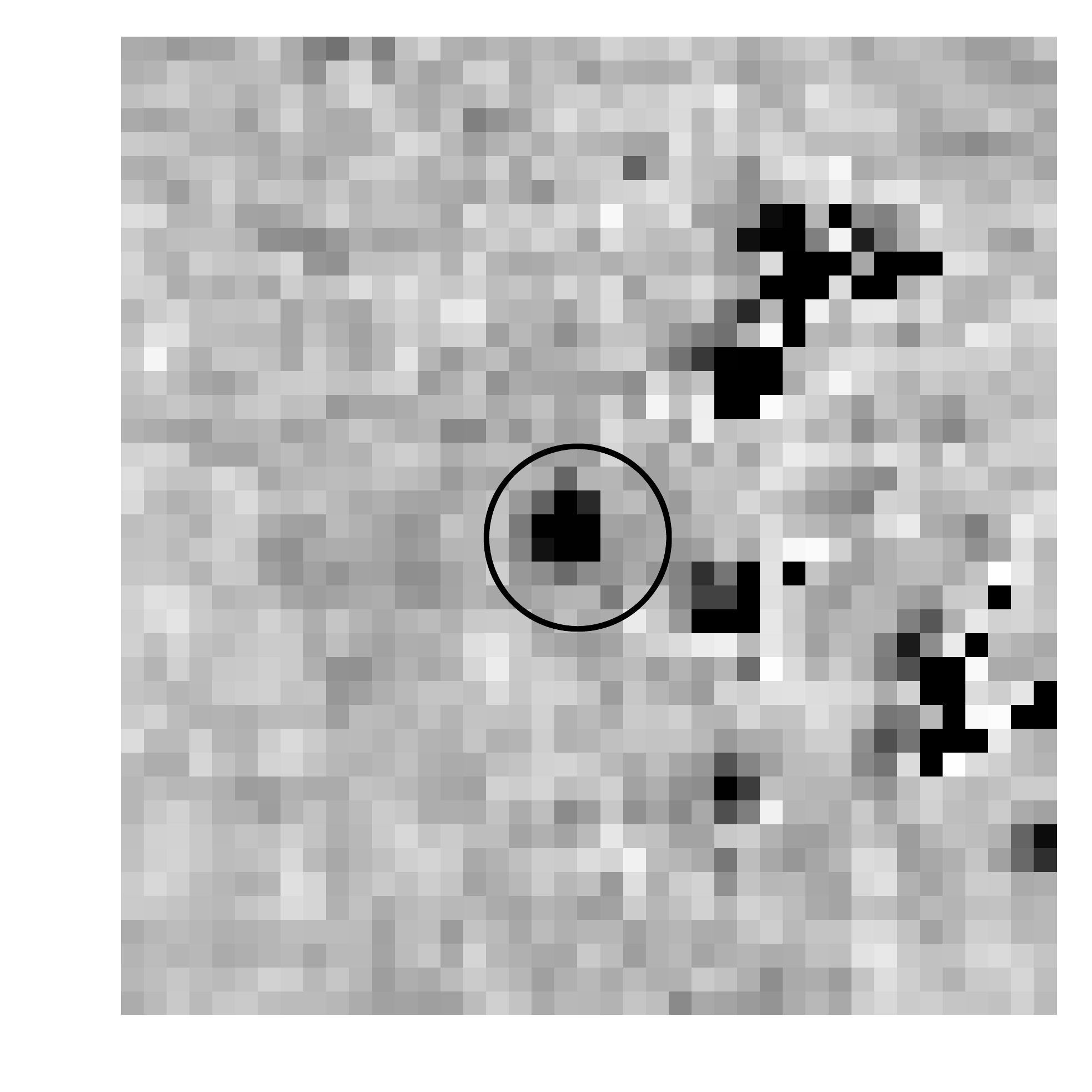}}

\caption{Source \# 1012 in F469N, F435W, F555W, F814W and continuum
  subtracted F469N filters, respectively}

\end{figure}

\begin{figure}

\subfigure[]{\includegraphics[width=0.19\columnwidth]{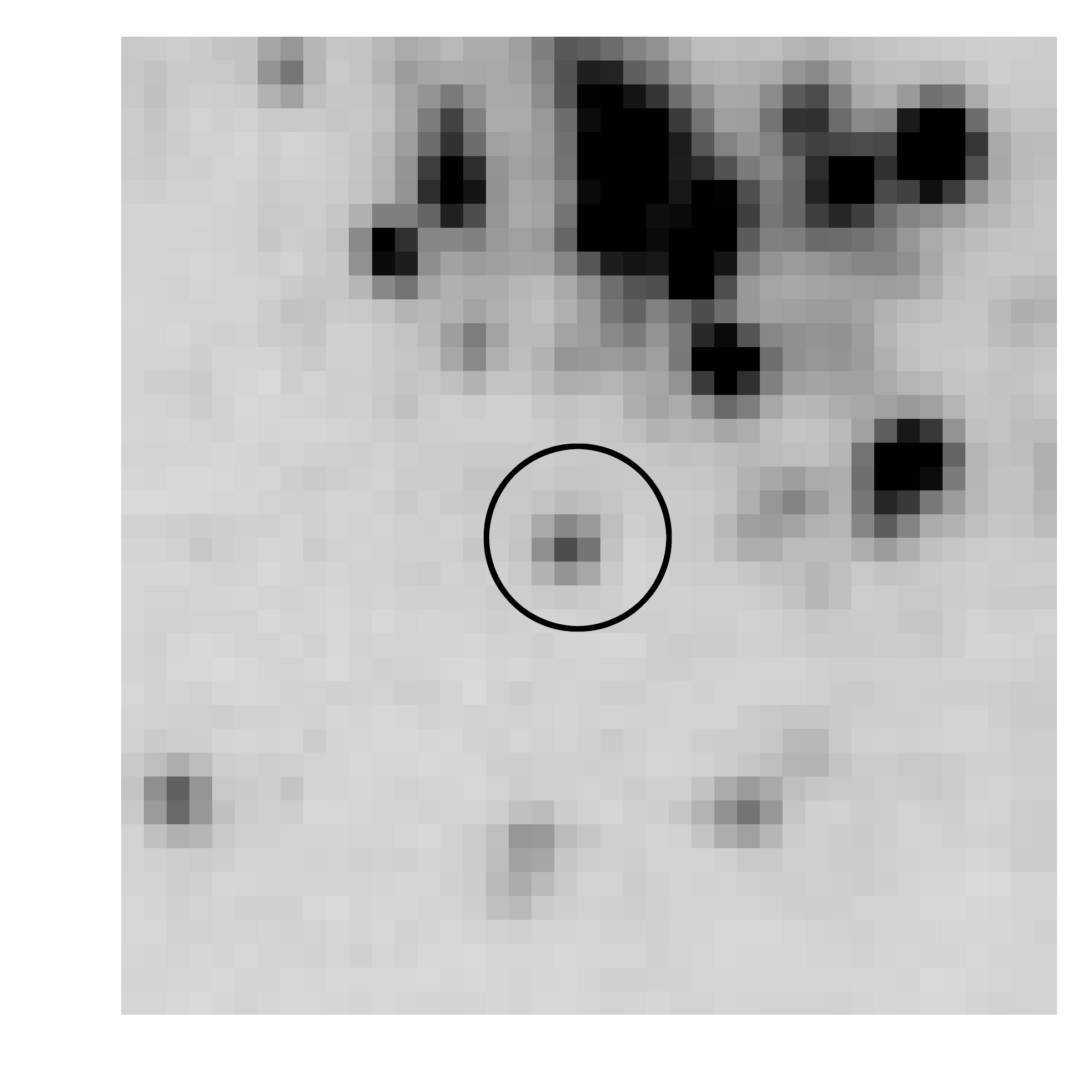}}
\subfigure[]{\includegraphics[width=0.19\columnwidth]{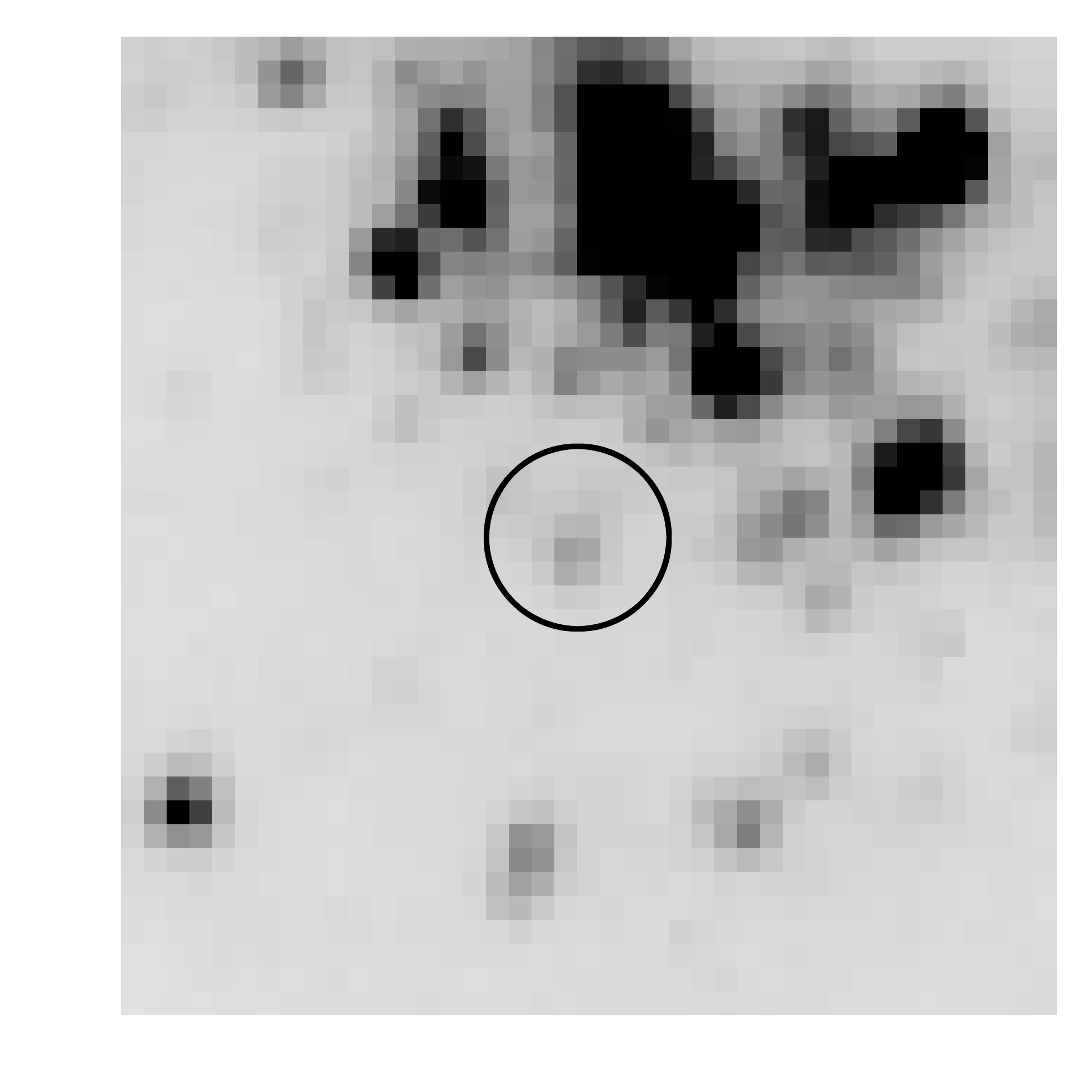}}
\subfigure[]{\includegraphics[width=0.19\columnwidth]{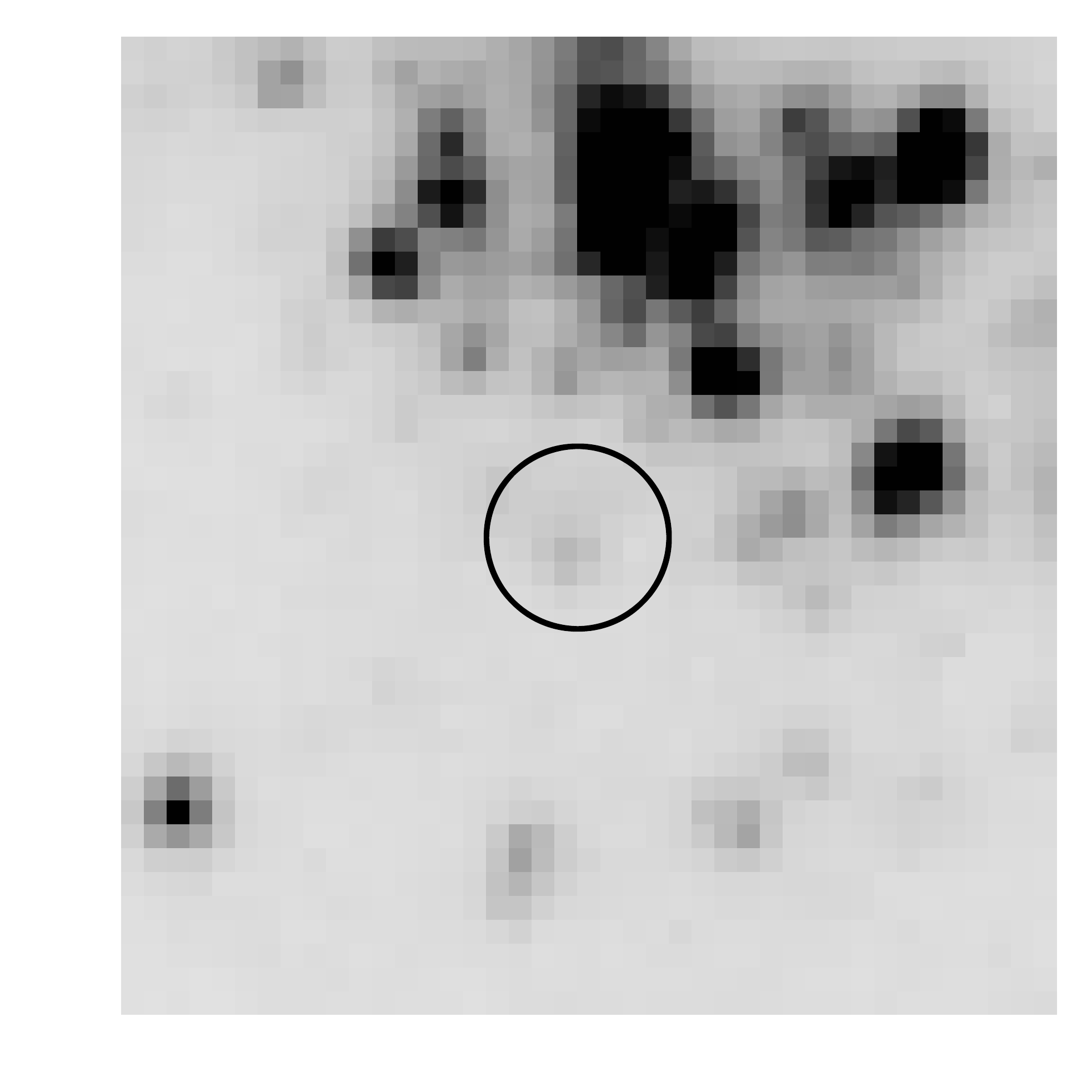}}
\subfigure[]{\includegraphics[width=0.19\columnwidth]{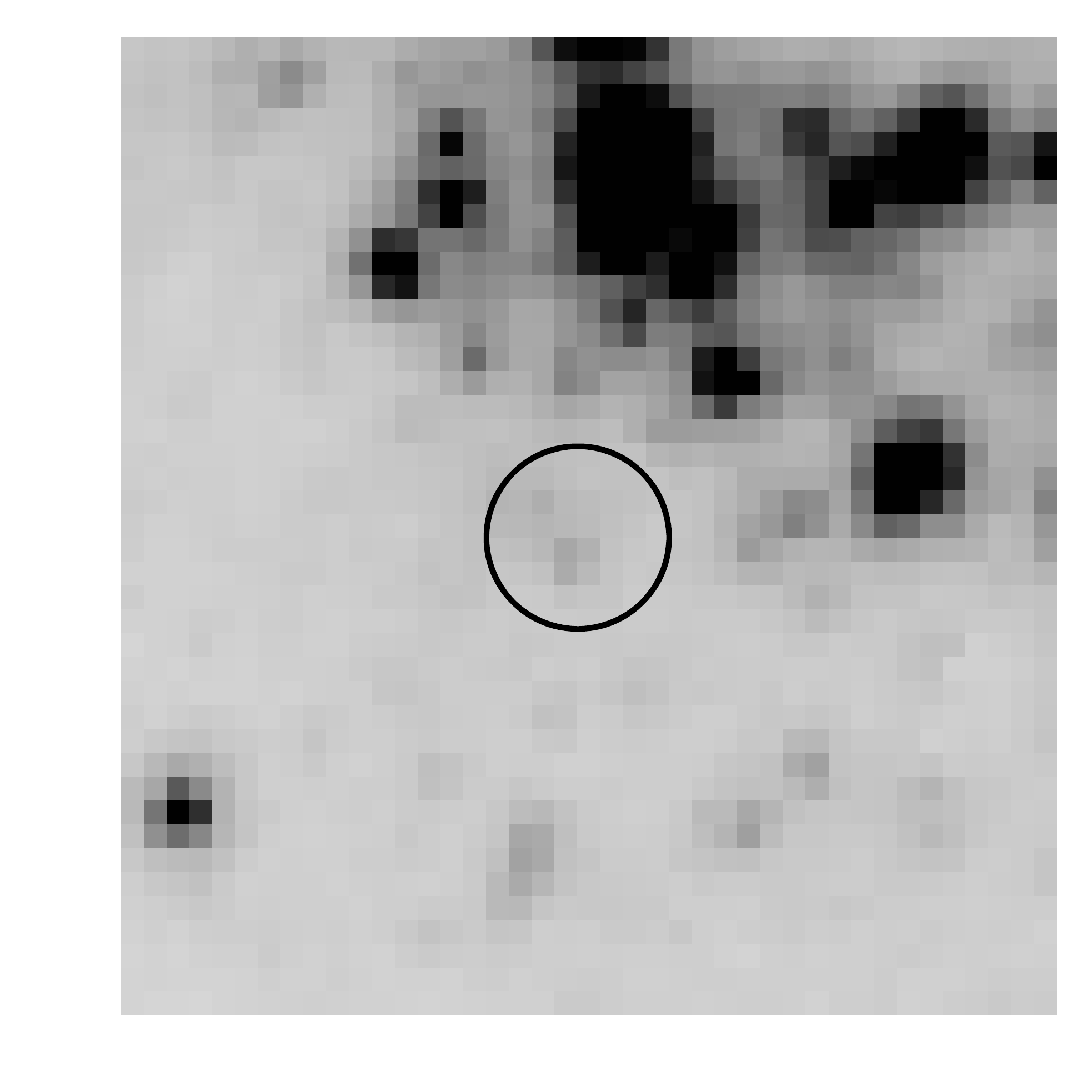}}
\subfigure[]{\includegraphics[width=0.19\columnwidth]{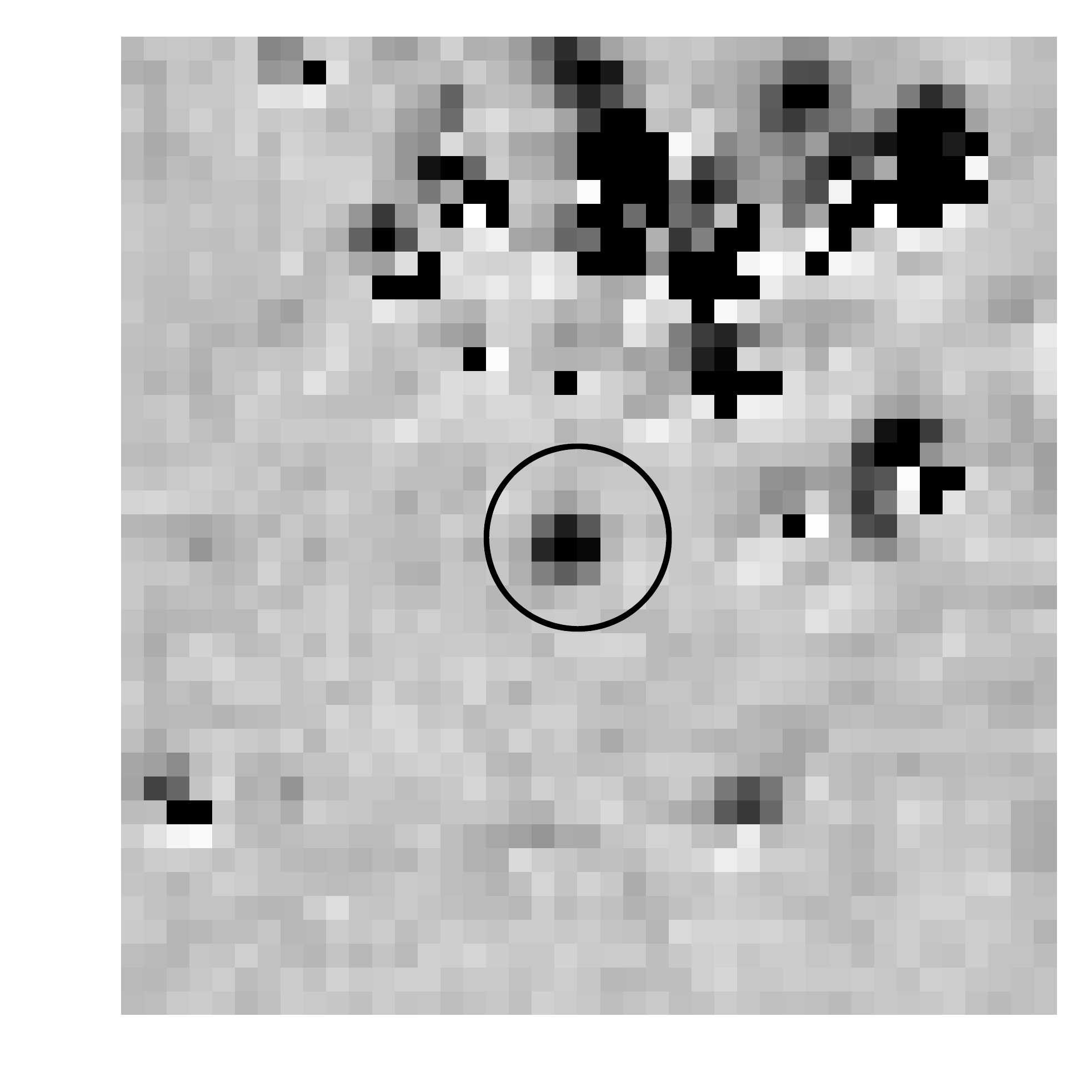}}

\caption{Source \# 1016 in F469N, F435W, F555W, F814W and continuum
  subtracted F469N filters, respectively}

\end{figure}

\begin{figure}

\subfigure[]{\includegraphics[width=0.19\columnwidth]{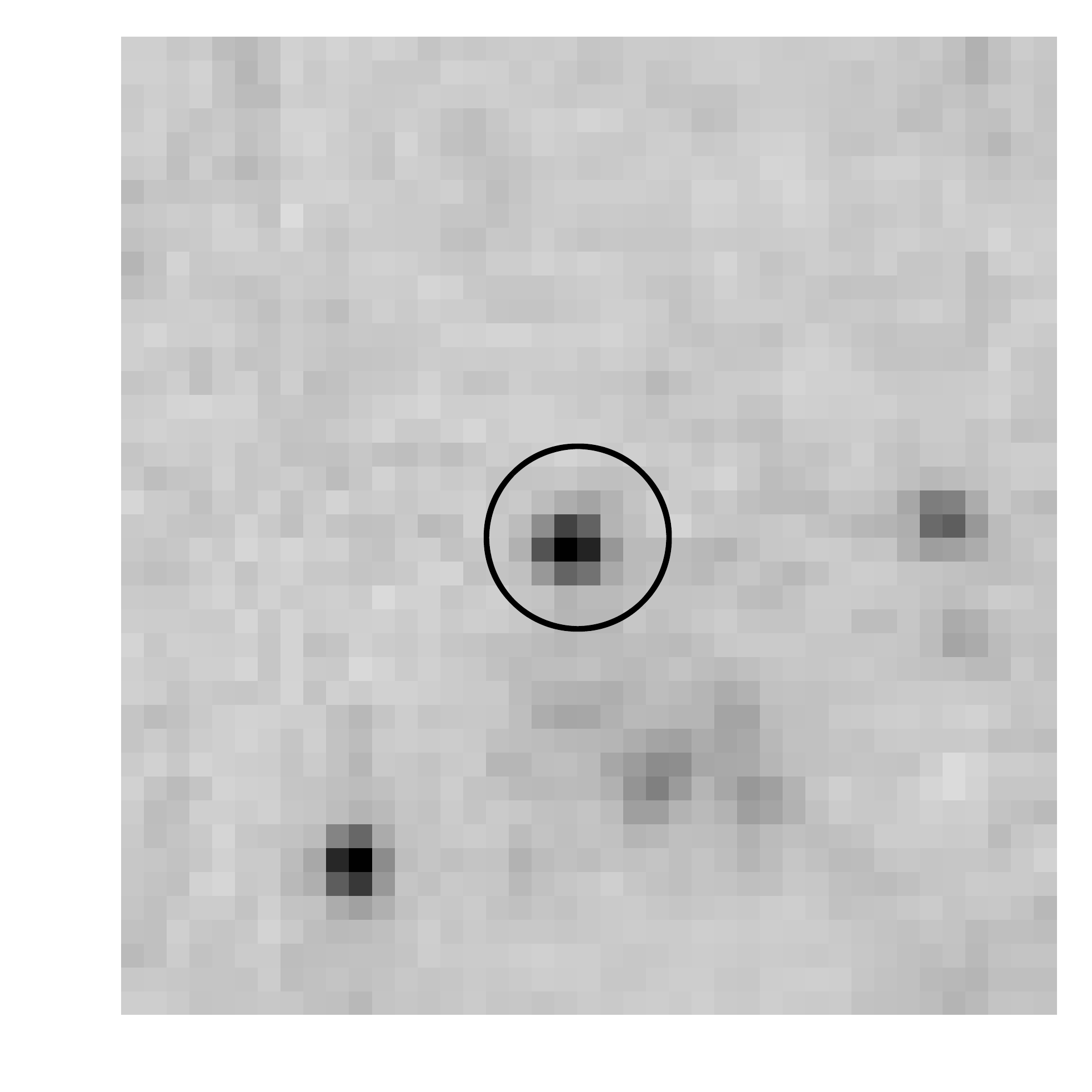}}
\subfigure[]{\includegraphics[width=0.19\columnwidth]{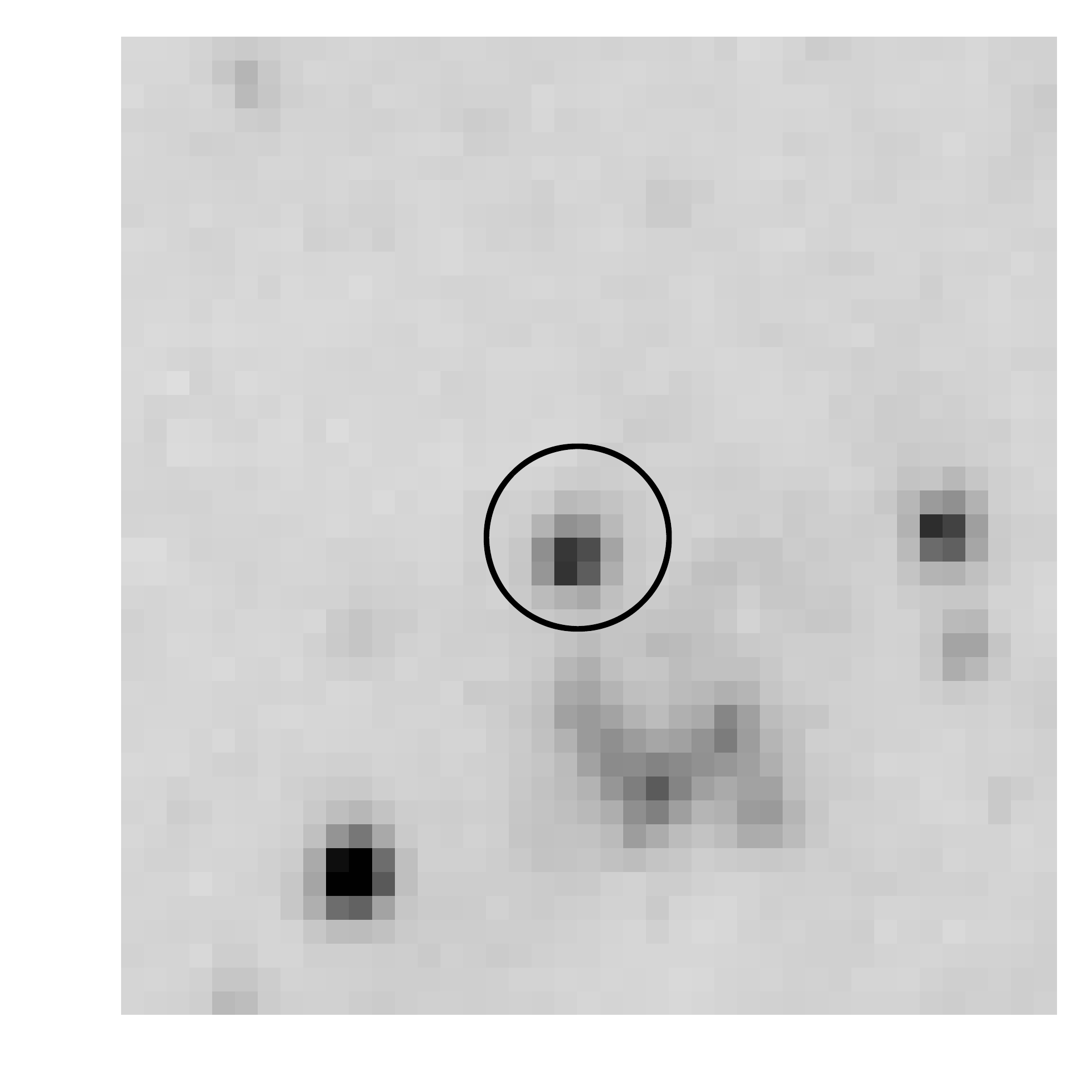}}
\subfigure[]{\includegraphics[width=0.19\columnwidth]{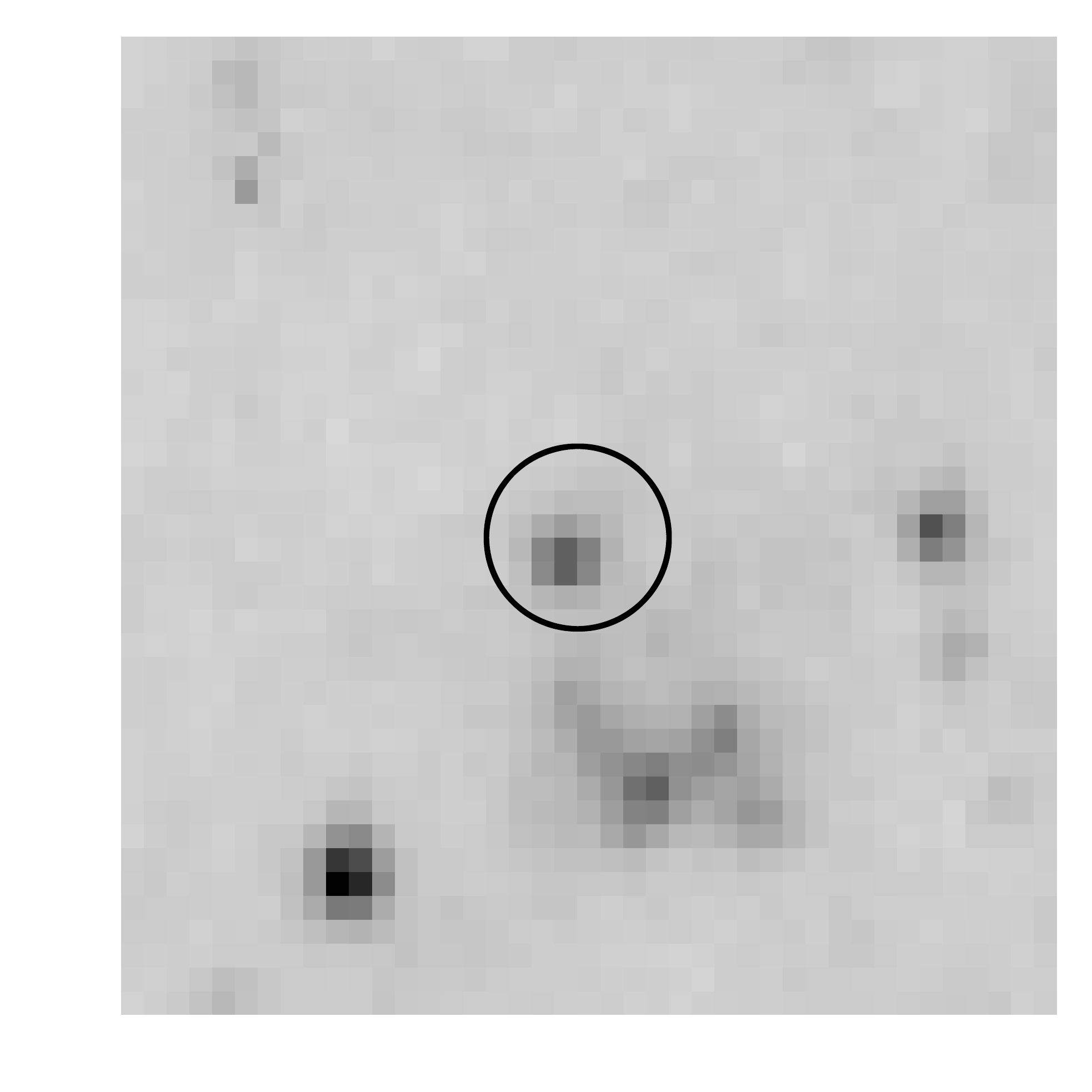}}
\subfigure[]{\includegraphics[width=0.19\columnwidth]{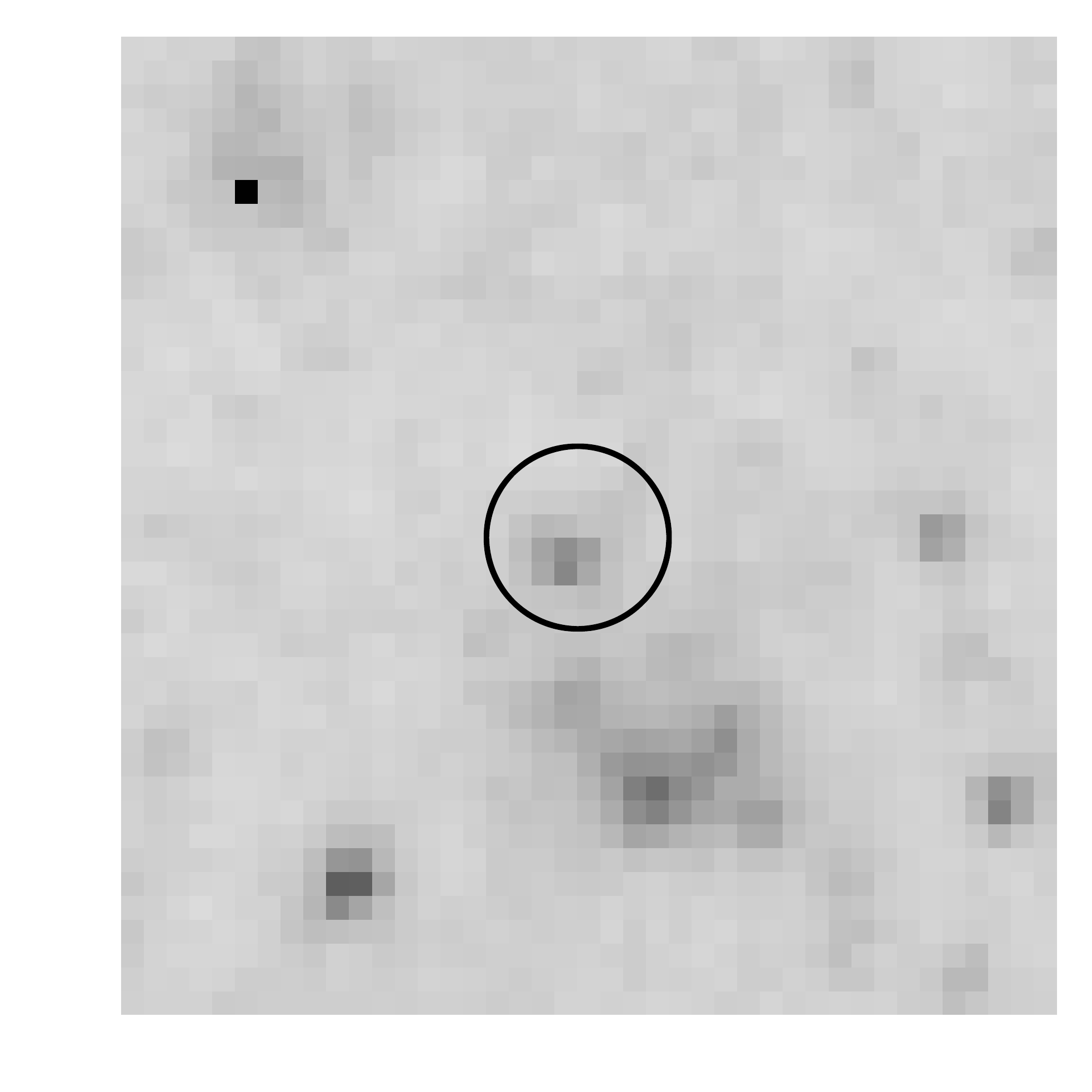}}
\subfigure[]{\includegraphics[width=0.19\columnwidth]{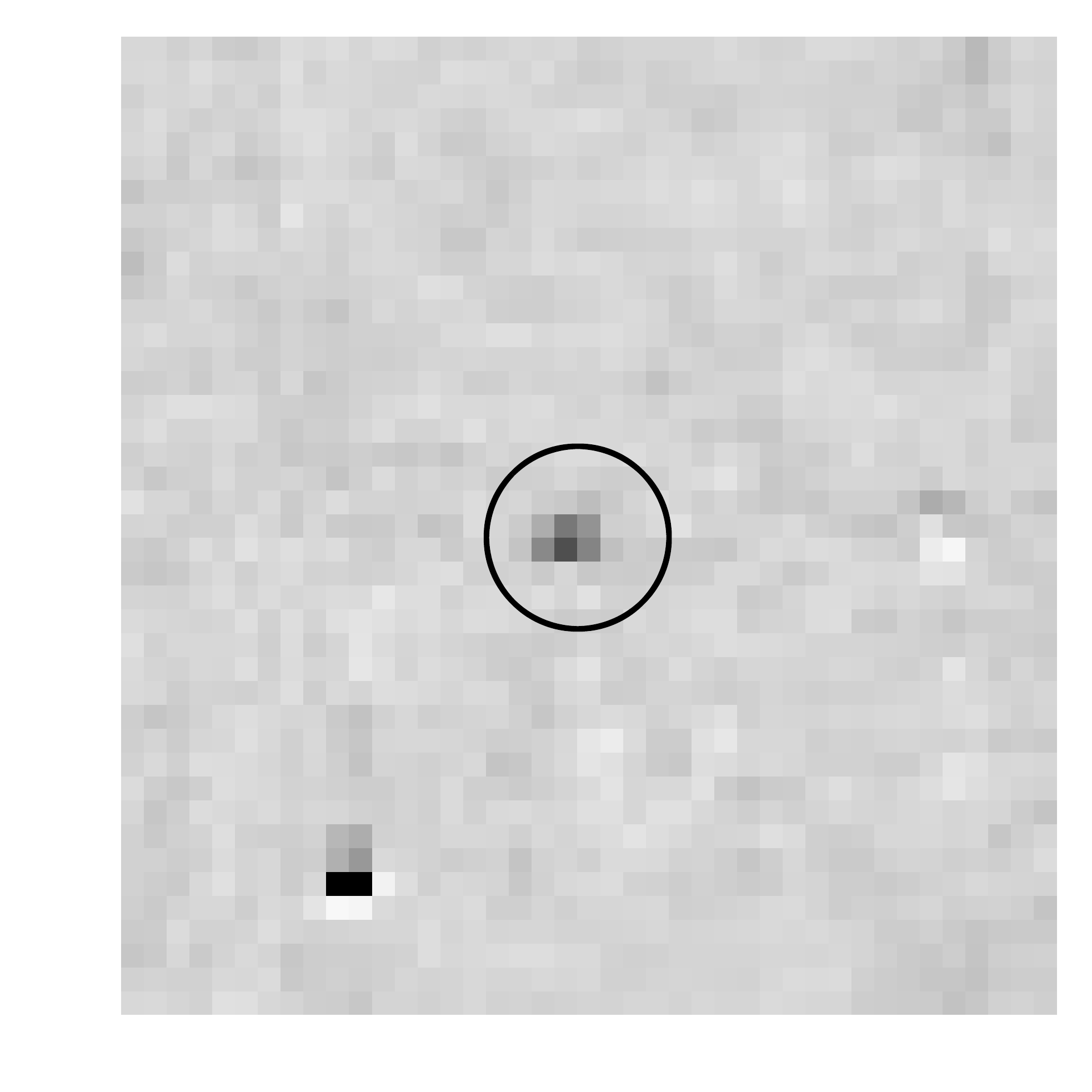}}

\caption{Source \# 1024 in F469N, F435W, F555W, F814W and continuum
  subtracted F469N filters, respectively}

\end{figure}

\begin{figure}

\subfigure[]{\includegraphics[width=0.19\columnwidth]{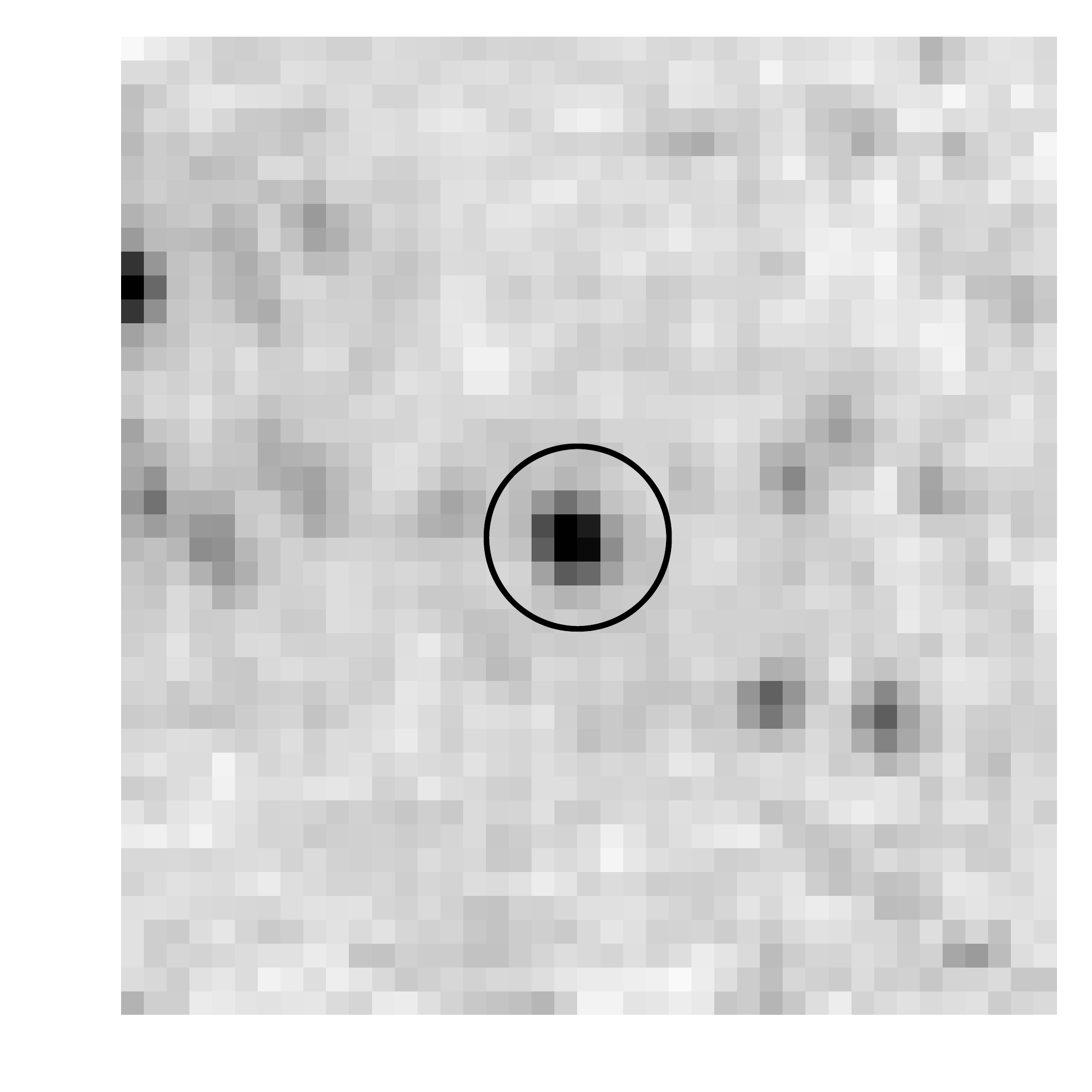}}
\subfigure[]{\includegraphics[width=0.19\columnwidth]{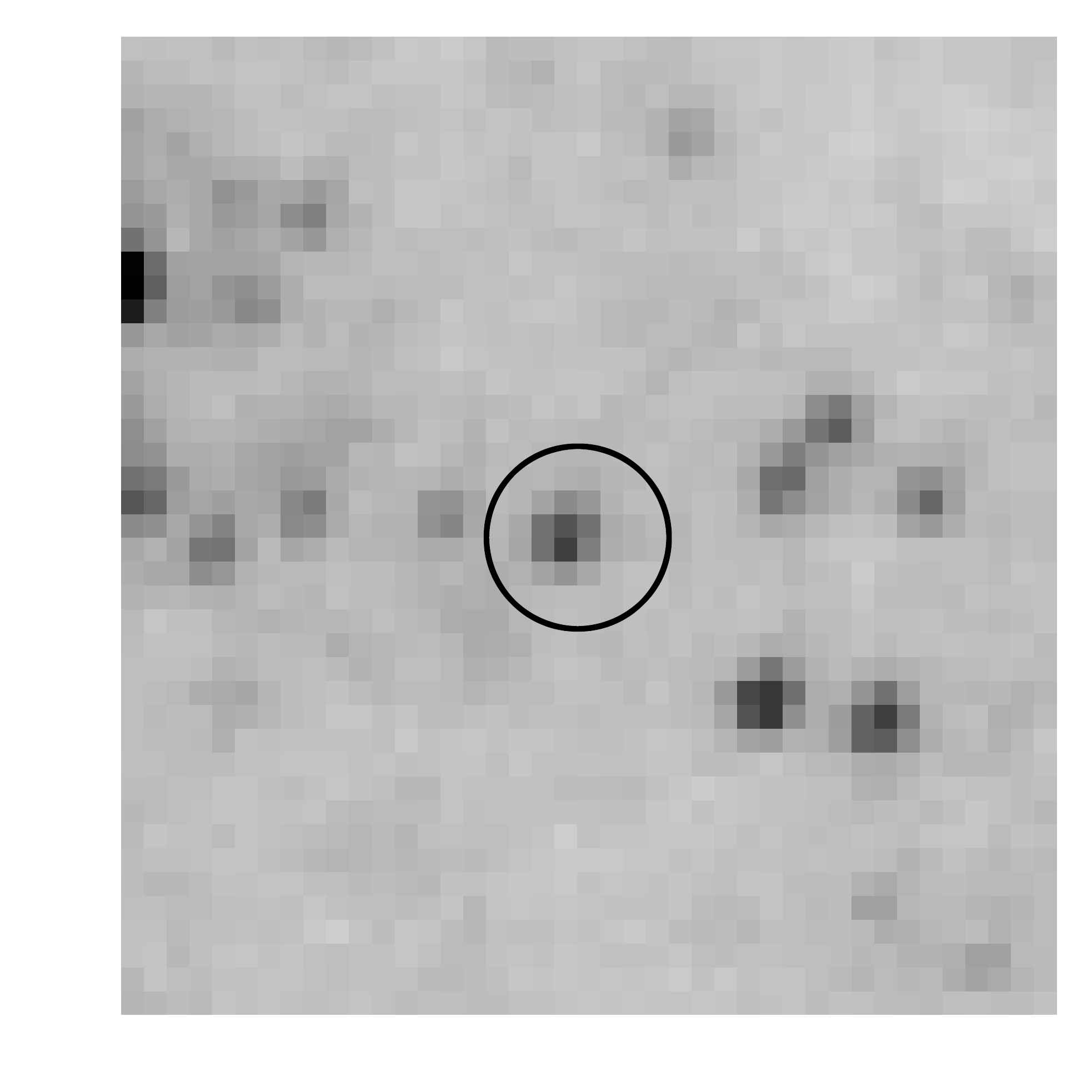}}
\subfigure[]{\includegraphics[width=0.19\columnwidth]{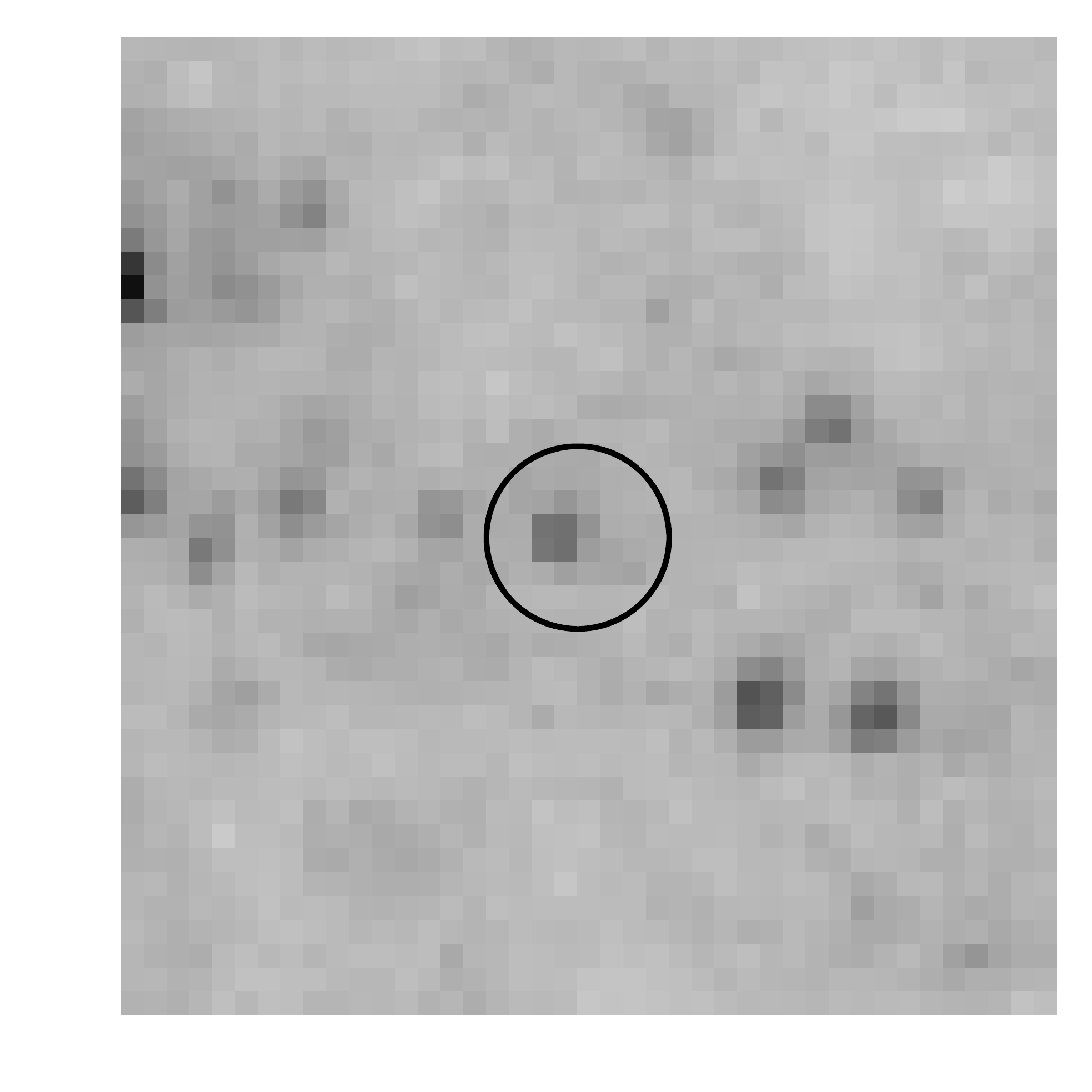}}
\subfigure[]{\includegraphics[width=0.19\columnwidth]{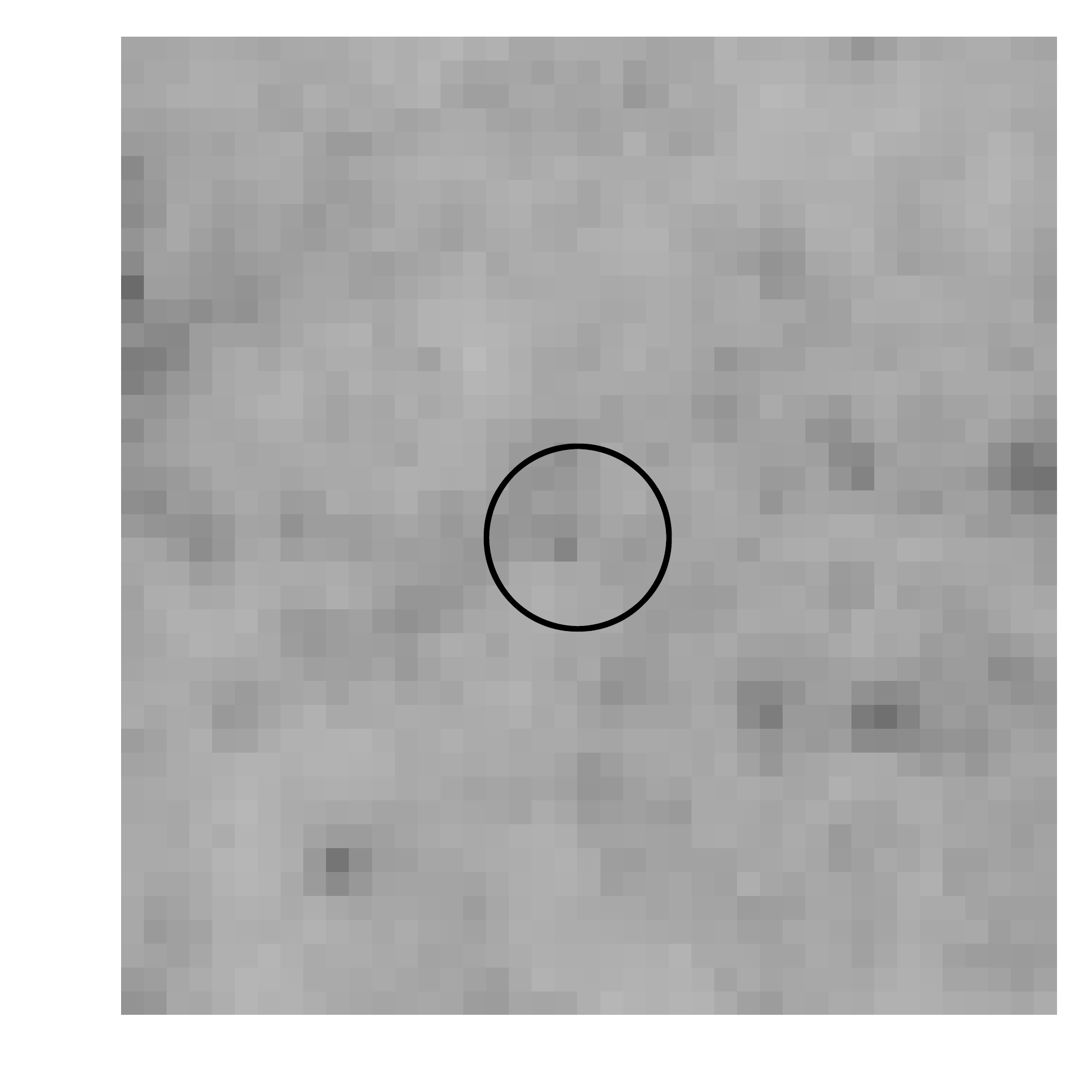}}
\subfigure[]{\includegraphics[width=0.19\columnwidth]{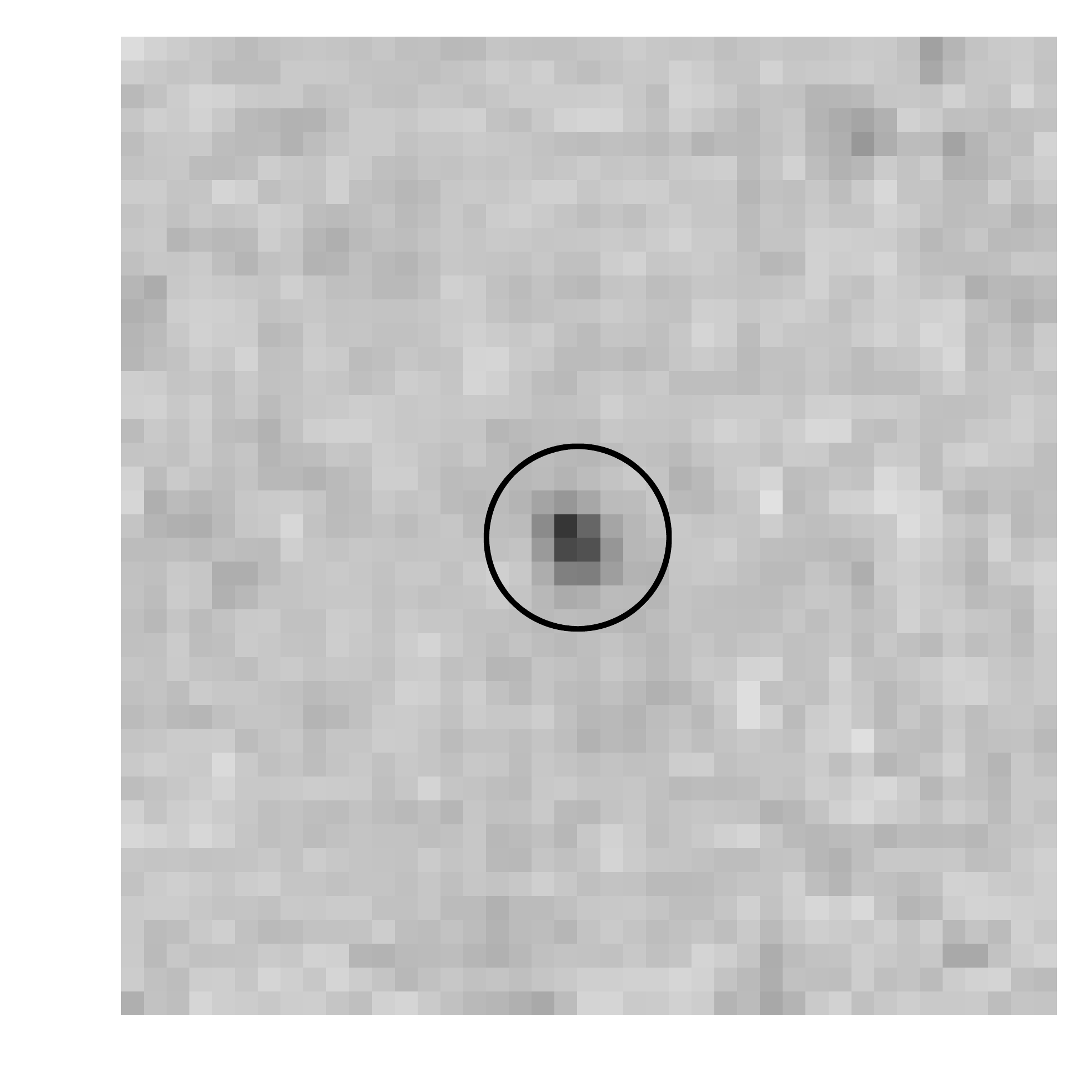}}

\caption{Source \# 1030 in F469N, F435W, F555W, F814W and continuum
  subtracted F469N filters, respectively}

\end{figure}

\begin{figure}

\subfigure[]{\includegraphics[width=0.19\columnwidth]{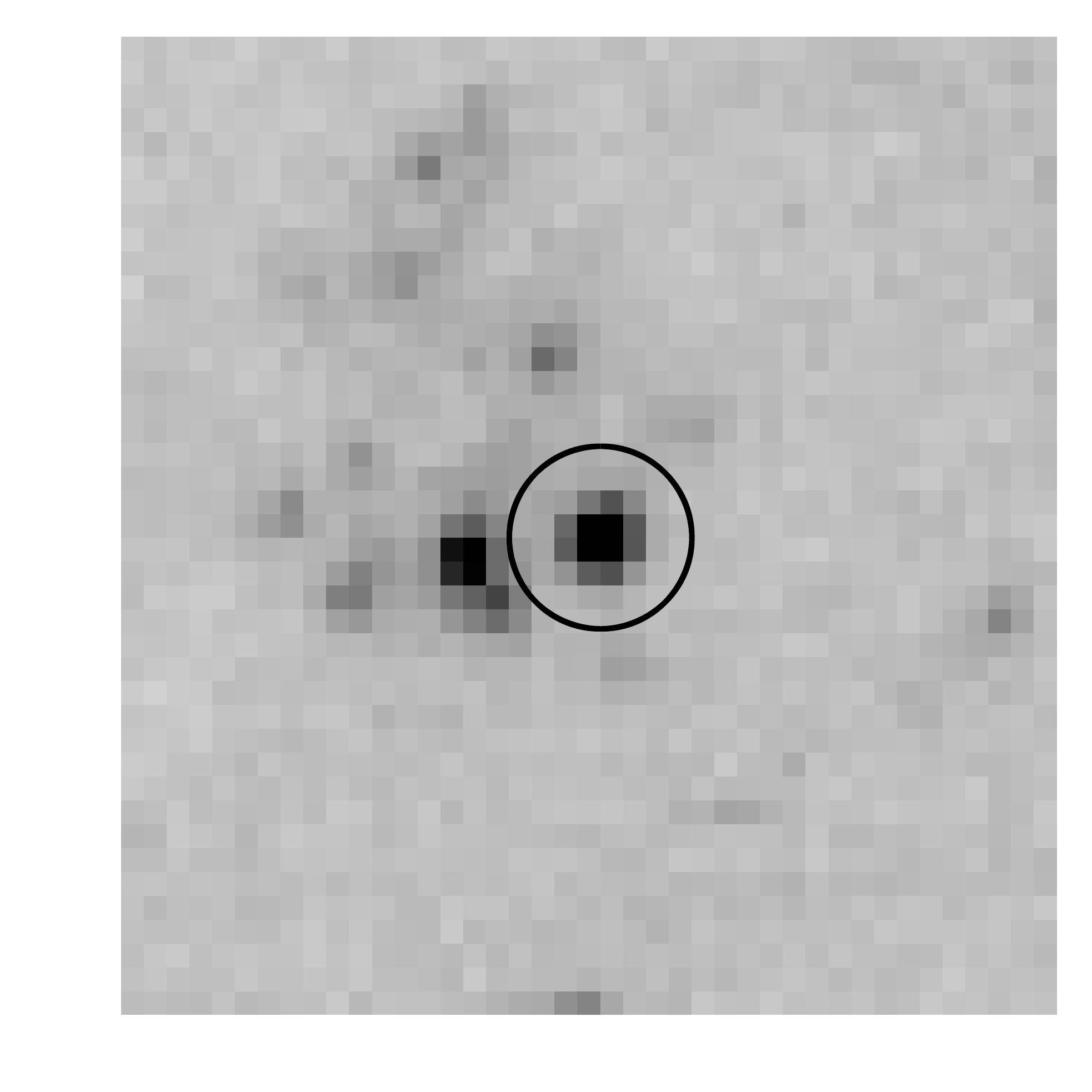}}
\subfigure[]{\includegraphics[width=0.19\columnwidth]{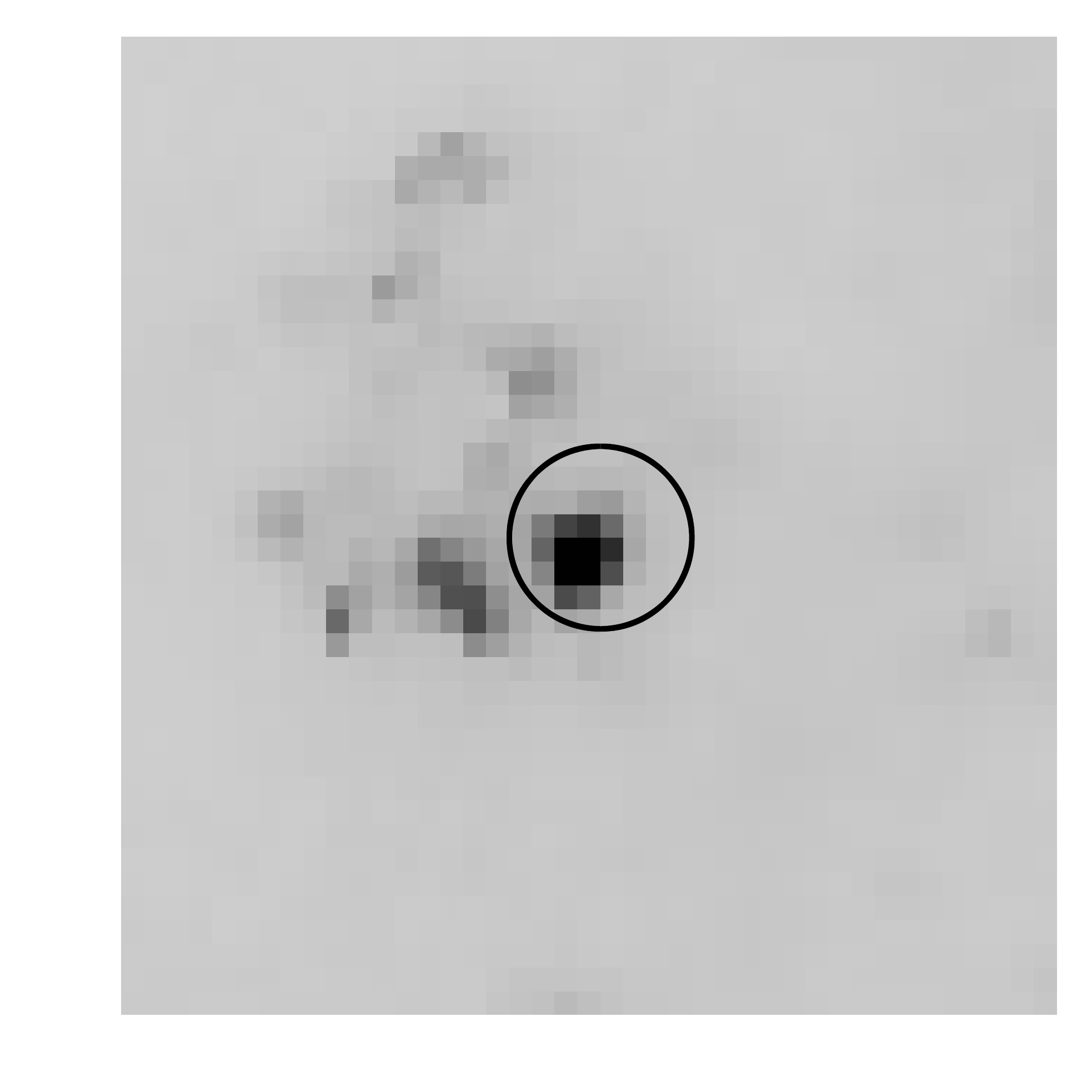}}
\subfigure[]{\includegraphics[width=0.19\columnwidth]{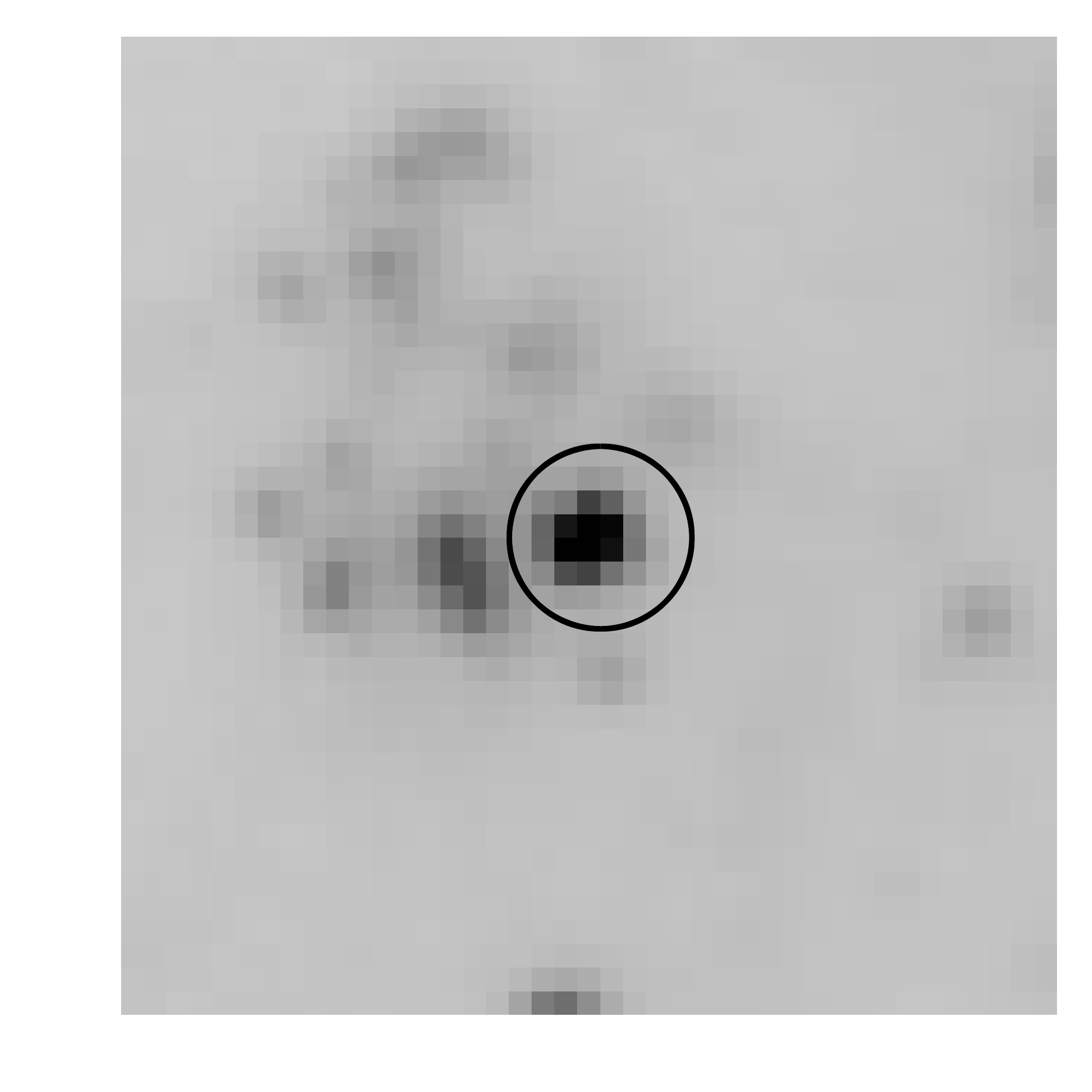}}
\subfigure[]{\includegraphics[width=0.19\columnwidth]{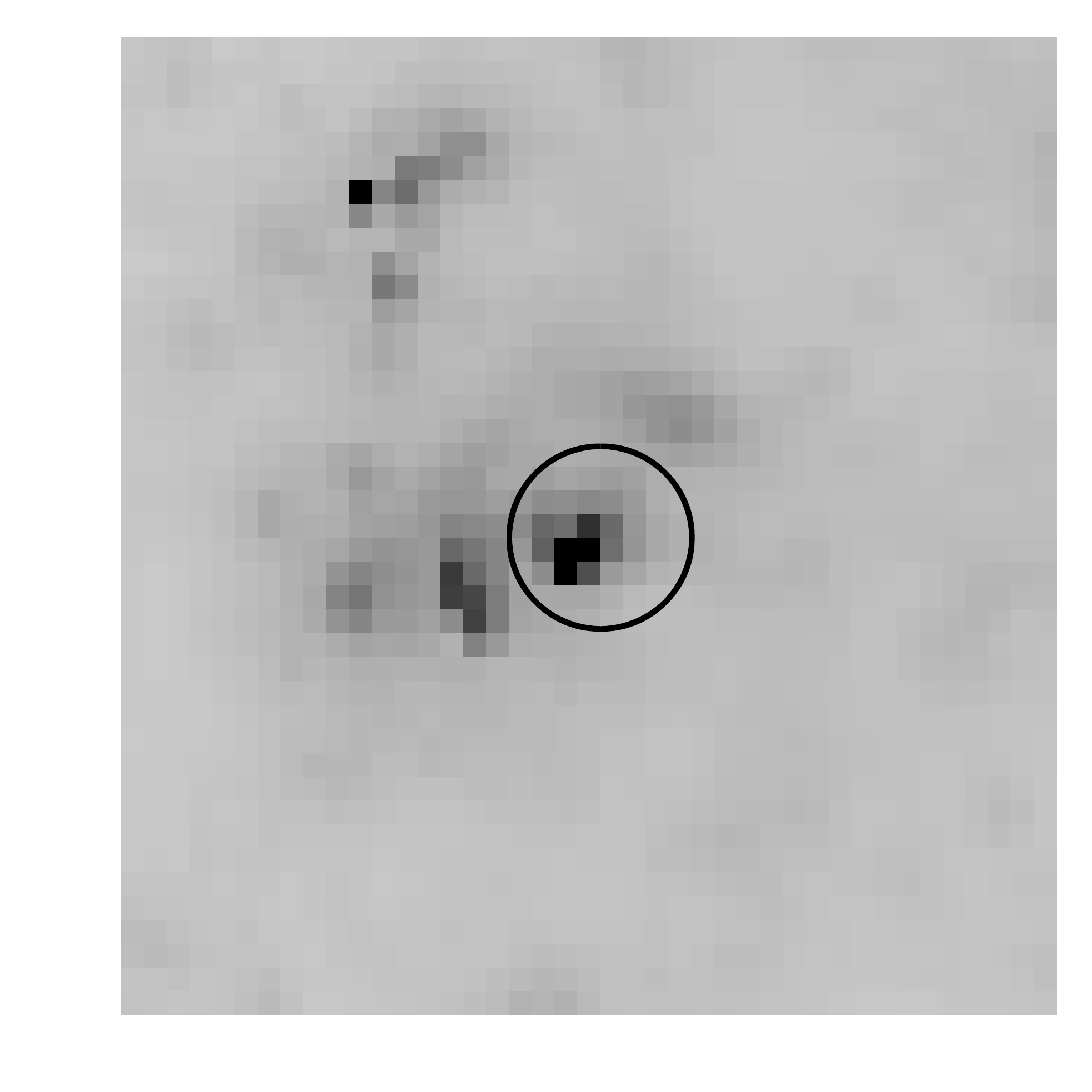}}
\subfigure[]{\includegraphics[width=0.19\columnwidth]{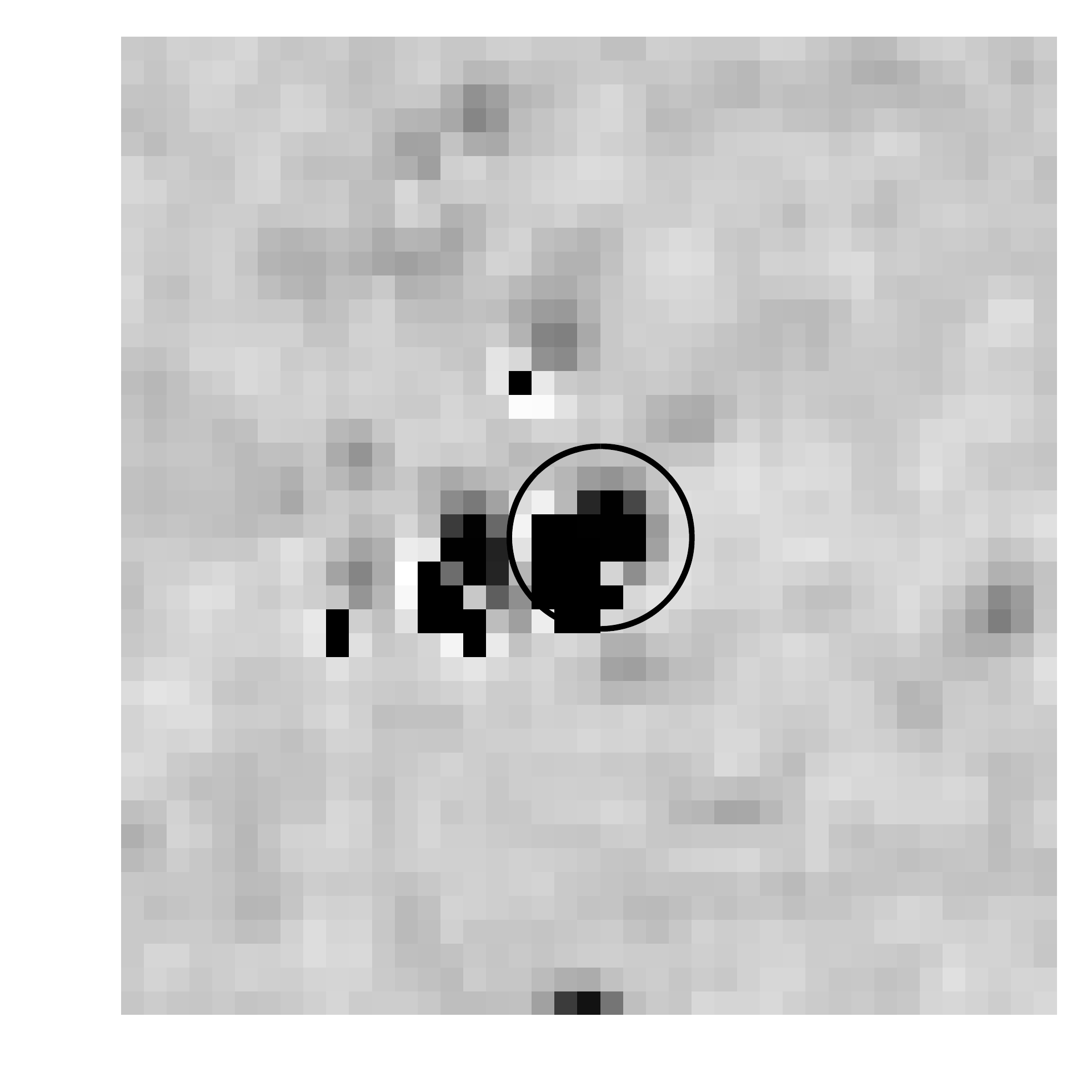}}

\caption{Source \# 112 in F469N, F435W, F555W, F814W and continuum
  subtracted F469N filters, respectively}

\end{figure}

\begin{figure}

\subfigure[]{\includegraphics[width=0.19\columnwidth]{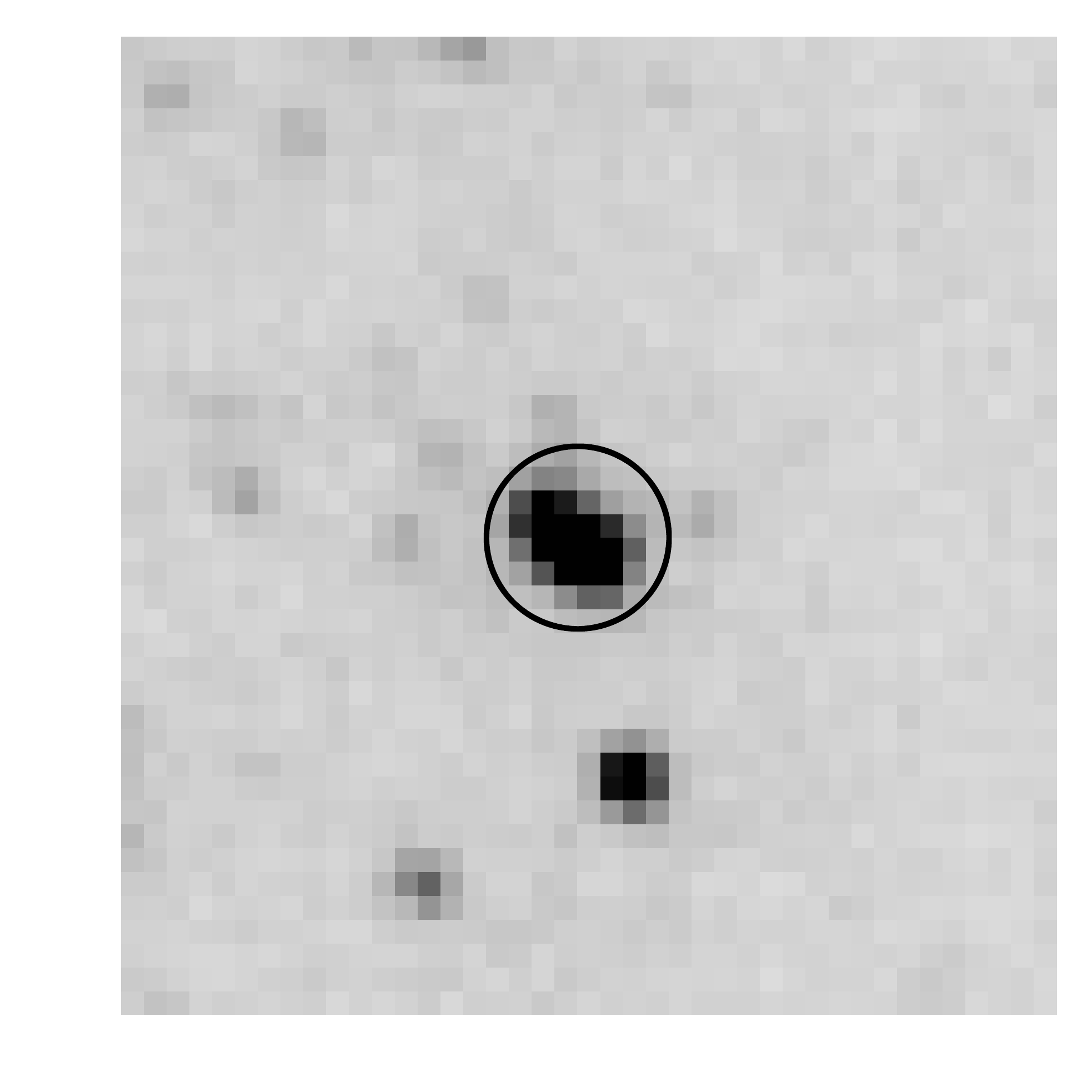}}
\subfigure[]{\includegraphics[width=0.19\columnwidth]{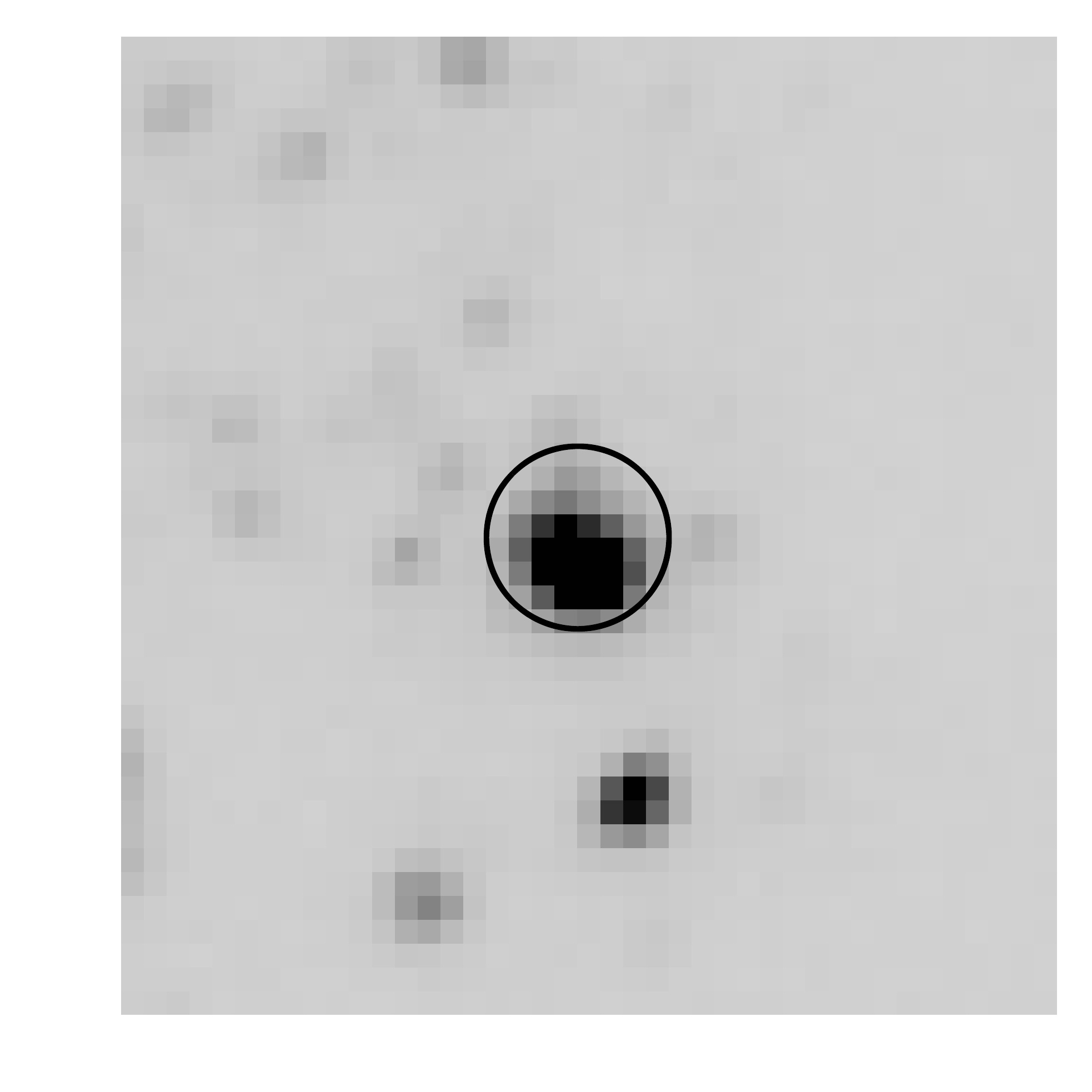}}
\subfigure[]{\includegraphics[width=0.19\columnwidth]{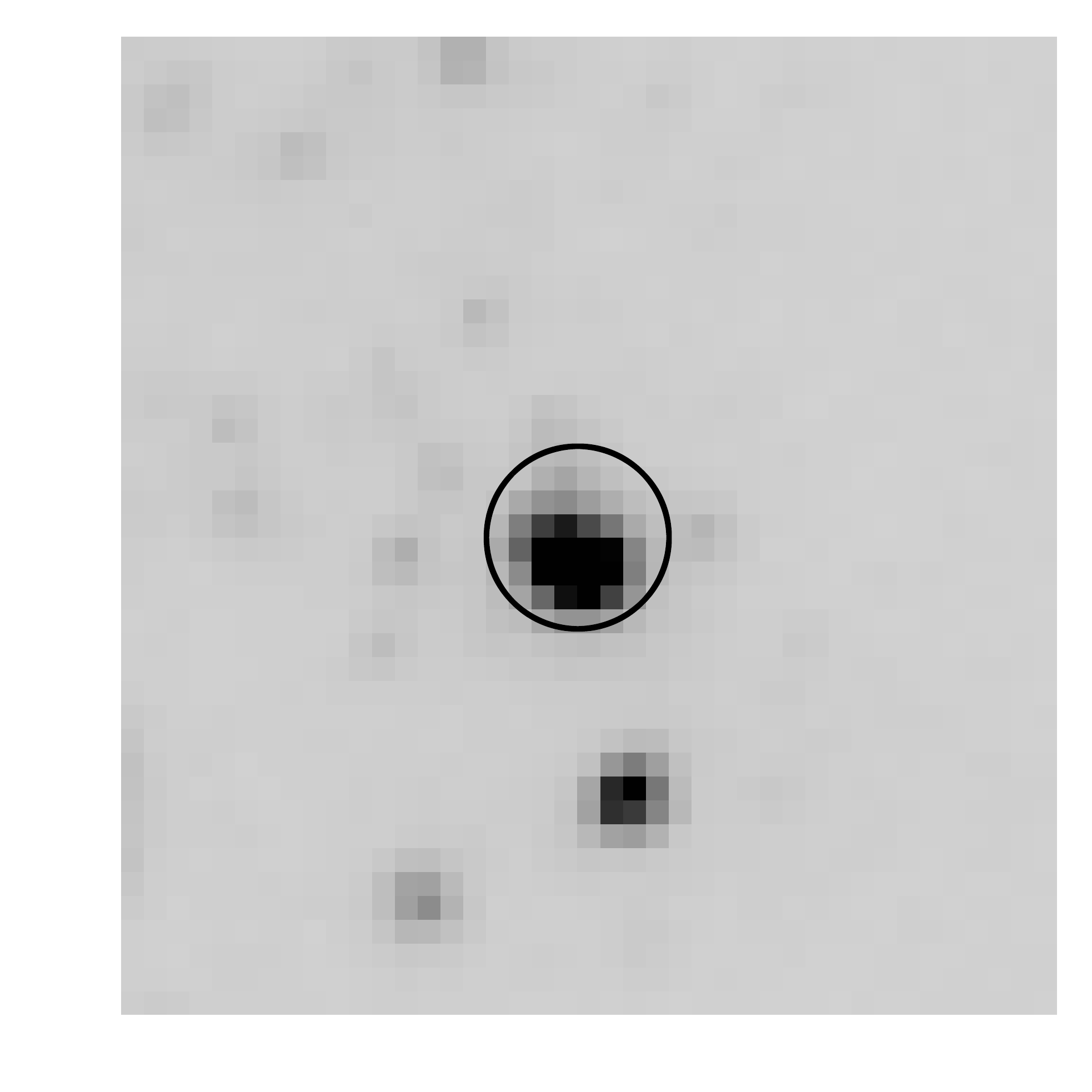}}
\subfigure[]{\includegraphics[width=0.19\columnwidth]{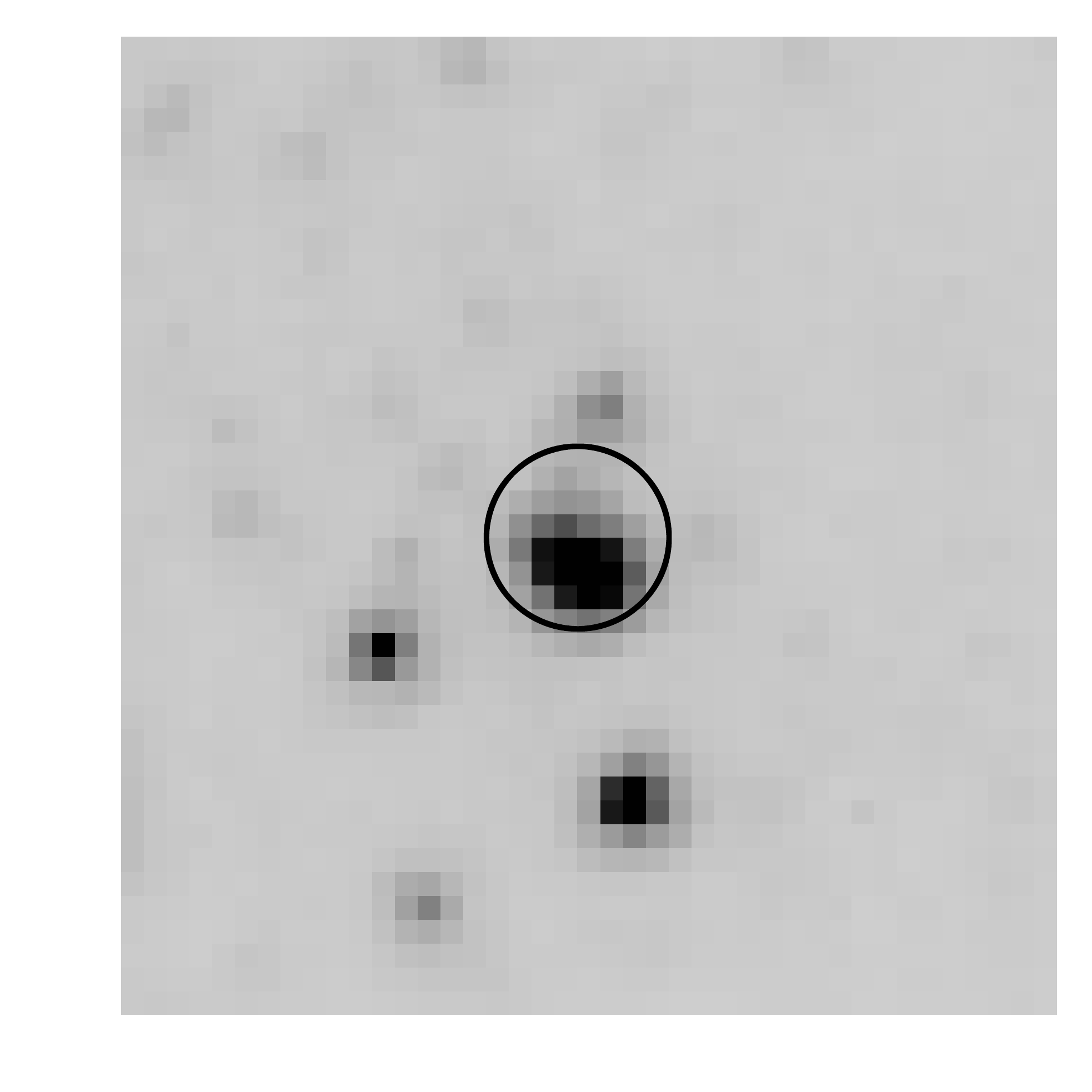}}
\subfigure[]{\includegraphics[width=0.19\columnwidth]{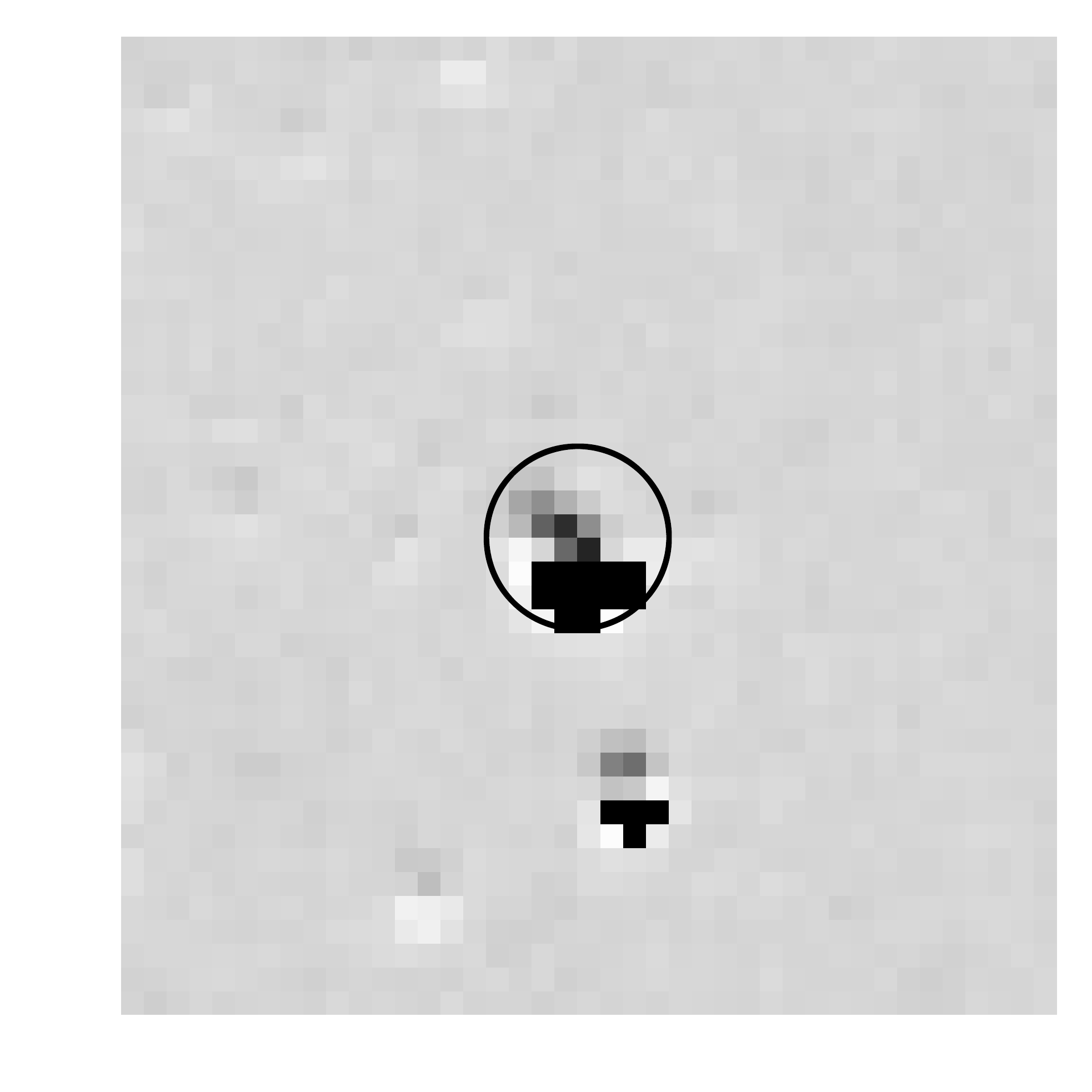}}

\caption{Source \# 114 in F469N, F435W, F555W, F814W and continuum
  subtracted F469N filters, respectively}

\end{figure}

\begin{figure}

\subfigure[]{\includegraphics[width=0.19\columnwidth]{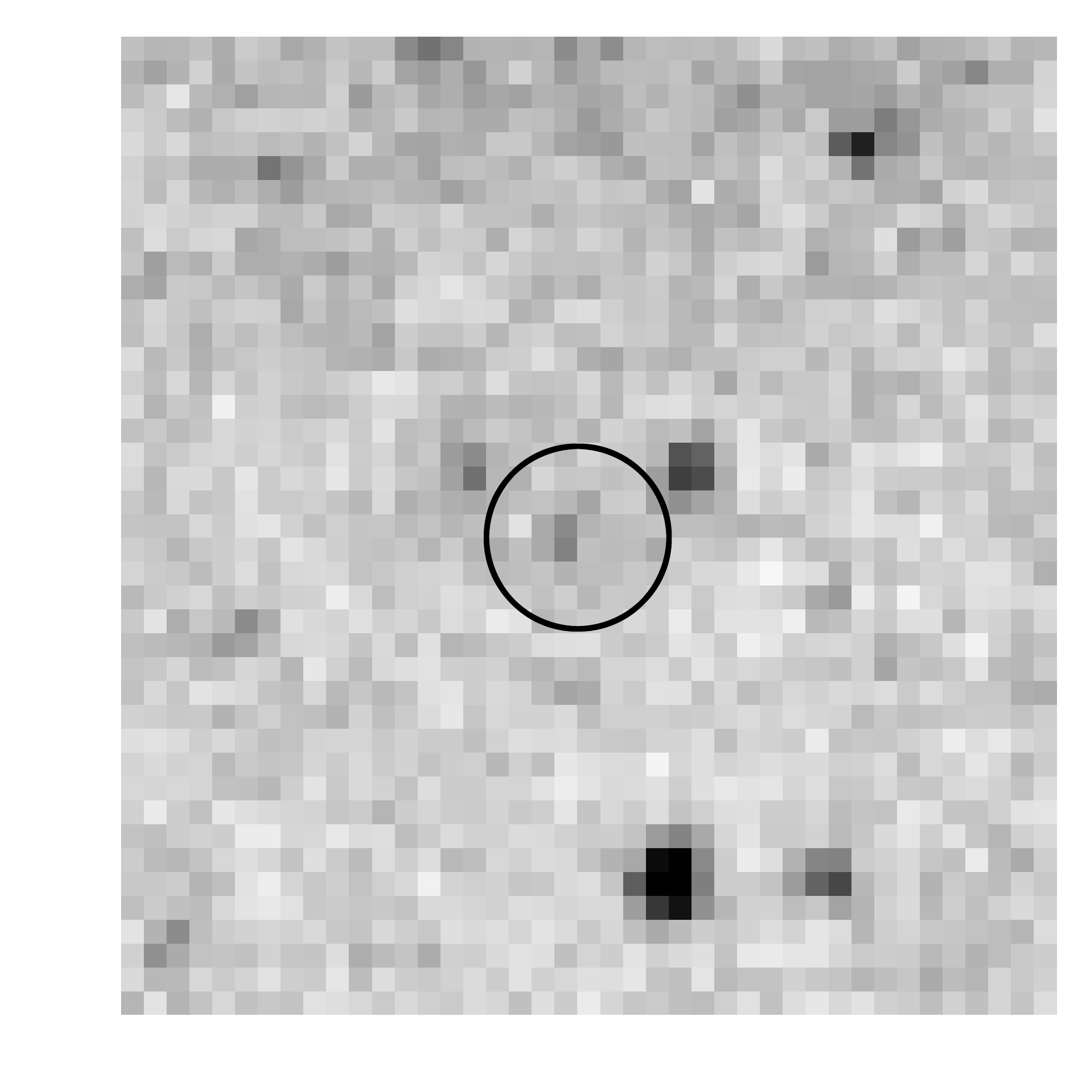}}
\subfigure[]{\includegraphics[width=0.19\columnwidth]{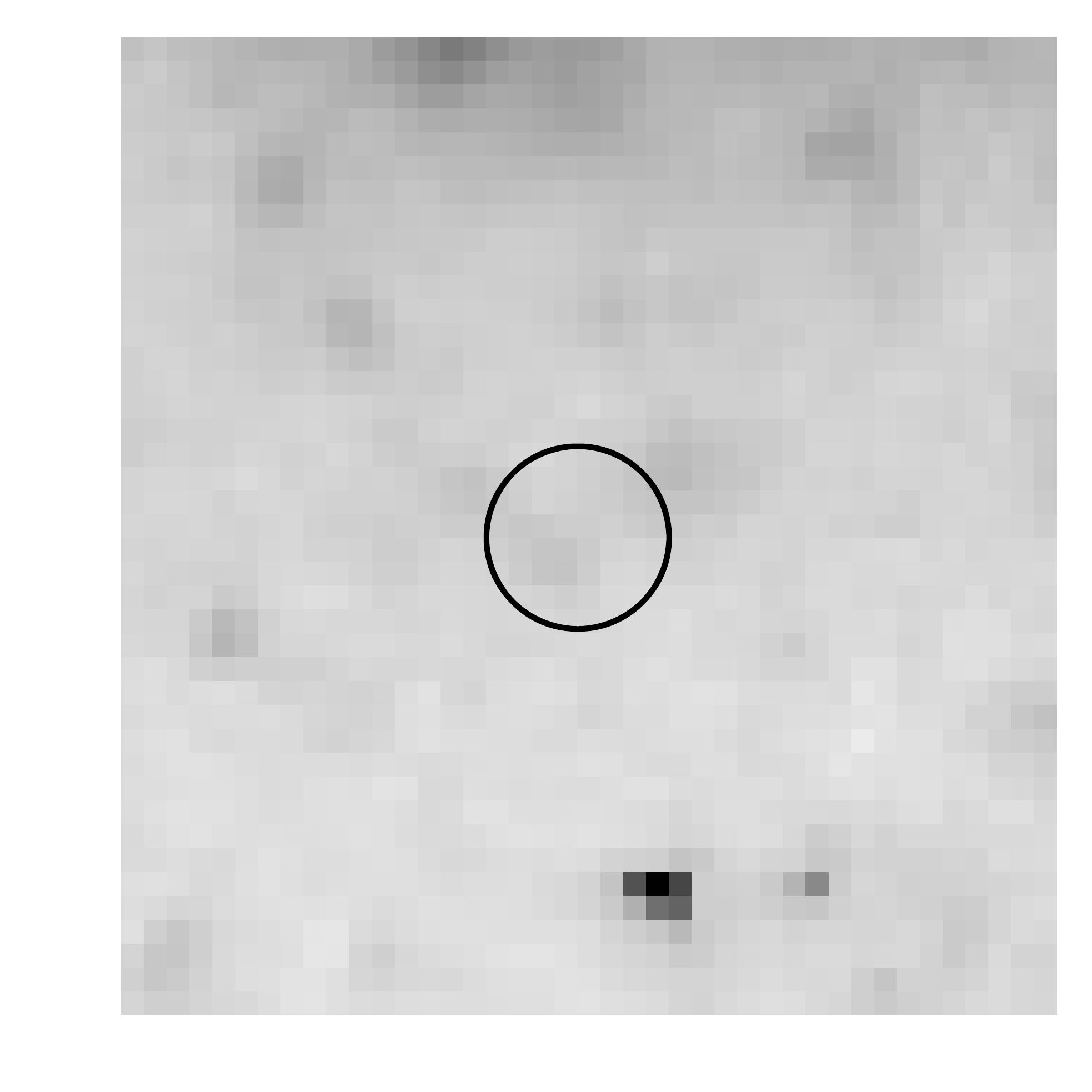}}
\subfigure[]{\includegraphics[width=0.19\columnwidth]{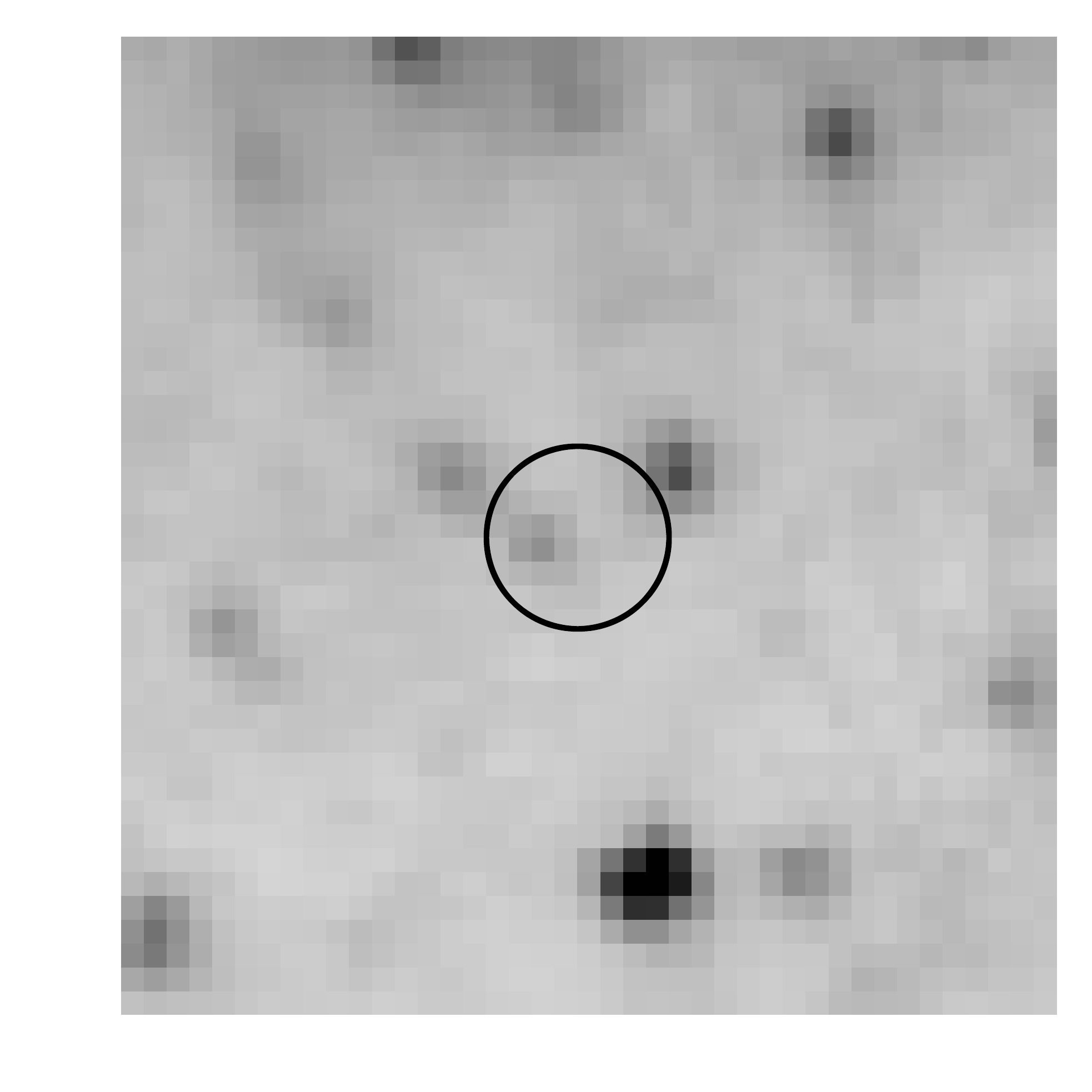}}
\subfigure[]{\includegraphics[width=0.19\columnwidth]{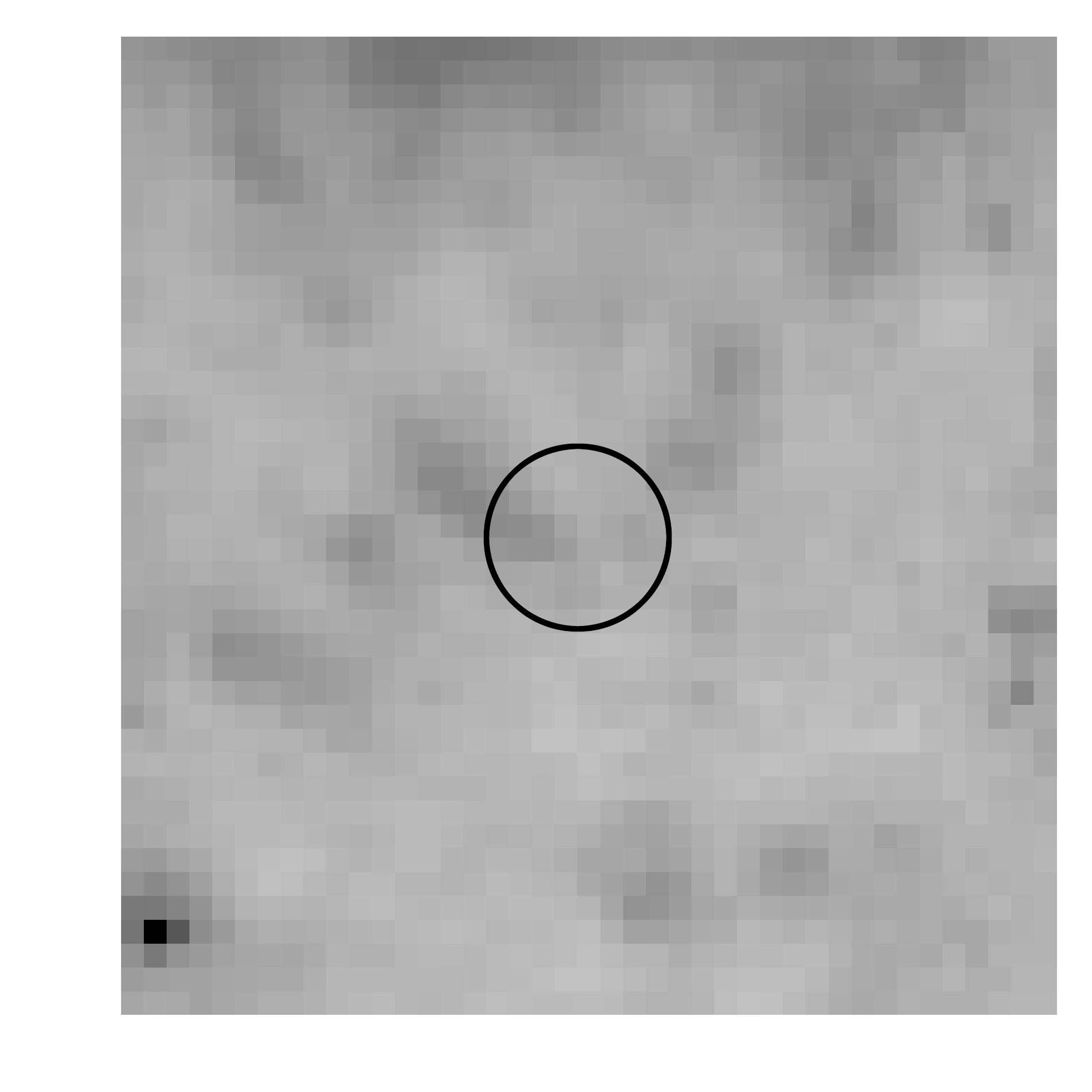}}
\subfigure[]{\includegraphics[width=0.19\columnwidth]{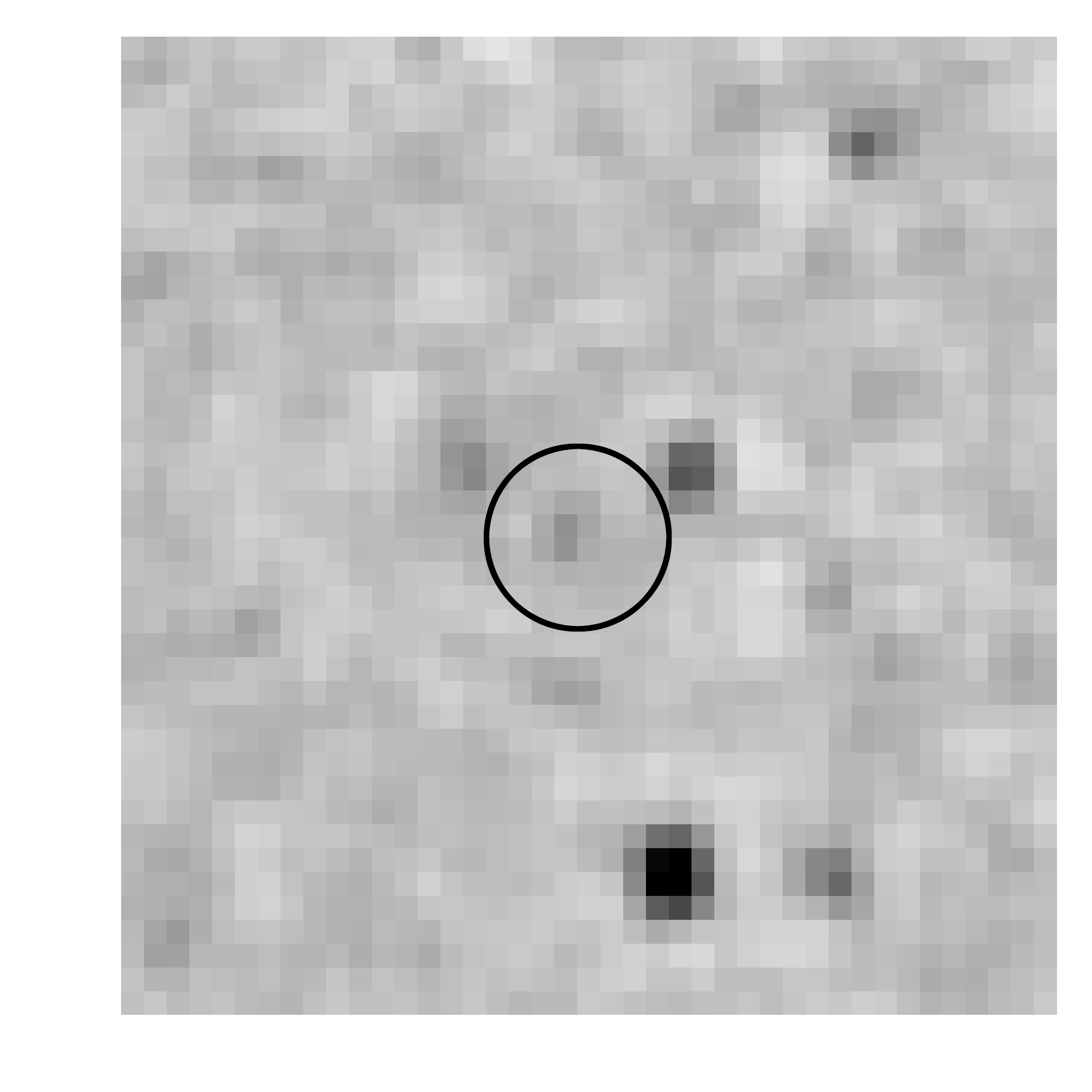}}

\caption{Source \# 2053 in F469N, F435W, F555W, F814W and continuum
  subtracted F469N filters, respectively}

\end{figure}

\begin{figure}

\subfigure[]{\includegraphics[width=0.19\columnwidth]{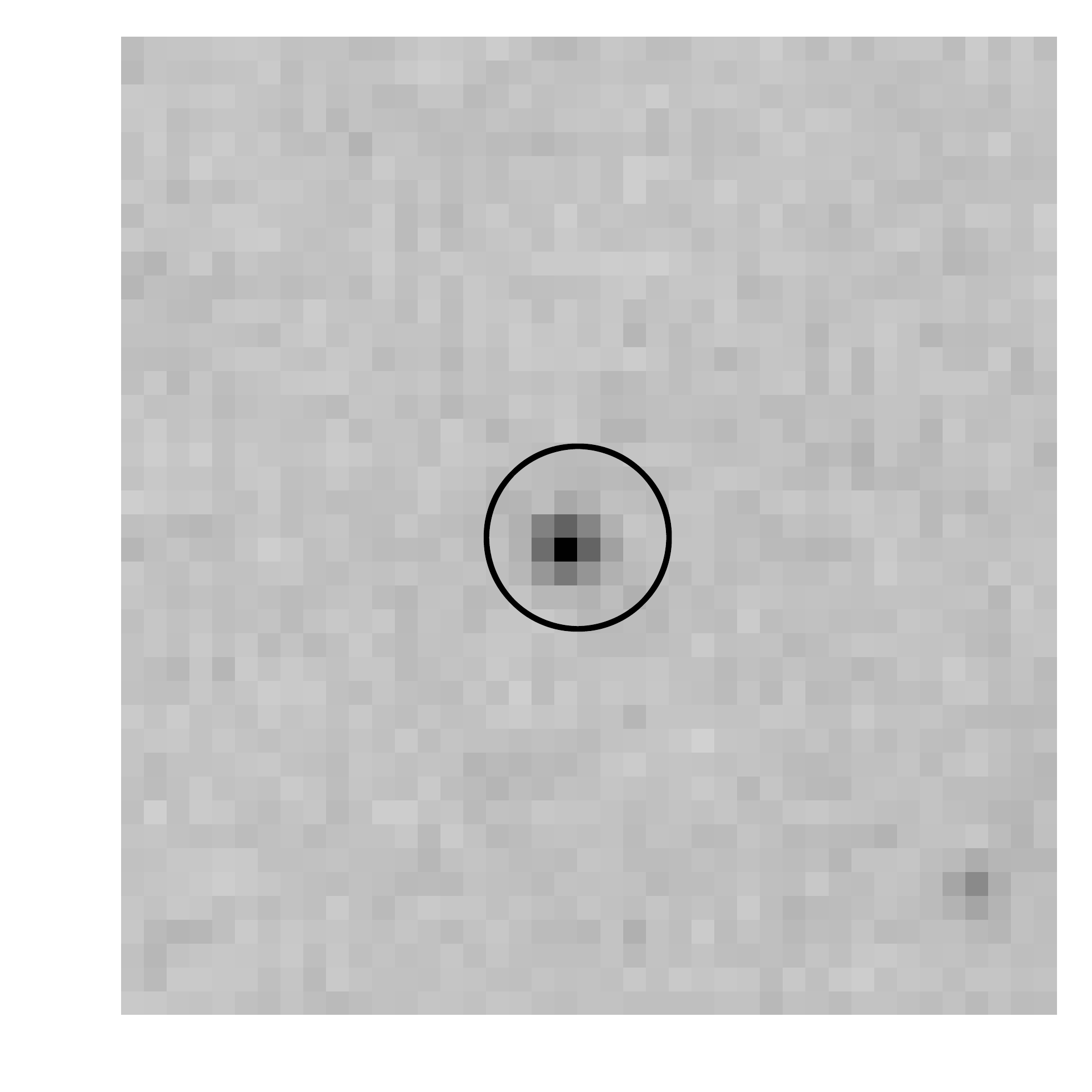}}
\subfigure[]{\includegraphics[width=0.19\columnwidth]{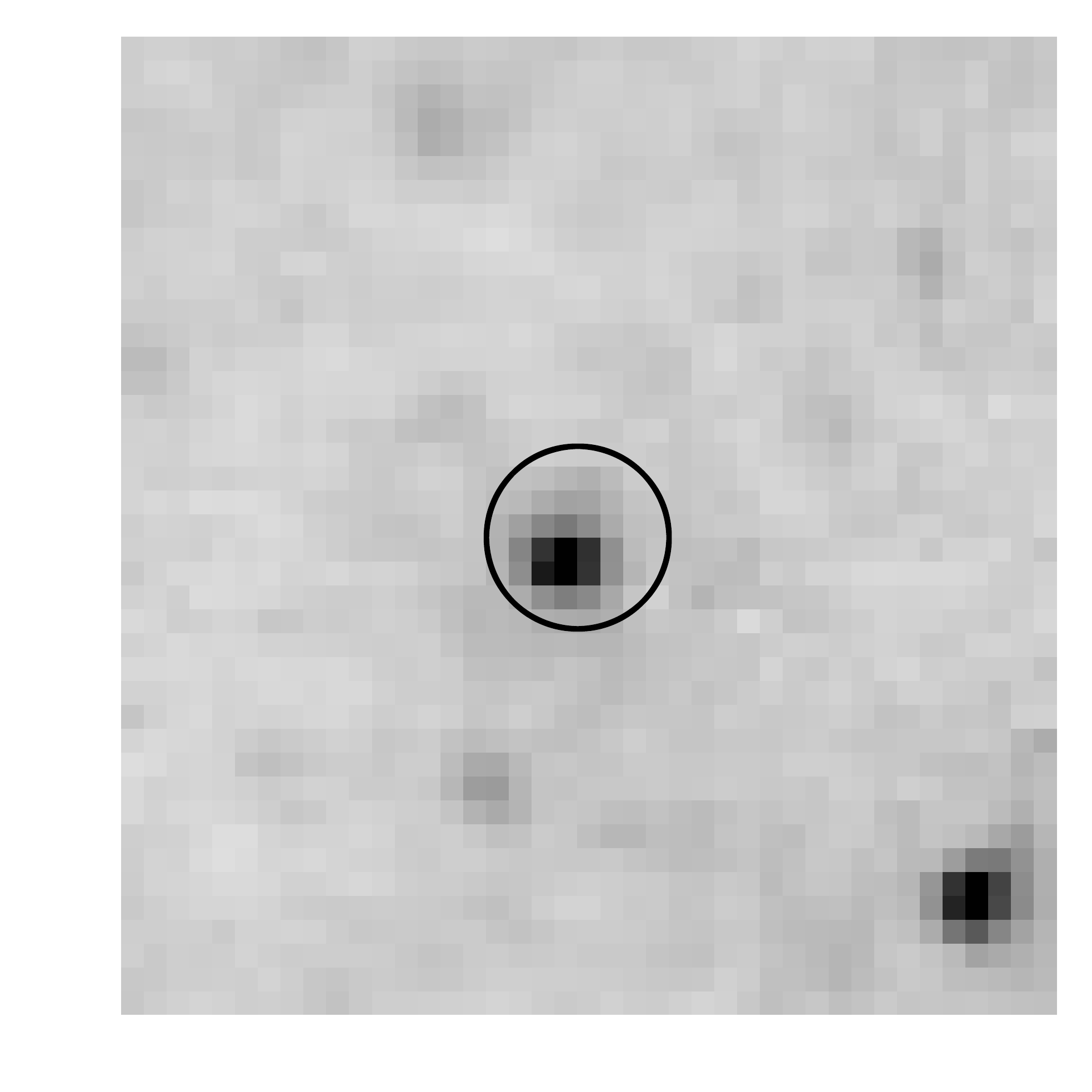}}
\subfigure[]{\includegraphics[width=0.19\columnwidth]{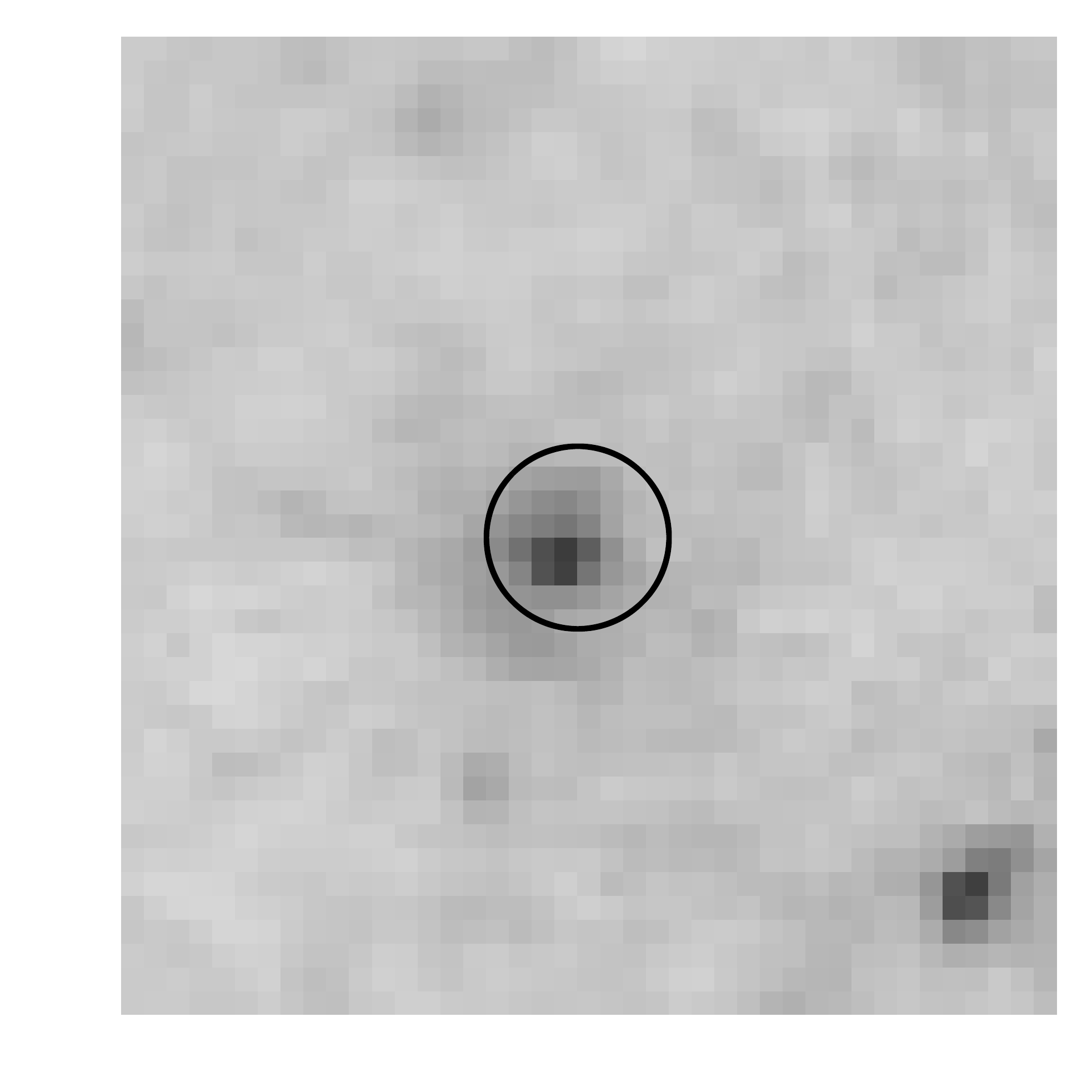}}
\subfigure[]{\includegraphics[width=0.19\columnwidth]{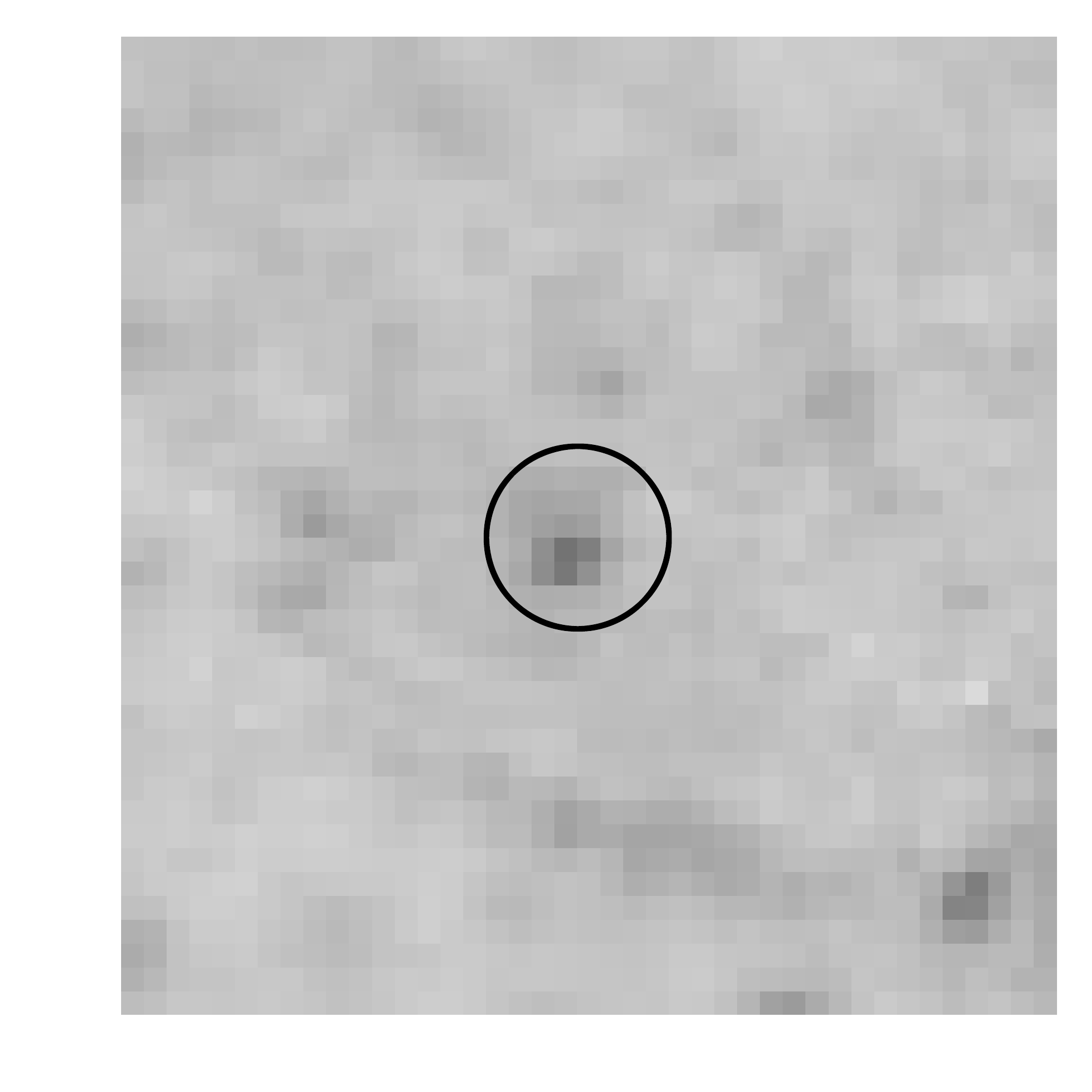}}
\subfigure[]{\includegraphics[width=0.19\columnwidth]{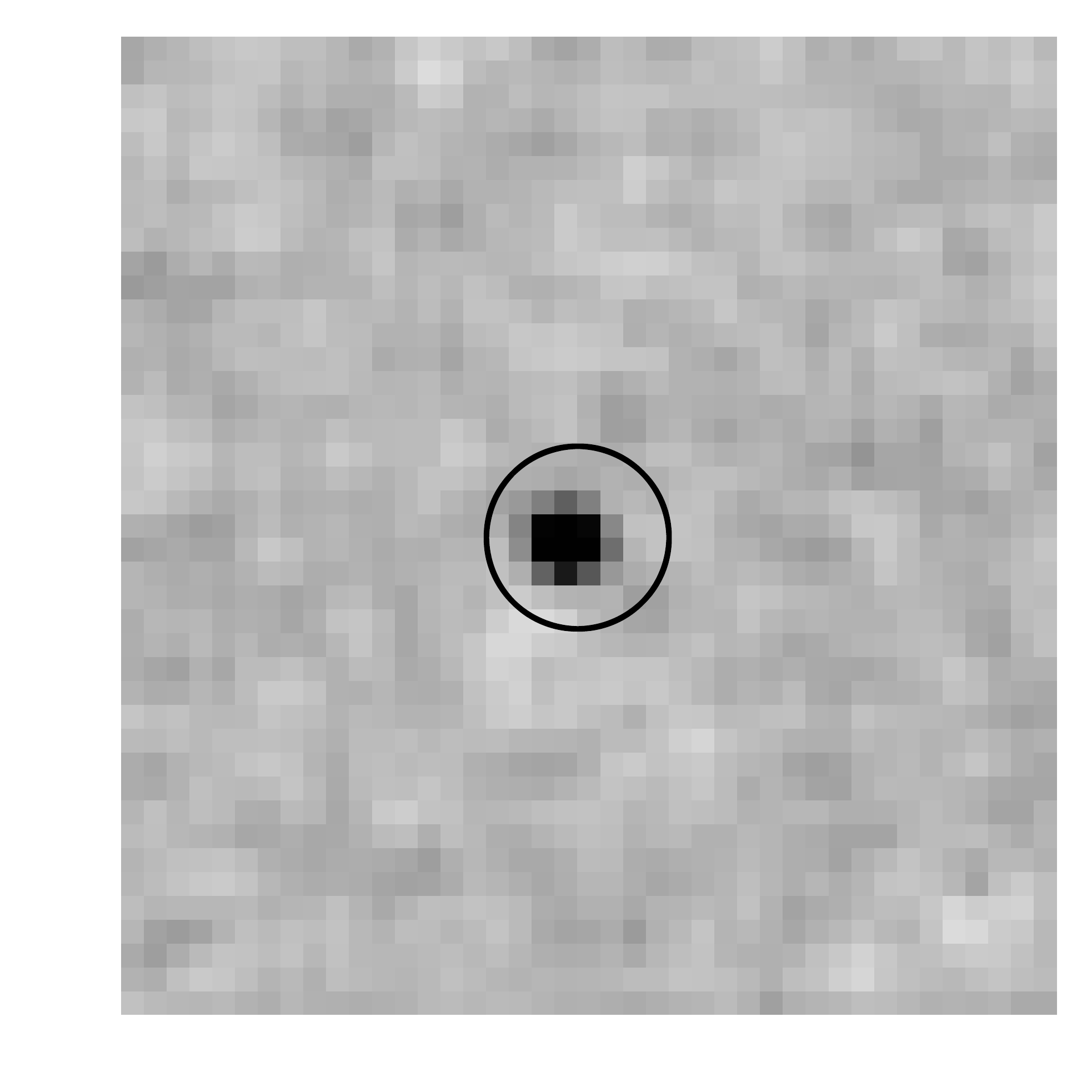}}

\caption{Source \# 48 in F469N, F435W, F555W, F814W and continuum
  subtracted F469N filters, respectively}

\end{figure}

\begin{figure}

\subfigure[]{\includegraphics[width=0.19\columnwidth]{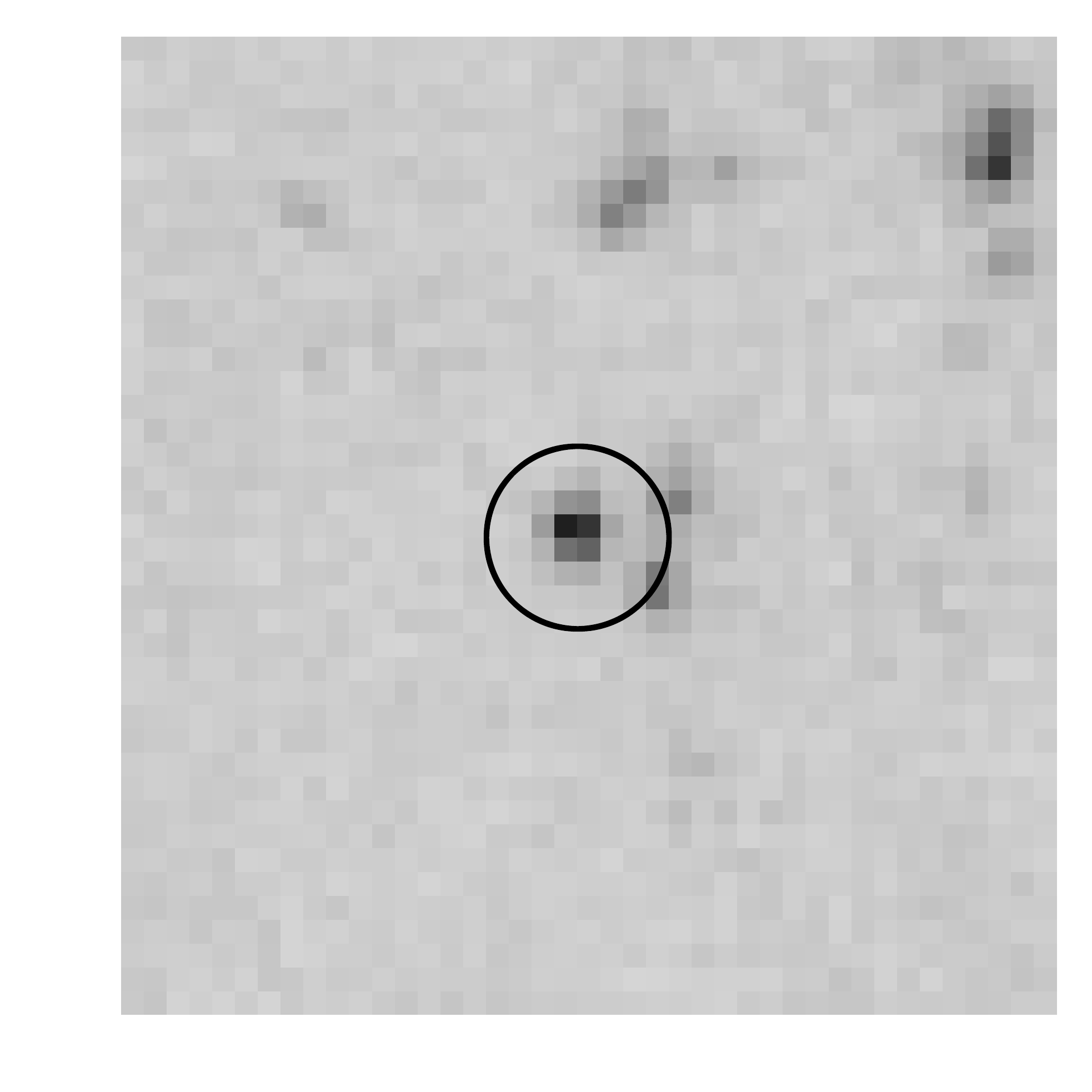}}
\subfigure[]{\includegraphics[width=0.19\columnwidth]{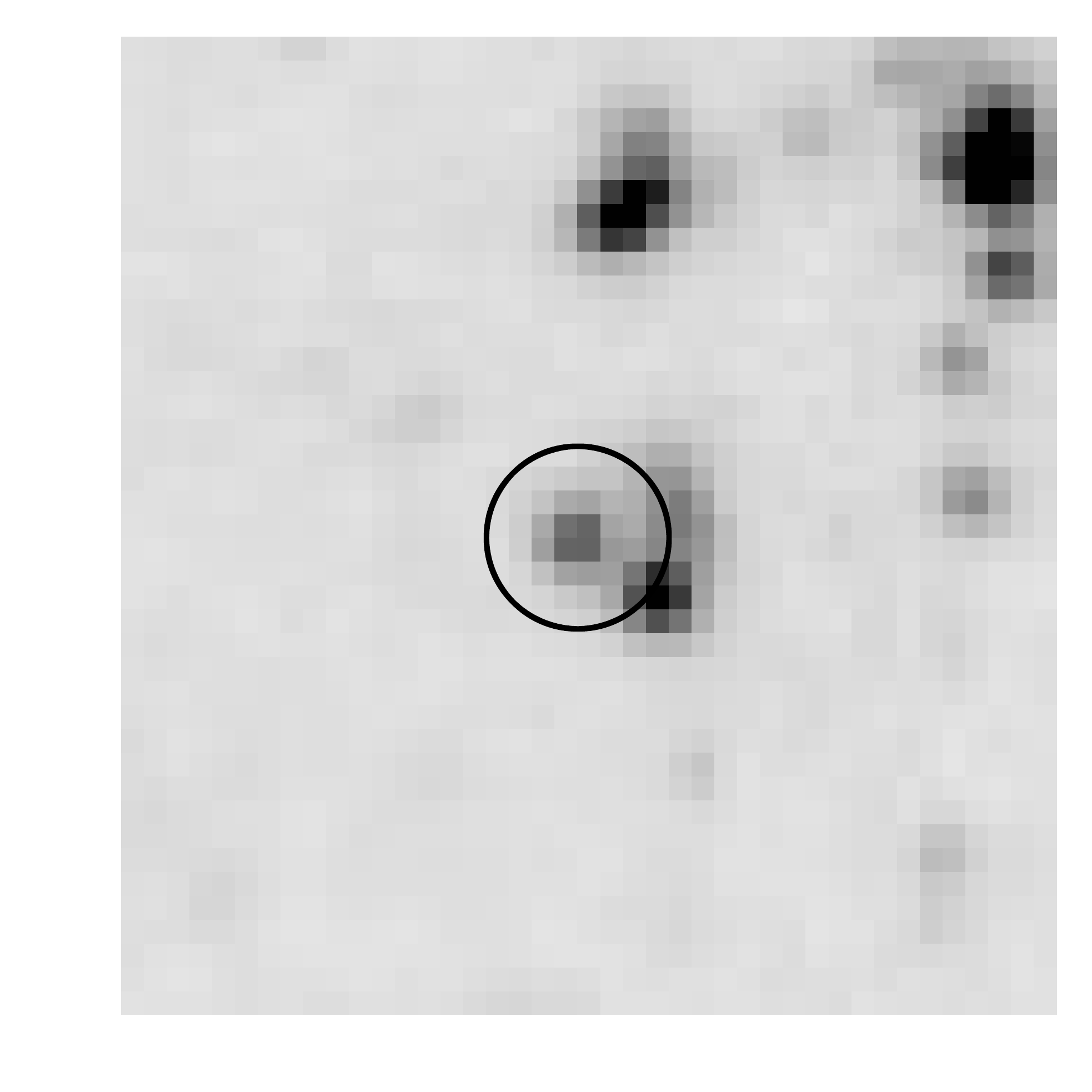}}
\subfigure[]{\includegraphics[width=0.19\columnwidth]{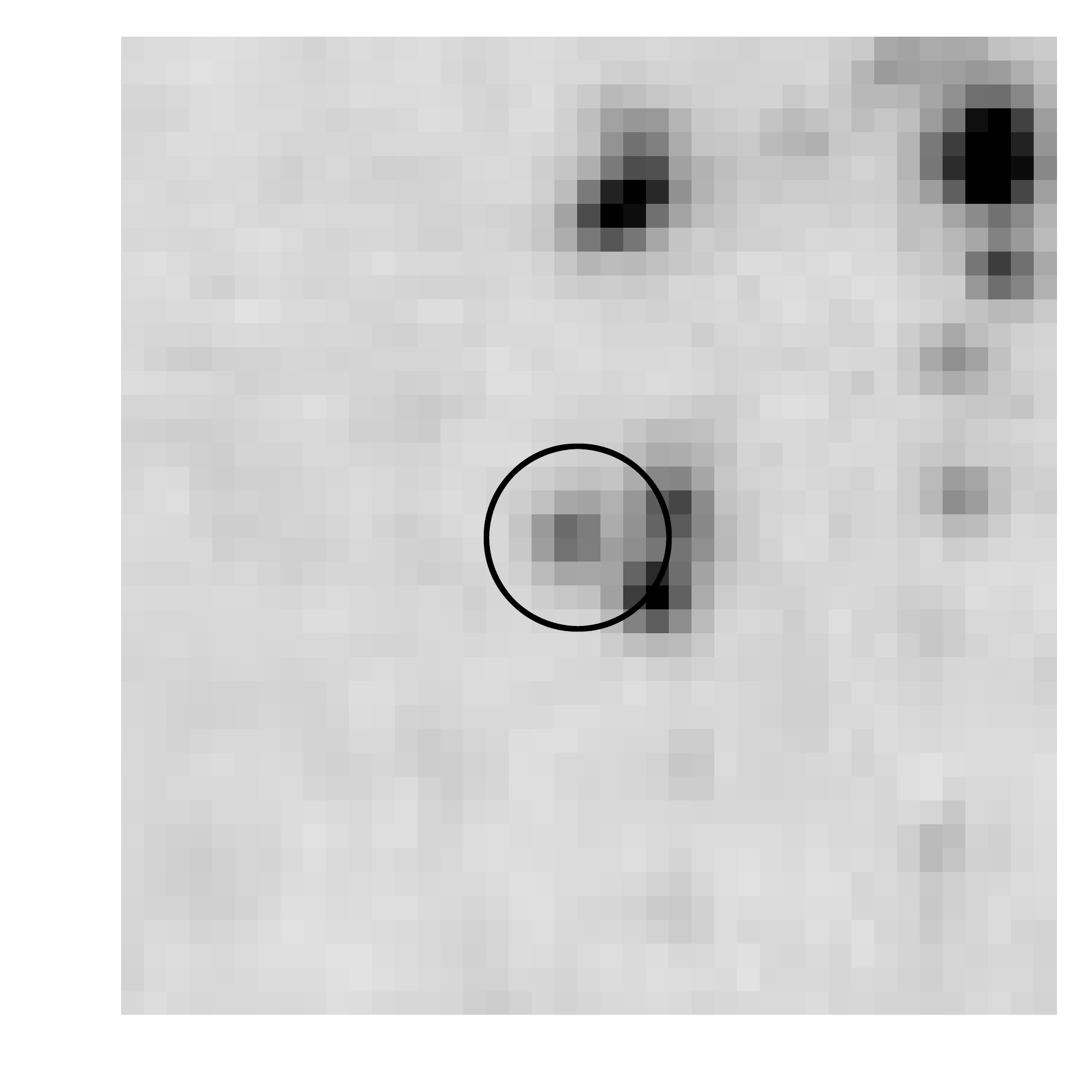}}
\subfigure[]{\includegraphics[width=0.19\columnwidth]{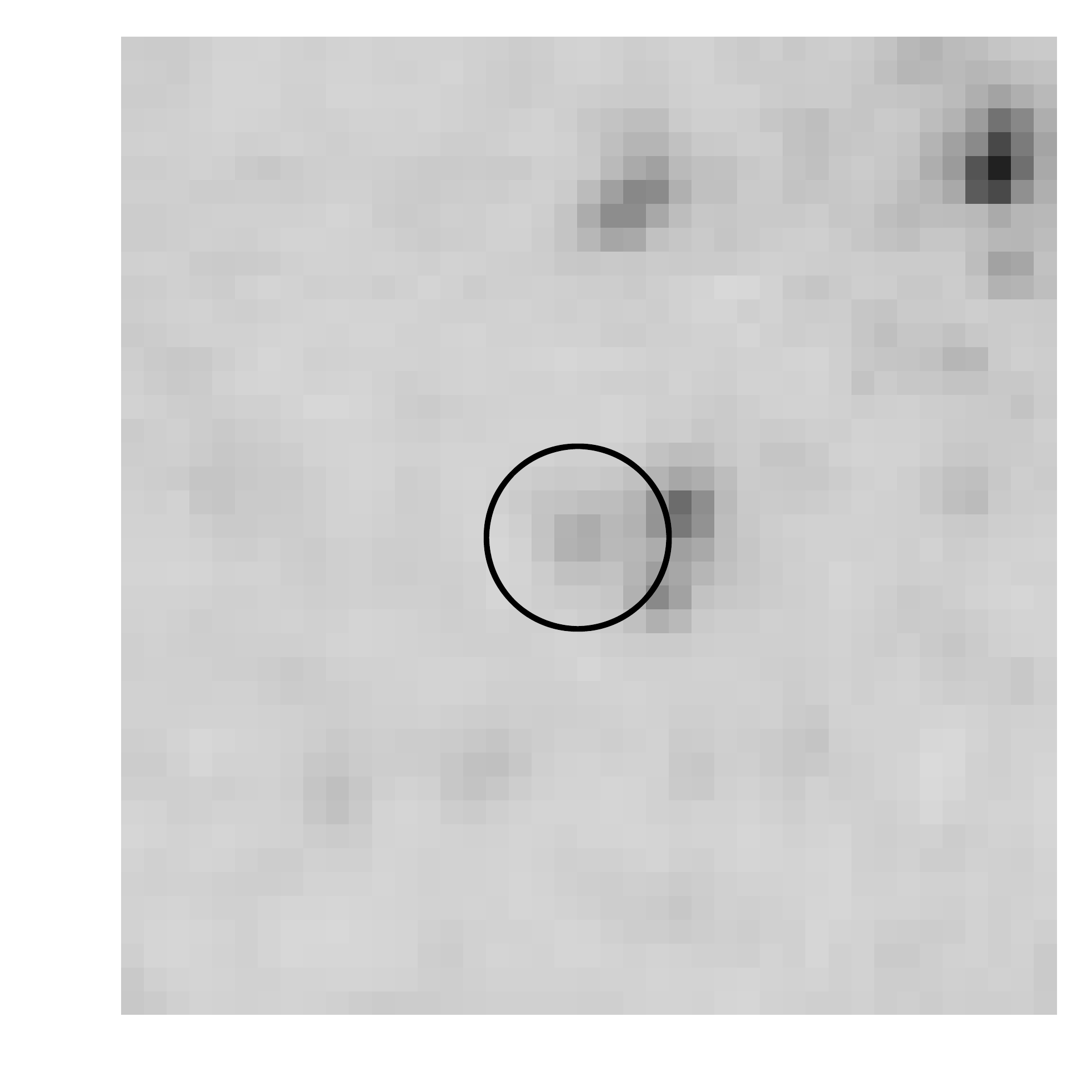}}
\subfigure[]{\includegraphics[width=0.19\columnwidth]{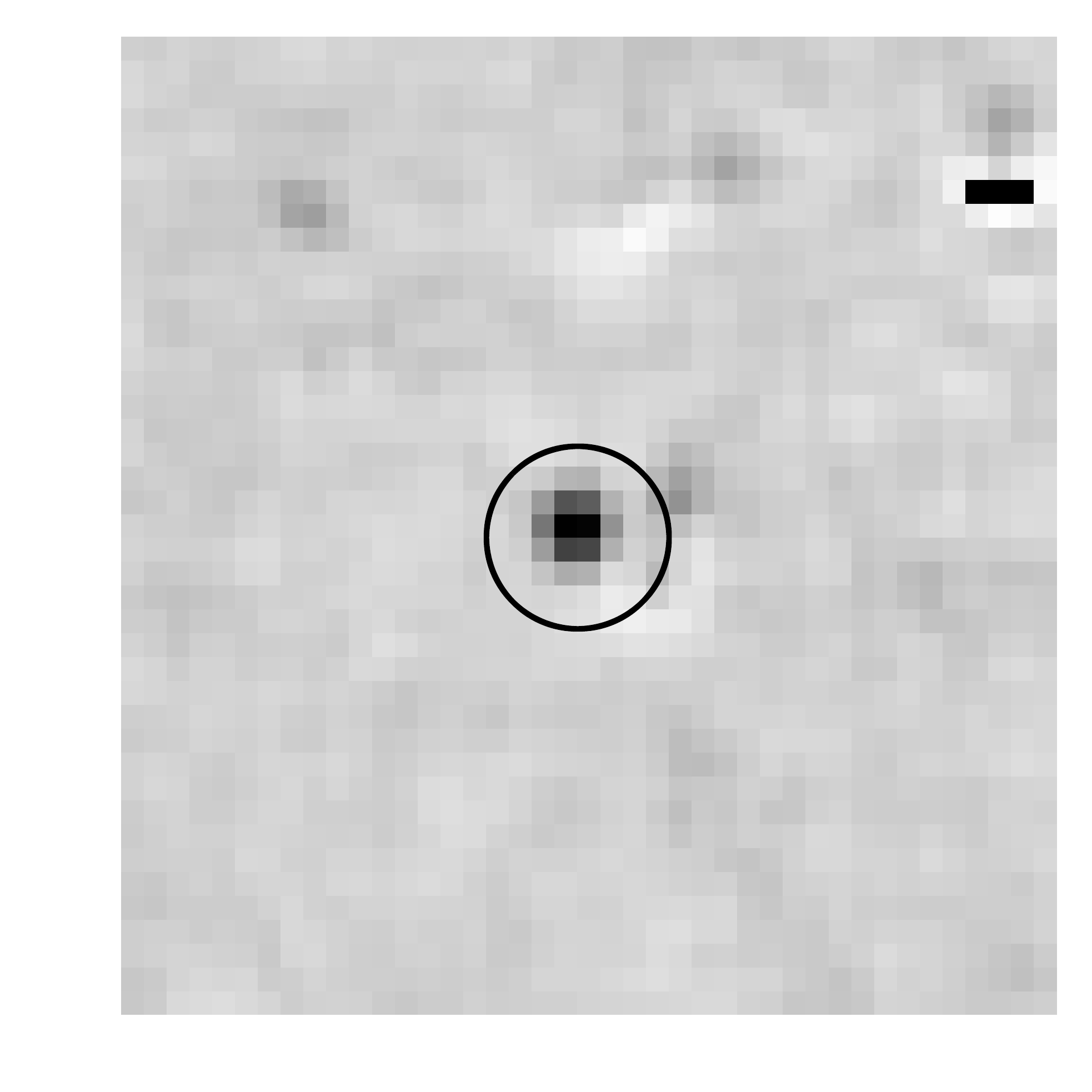}}

\caption{Source \# 49 in F469N, F435W, F555W, F814W and continuum
  subtracted F469N filters, respectively}

\end{figure}

\begin{figure}

\subfigure[]{\includegraphics[width=0.19\columnwidth]{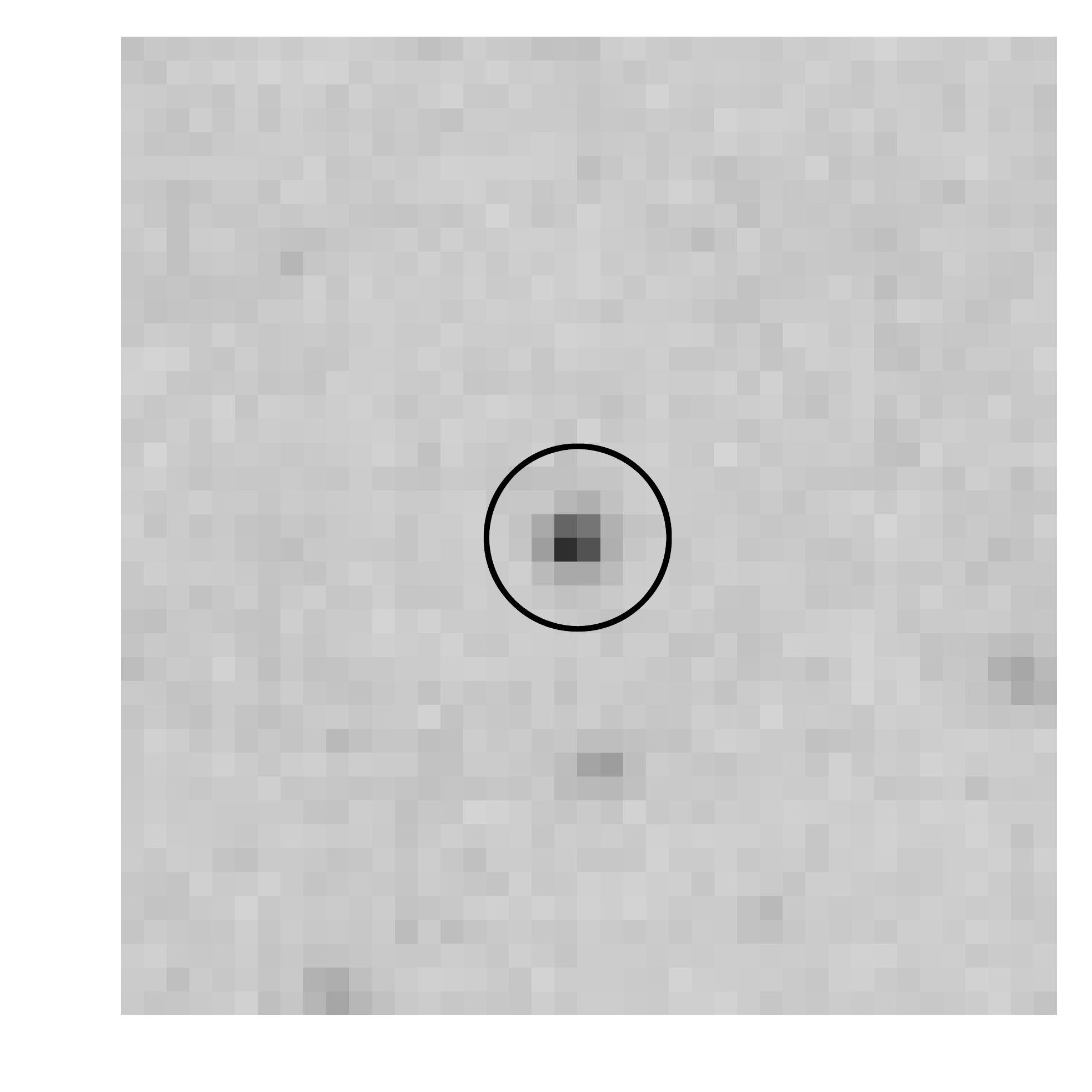}}
\subfigure[]{\includegraphics[width=0.19\columnwidth]{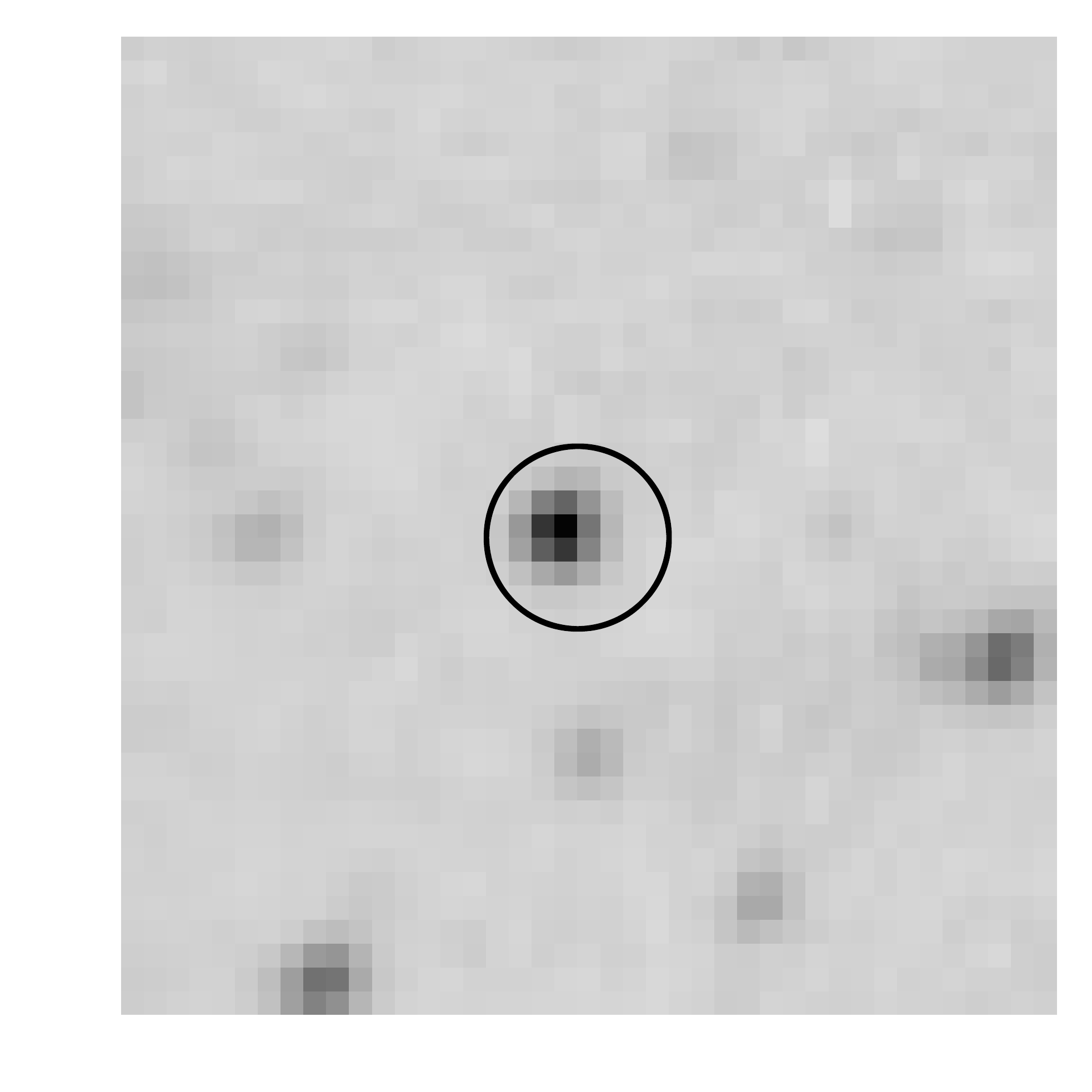}}
\subfigure[]{\includegraphics[width=0.19\columnwidth]{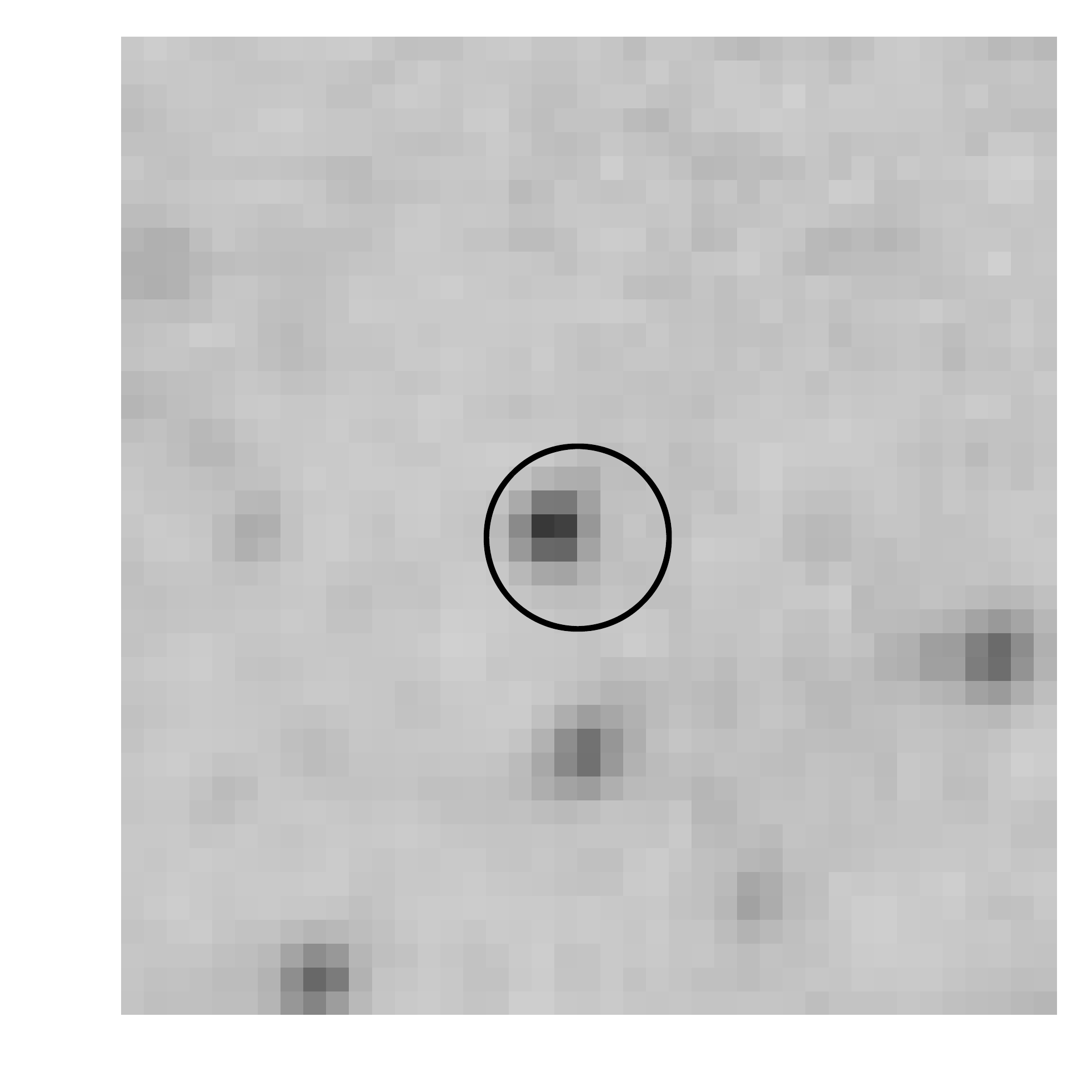}}
\subfigure[]{\includegraphics[width=0.19\columnwidth]{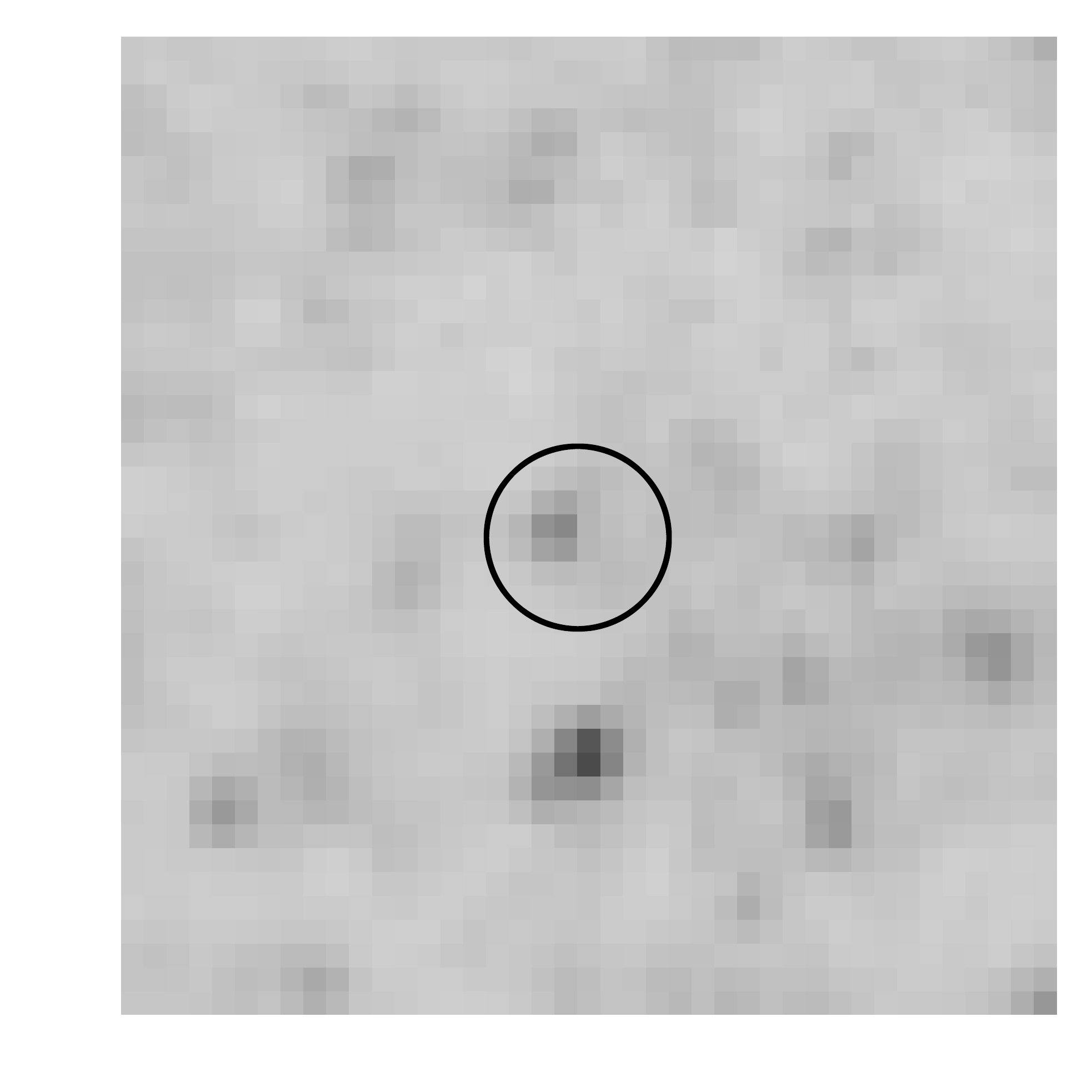}}
\subfigure[]{\includegraphics[width=0.19\columnwidth]{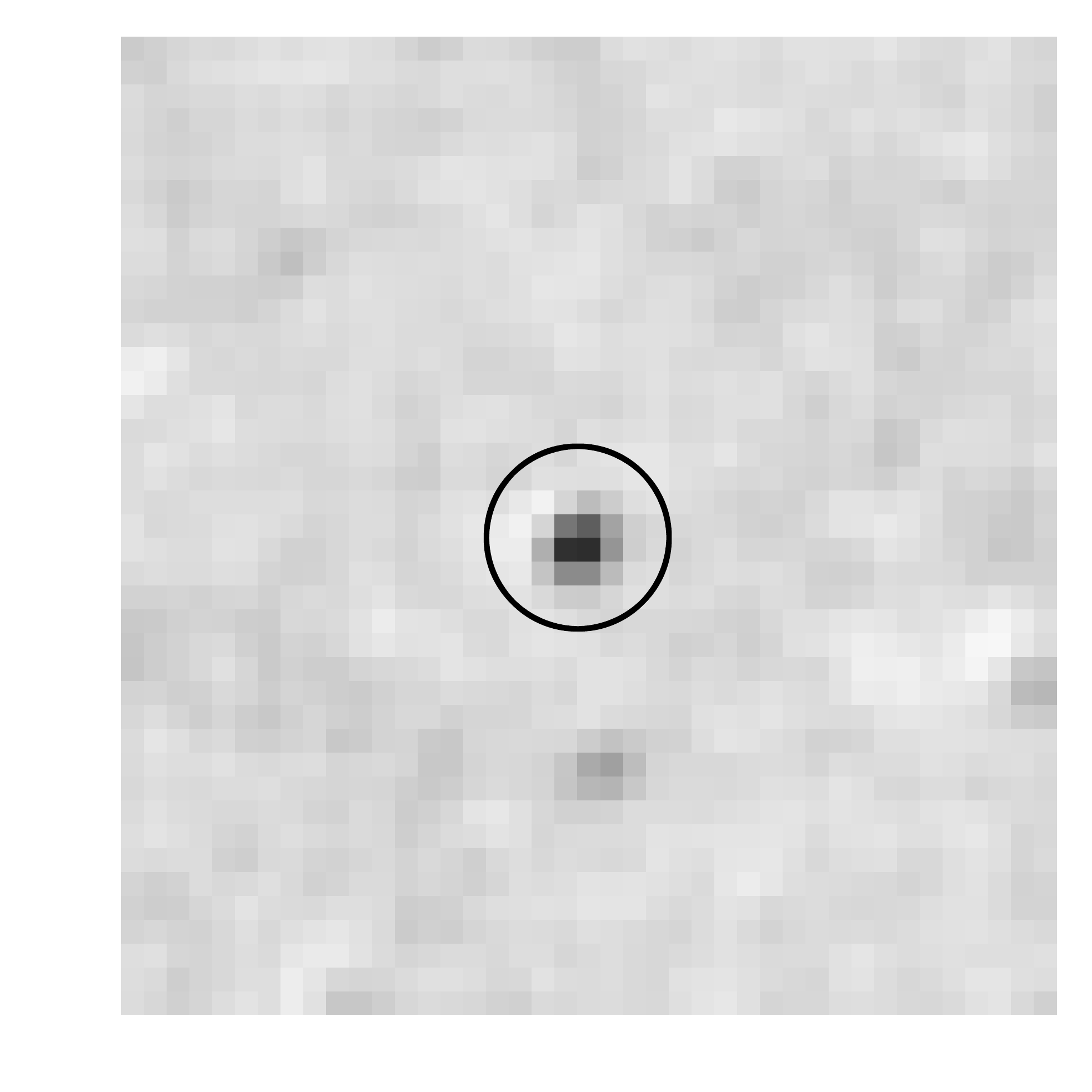}}

\caption{Source \# 56 in F469N, F435W, F555W, F814W and continuum
  subtracted F469N filters, respectively}

\end{figure}

\clearpage
\section[]{Data for Galactic Flux Templates}
\label{galactic}
\clearpage
\begin{table}
\caption{This appendix contains information regarding the Galactic WR stars
used to produce the flux templates used in this analysis, taken from
\citet{Rosslowe2015} and will be updated following the GAIA DR2
release.}
\begin{center}
\begin{tabular}{cccc}
ID & Subtype & Distance (kpc) & E(B-V) \\
\hline
\multicolumn{4}{l}{WN4-6 template} \\
\hline
WR1     & WN4b  & 2.3 & 1.09 \\
WR6     & WN4b  & 1.8 & 0.17 \\
WR7     & WN4b  & 5.5 & 0.69 \\
WR18   & WN4b  & 2.3 & 0.91 \\
WR134 & WN6b  & 1.9 & 0.50 \\
WR136 & WN6b  & 1.3 & 0.59 \\
\hline
\multicolumn{4}{l}{WN7-8 template} \\
\hline
WR12   & WN7  & 4.2 & 0.72 \\
WR66   & WN8  & 3.6 & 1.12 \\
\hline
\multicolumn{4}{l}{WC4-6} \\
\hline
WR144 & WC4  & 1.4 & 2.65 \\
WR111 & WC5  & 1.9 & 0.34 \\
WR114 & WC5  & 2.05 & 1.45 \\
WR23   & WC6  & 2.3 & 0.42 \\
WR154 & WC6  & 3.5 & 0.78 \\
\hline
\multicolumn{4}{l}{WC7-8 template} \\
\hline
WR14   & WC7  & 2.2 & 0.57 \\
WR68   & WC7  & 3.6 & 1.55 \\
WR135 & WC8  & 1.9 & 0.28 \\
WR53   & WC8d& 4.2 & 0.58 \\
\end{tabular}
\end{center}
\end{table}

\clearpage

\section[]{Spectra for each  WR source.}
\label{WR_spectra_templates}
This appendix contains the flux calibrated Gemini/GMOS spectra of each
WR source. Also shown is the best fitting Galactic WR template spectra from \citet{Rosslowe2015}.
\clearpage

\begin{figure}
\includegraphics[width=0.6\columnwidth, angle=-90]{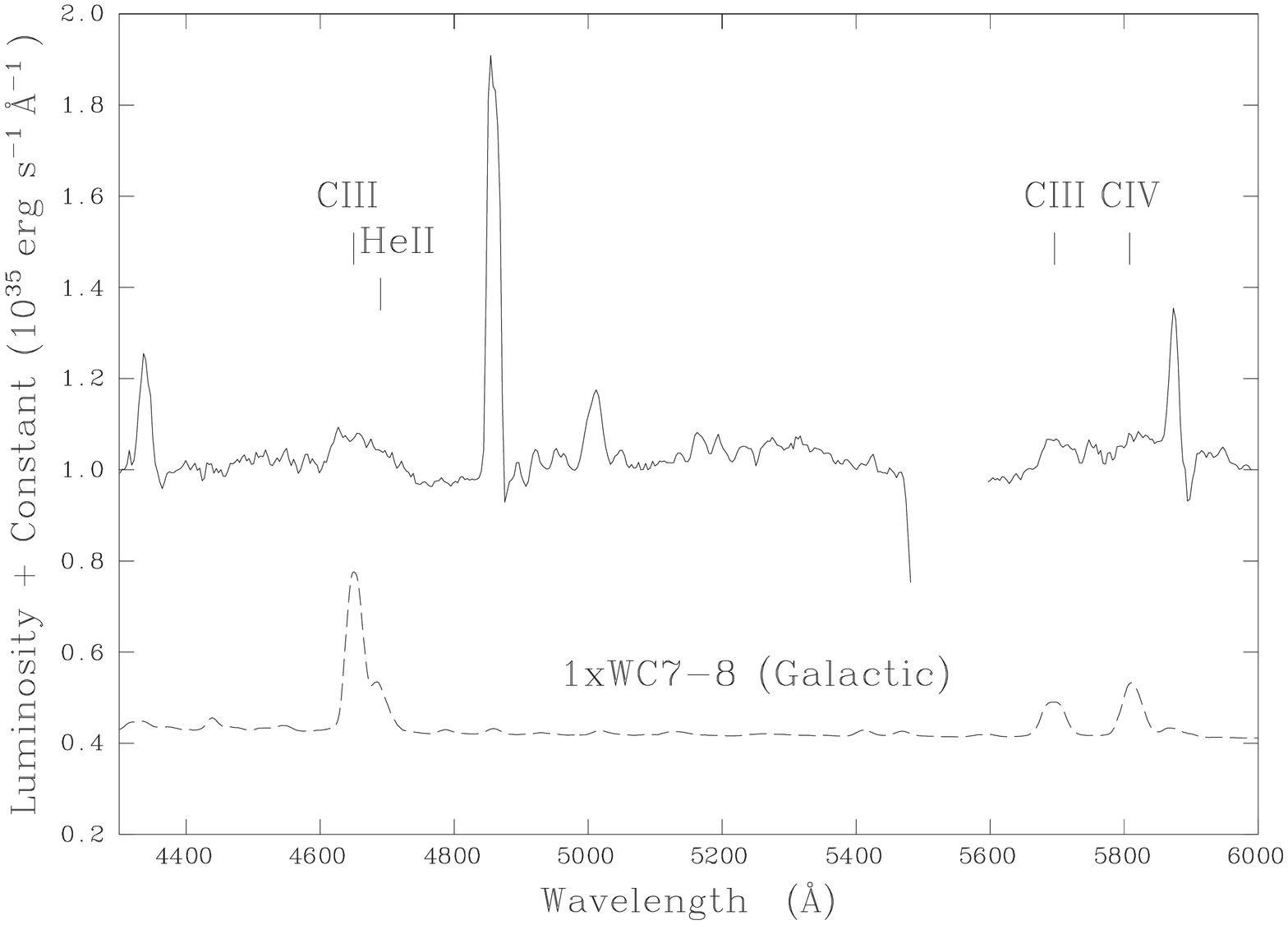}
\caption{Spectrum of Source \# 1016.}
\end{figure}

\begin{figure}
\includegraphics[width=0.6\columnwidth, angle=-90]{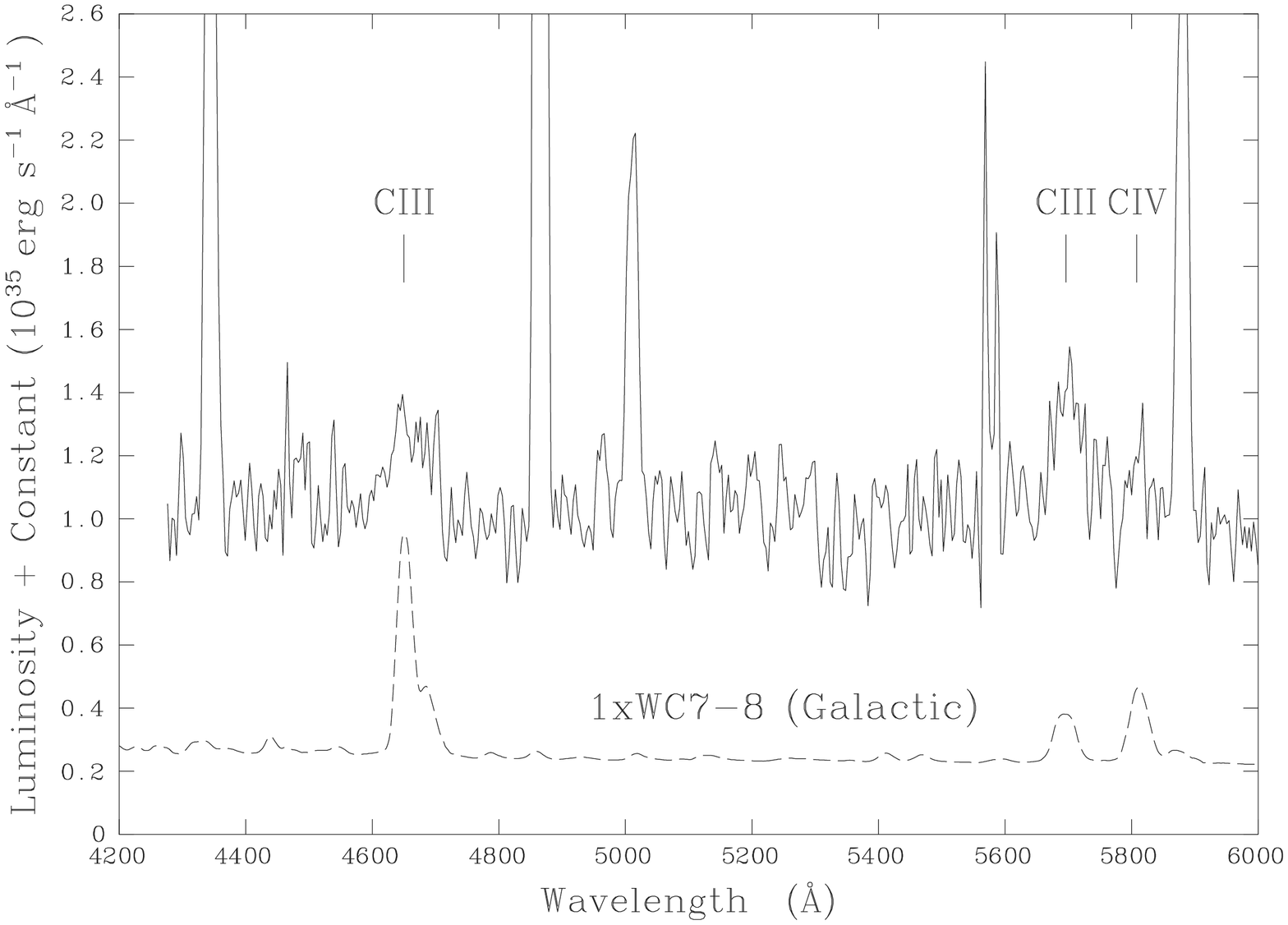}
\caption{Spectrum of Source \# 1024.}
\end{figure}

\begin{figure}
\includegraphics[width=0.6\columnwidth, angle=-90]{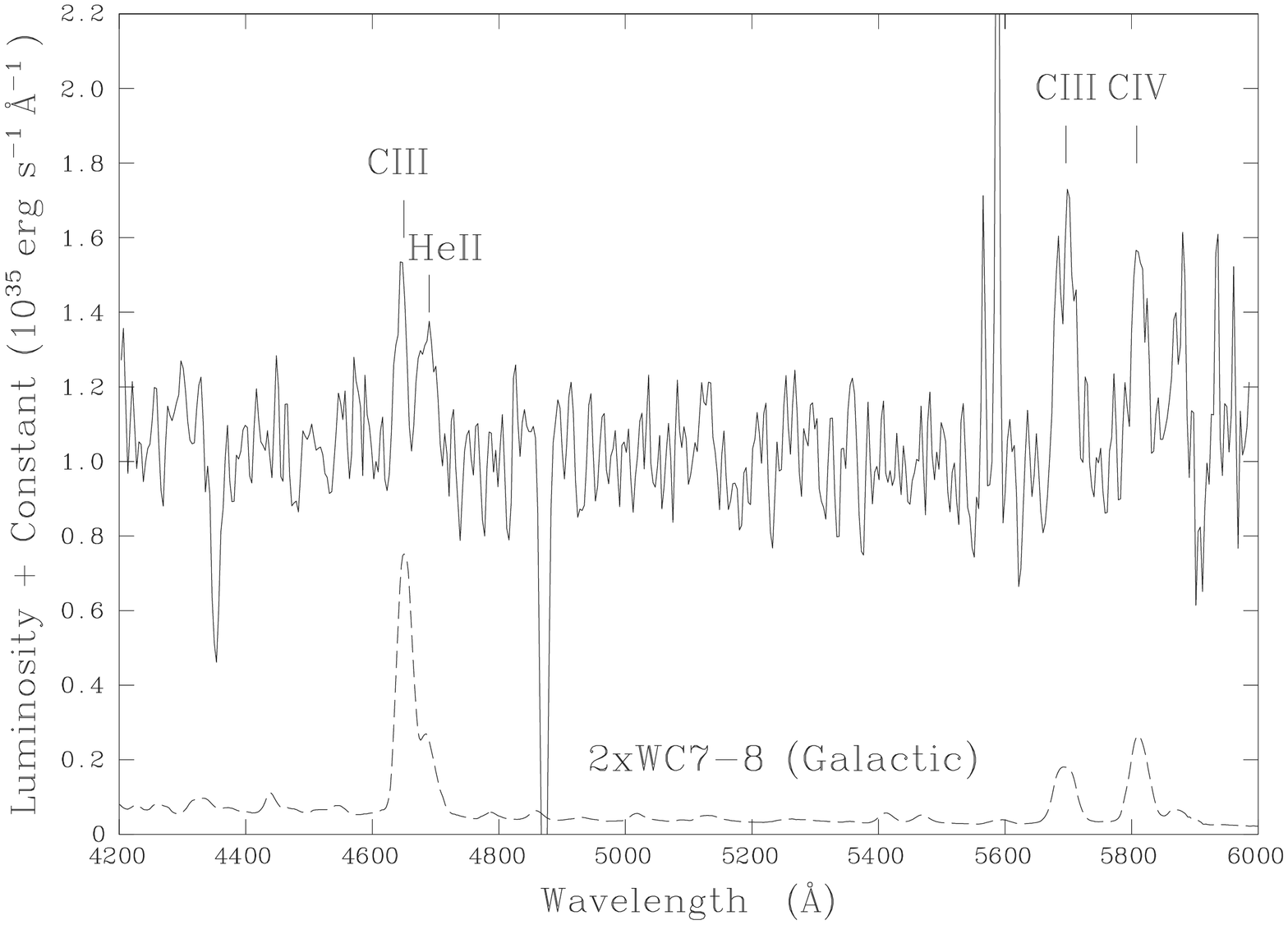}
\caption{Spectrum of Source \# 1030.}
\end{figure}


\begin{figure}
\includegraphics[width=0.6\columnwidth, angle=-90]{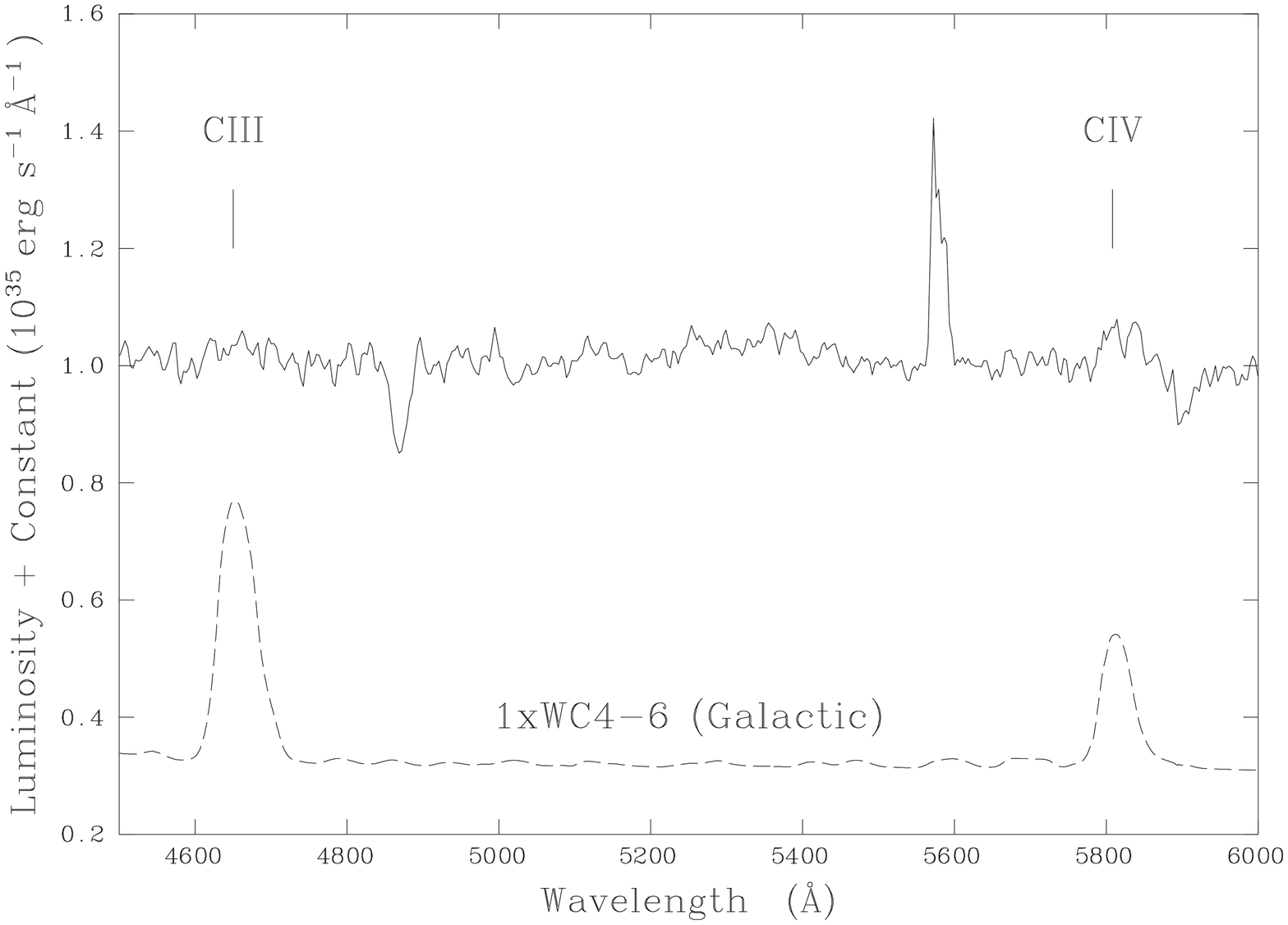}
\caption{Spectrum of Source \# 114.}
\end{figure}

\begin{figure}
\includegraphics[width=0.6\columnwidth, angle=-90]{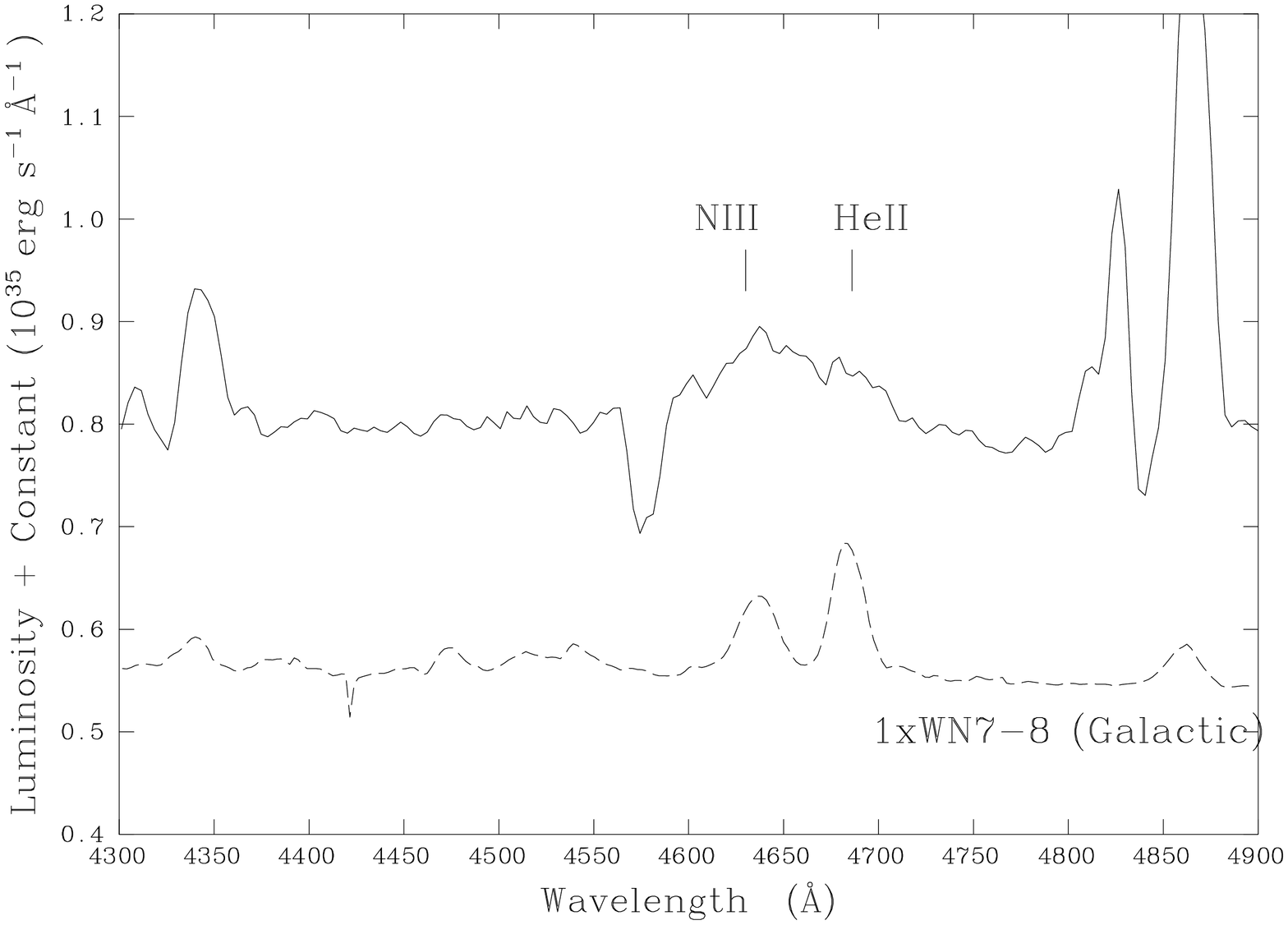}
\caption{Spectrum of Source \# 2053.}
\end{figure}


\begin{figure}
\includegraphics[width=0.6\columnwidth, angle=-90]{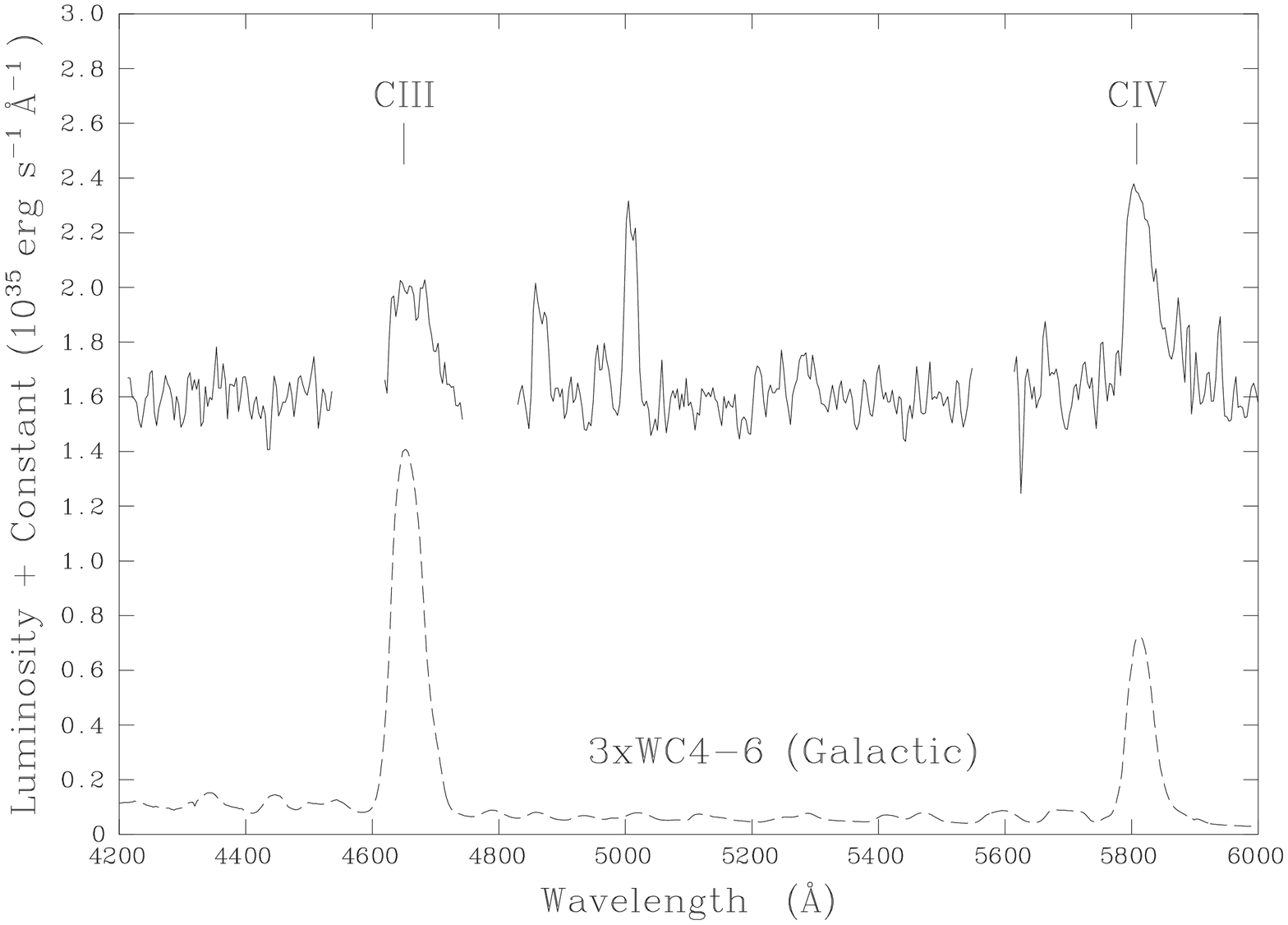}
\caption{Spectrum of Source \# 49.}
\end{figure}


\bsp

\label{lastpage}


\begin{thebibliography}{99}




\bibitem[Anderson et al.(2012)]{Anderson2012} Anderson, J.~P., Habergham, S.M., James, P.A, \& Hamuy, M.\ 2012, MNRAS, 424, 1372

\bibitem[Anderson et al.(2010)]{Anderson2010} Anderson, J.~P., Covarrubias, R.A., James, P.A., Hamuy, M., Habergham, S.M., \ 2010, MNRAS, 407, 2660





\bibitem[Bibby \& Crowther(2010)]{Bibby2010} Bibby, J.L., \& Crowther, P.A.\ 2010, MNRAS, 405, 2737

\bibitem[Bibby \& Crowther(2012)]{Bibby2012} Bibby, J.L., \& Crowther, P.A.\ 2012, MNRAS, 420, 3483

\bibitem[Bresolin(2007)]{Bresolin2007} Bresolin, F. \ 2007, ApJ, 656, 186




\bibitem[Cao et al.(2013)]{Cao2013} Cao, Y., Kasliwal, M.M., Arcavi,
  I., Horesh, A., Hancock, P., Valenti, S., Cenko, S.B., Kulkarni,
  S.R., et al.\ 2013, AJ, 775, L7

\bibitem[Cedr\'es \& Cepa(2002)]{Cedres2002} Cedr\'es, B., \& Cepa, J.\
  2002, A\&A, 391, 809

\bibitem[Cedr\'es, Urbaneja \& Cepa(2004)]{Cedres2004} Cedr\'es, B., Urbaneja, M.A.,~ \& Cepa, J.\
  2004, A\&A, 422, 514


\bibitem[Chomiuk \& Povich(2011)]{Chomiuk2011} Chomiuk, L., \& Povich,
  M.~S.\ 2011, AJ, 142, 197

\bibitem[Conti \& Vacca(1990)]{Conti1990} Conti, P.S., \& Vacca, W.D.\ 1990, AJ, 100, 431

\bibitem[Crockett et al.(2007)]{Crockett2007} Crockett, R.M., Smartt, S.J., Eldridge, J.J., Mattila, S., Young, D.R., Pastorello, A., Maund, J.R., Benn, C.R., \& Skillen, I.\ 2007, MNRAS, 381, 835


\bibitem[Crowther \& Hadfield (2006)]{Crowther2006} Crowther, P.~A., \& Hadfield, L.~J.\ 2006, A\&A, 449, 711


\bibitem[Crowther et al.(2010)]{Crowther2010} Crowther, P.~A.,
  Barnard, R., Carpano, S., et al.\ 2010, MNRAS, 403, L41

\bibitem[Crowther(2013)]{Crowther2013} Crowther, P.~A.\ 2013, MNRAS, 428, 1927 

\bibitem[Crowther(2015)]{Crowther2015a} Crowther, P.~A.\ 2015,
  Proceedings of an International Workshop help in Potsdam, Edited by
  W.R. Hamann, A. Sander \& H. Todt.






\bibitem[Drissen et al.(1999)]{Drissen1999} Drissen, L., Roy, J.-R., Moffat, A.~F.~J., \& Shara M.M.\ 1999, AJ, 117, 1249


\bibitem[Eldridge \& Vink(2006)]{Eldridge2006} Eldridge, J.~J., Vink, J.S. \ 2006, A\&A, 452, 295

\bibitem[Eldridge, Langer, Tout(2011)]{Eldridge2011} Eldridge, J.~J., Langer, N., \& Tout, C.A. \ 2011, MNRAS, 414, 3501

\bibitem[Eldridge et al.(2013)]{Eldridge2013} Eldridge, J.~J., Fraser, M., Smartt, S.J., Maund, J.R., \& Crockett, M.R., Submitted to MNRAS, arXiv:1301.1975 


\bibitem[Fruchter et al.(2006)]{Fruchter2006} Fruchter, A.~S., Levan, A.~J., Strolger, L., Vreeswikj, P.M., Thorsett, S.E., Bersier, D., Burud, I., Castro C$\acute{e}$ron, J.M., et al., \ 2006, Nature, 441, 463 

\bibitem[Galama et al.(1998)]{Galama1998} Galama, T.J., Vreeswijk, P.M., van Paradijs, J., Kouveliotou, C., Augusteijn, T., B$\ddot{o}$hnhardt, H., Brewer, J.P., Doublier, V., et al. \ 1998, Nature, 395, 670

\bibitem[Gies et al.(2003)]{Gies2003} Gies, D.R., Bolton, C.T.,
  Thomson, J.R., Huang, W., McSwain, M.V., Riddle, R.L., Wang, Z.,
  Wiita, P.J., et al. 2003, ApJ, 583, 424

\bibitem[Groh et al.(2013)]{Groh2013} Groh, J.H., Georgy, C., \&
  Ekstr{\"o}m, S.\ 2013, A\&A, 558, L1

\bibitem[Gvaramadze et al.(2012)]{Gvaramadze2012} Gvaramadze, V.V., Weidner, C., Kroupa, P., \& Pflamm-AAltenburg, J.\ 2012, MNRAS, 424, 3037

\bibitem[Hadfield et al.(2005)]{Hadfield2005} Hadfield, L.~J., Crowther, P.~A., Schild, H., \& Schmutz, W.\ 2005, A\&A, 439, 265 





\bibitem[Hodge et al.(1990)]{Hodge1990} Hodge, P.W., Gurwell, M.,
  Goldader, J.D., \& Kennicutt, R.C.Jr, 1990, ApJS, 73, 661

\bibitem[Howarth et al.(2004)]{Howarth2004} Howarth, I.D., Murray, J.,
  Mills, D., \& Berry, D.S., 2004, Starlink User Note 50, Starlink Project



\bibitem[Jarrett et al.(2013)]{Jarrett2013} Jarrett, T.~H., Masci, F.,
  Tsai, C.~W., Petty, S., Cluver, M.~E., Assef, R.~J., Benford, D.,
  Blain, A., et al.\ 2013, AJ, 145, 6

\bibitem[Karamehmetoglu et al.(2017)]{Karamehmetoglu2017}
  Karamehmetoglu, E., Taddia, F., Sollerman, J, Wyrzykowski, L.,
  Schmidl, S., Fraser, M., Fremling, et al., 2017, arXiv:1703.08222

\bibitem[Kelly et al.(2008)]{Kelly2008} Kelly, P.~L., Kirshner, R.~P., \& Pahre, M.\ 2008, ApJ, 687, 1201 


\bibitem[Kenney et al. (1991)]{Kenney1991} Kenney, J.D.P. \& Lord, S.D.,
  1991, ApJ, 381, 118

\bibitem[Kennicutt (1984)]{Kennicutt1984} Kennicutt, R.~C., Jr., 1984, ApJ, 287, 116


\bibitem[Kennicutt (1998)]{Kennicutt1998} Kennicutt, R.~C., Jr., 1998, ARA\&A, 36, 189

\bibitem[Kennicutt et al. (2008)]{Kennicutt2008} Kennicutt, R.~C.,
  Jr., Lee, J.C., Funes, S.J., Jos\'e , G., Sakai, S., \& Akiyama, S.,
  2008, ApJS, 178, 247

\bibitem[Knapen et al. (2004)]{Knapen2004} Knapen, J.H., Stedman, S.,
  Bramich, D.M., Folkes, S.F., \& Bradley, T.R., 2004, A\&A, 426, 1135


\bibitem[Lee et al.(2009)]{Lee2009} Lee, J.C., et al., 2009, \apj, 706, 599

\bibitem[Leitherer(1997)]{Leitherer1997} Leitherer, C.\ 1997, Revista Mexicana de Astronomia y Astrofisica Conference Series, 6, 114 


\bibitem[Leloudas et al.(2011)]{Leloudas2011} Leloudas, G., Gallazzi, A., Sollerman, J., Strizinger, M.D., Fynbo, J.P.U., Hjorth, J., Malasani, D., et al., \ 2011, A\&A, 530, 95

\bibitem[Leloudas et al.(2010)]{Leloudas2010} Leloudas, G., Sollerman, J., Levan, A.~J., et al.\ 2010, A\&A, 518, A29 



\bibitem[Levesque et al.(2010)]{Levesque2010} Levesque, E.M., Kewley, L.J., Graham, J.F., Fruchter, A.S. \ 2010, ApJ, 712, 26

\bibitem[Liu et al.(2013)]{Liu2013} Liu, J., Bregman, J.N., Bai, Y.,
  Justham, S. \& Crowther, P.A.\ 2013, Nature, 503, 500



\bibitem[Meynet \& Maeder(2005)]{Meynet2005} Meynet G., Maeder A., A\&A, 429, 581




\bibitem[Martins, Schaerer \& Hillier (2005)]{Martins2005} Martins, F., Schaerer, D., \& Hillier, D.J.,\ 2005, A\&A, 436, 1049


\bibitem[Massey \& Johnson(1998)]{Massey1998} Massey, P., \& Johnson, O.\ 1998, ApJ, 505, 793 


\bibitem[Mattila et al.(2010)]{Mattila2010} Mattila, S., Smartt, S.J., Maund, J., Benetti, S., \& Ergon, M.\ 2010, arXiv, 1011.5494



\bibitem[Modjaz et al.(2011)]{Modjaz2011} Modjaz, M., Kewley, L., Bloom, J.S., Filippenko, A.V., Perley, D., Silverman, J.M. \ 2011, ApJ, 731, 4

\bibitem[Modjaz et al.(2008)]{Modjaz2008} Modjaz, M., Kewley, L., Kirshner, R.P., Stanek, K.Z., Challis, P., Garnavich, P.M., Greene, J.E., Kelly, P.L., Prieto, J.L. \ 2008, AJ, 135, 1136


\bibitem[Moffat \& Shara(1983)]{Moffat1983} Moffat, A..J., \& Shara, M.M.\ 1983, ApJ, 273, 544
\bibitem[Neugent \& Massey(2011)]{Neugent2011} Neugent, K.~F., \& Massey, P.\ 2011, ApJ, 733, 123 


\bibitem[Oke(1990)]{Oke1990} Oke, J.~B.\ 1990, AJ, 99, 1621

\bibitem[Orosz et al.(2007)]{Orosz2007} Orosz, J.A., McClintock, J.E.,
  Narayan, R., Bailyn, C.D., Hartman, J.D., Macri, L., Liu, J., et
  al. 2007, Nature, 449, 872

\bibitem[Osterbrock \& Ferland, (2006)]{Osterbrock2006} Osterbrock,
  D.E., \& Ferland, G.J., 2006, ``Astrophysics of gaseous nebulae and
  active galactic nuclei'', 2nd.~ed. ~Sausalito, CA: University Science Books

\bibitem[Pettini \& Pagel, (2004)]{Pettini2004} Pettini, M., \& Pagel, B.E.J., 2004, MNRAS, 348, L59

\bibitem[Pilyugin \& Thuan, (2005)]{Pilyugin2005} Pilyugin, L.S., \& Thuan, T.X., 2005, ApJ, 631, 231

\bibitem[Prestwich et al.(2007)]{Prestwich2007} Prestwich, A.H.,
  Kilgard, R., Crowther, P.A., Carpano, S., Pollock, A.M.T., Zezas,
  A., Saar, S.H., Roberts, T.P., Ward, M.J, ApJ, 669, 21

\bibitem[Rosa \& Benvenuti(1994)]{Rosa1994} Rosa, M.R. \& 291, 1

\bibitem[Rosslowe \& Crowther(2015)]{Rosslowe2015} Rosslowe, C.~K., \&
  Crowther, P.~A.\ 2015, MNRAS, 447, 2322




\bibitem[Shappee \& Stanek(2011)]{Shappee11} Shappee, B.~J. \& Stanek, K.~Z.\ 2011, ApJ, 733, 124

\bibitem[Shara et al.(2013)]{Shara2013} Shara, M.~M., Bibby, J.~L.,
  Zurek, D., Crowther, P.~A., Moffat, A.~F.~J., Drissen, L., \ 2013,
  AJ, 146, 162

\bibitem[Shortridge et al.(2004)]{Shortridge2004} Shortridge, K.,
  Meyerdierks, H., Currie, M., Clayton, C., Lockley, J., Charles, A.,
  Davenhall, C., Taylor, M., et al. 2004, Starlink User Note 86,
  Starlink Project





\bibitem[Silverman \& Filippenko(2008)]{Silverman2008} Silverman, J.M., \& Filippenko, A.V.,
    ApJ, 678, 17

\bibitem[Smartt(2009)]{Smartt2009} Smartt, S.J.\ 2009, ARA\&A, 47, 63

\bibitem[Smith el al.(1990)]{Smith1990} Smith, L.F., Shara, M.M., \& Moffat, A.F.J., 1990, ApJ, 358, 229

\bibitem[Smith el al.(1996)]{Smith1996} Smith, L.F., Shara, M.M., \& Moffat, A.F.J., 1996, MNRAS, 281, 163

\bibitem[Smith et al.(2011)]{Smith2011} Smith, N., Li, W., Filippenko, A., Chornock, R.\ 2011, MNRAS, 412, 1522

\bibitem[Stanek et al.(2006)]{Stanek2006} Stanek, K.Z., Gnedin, O.Y., Beacom, J.F., Gould, A.P., Johnson, J.A., Kollmeier, J.A., Modjaz, M., Pinsonneault, M.H., et al., \ 2006, Acta Astronomica, 56, 333



\bibitem[Vacca(1994)]{Vacca1994} Vacca, W.~D. ApJ, 421, 140

\bibitem[Vacca \& Conti(1992)]{Vacca1992} Vacca, W.~D., \& Conti, P.~S., 1992, ApJ, 401, 543








\end{thebibliography}
\end{document}